\documentclass[twocolumn,aps,prl]{revtex4-2}
\usepackage{times}
\usepackage{amsmath,amsfonts,amssymb}
\usepackage{amsthm,mathrsfs}
\usepackage{soul,bm,array,graphicx,bbold,multirow}
\usepackage[normalem]{ulem}
\usepackage[makeroom]{cancel}
\usepackage[usenames,dvipsnames]{xcolor}
\usepackage[colorlinks=true,citecolor=blue,linkcolor=red]{hyperref}
\usepackage{pdfpages}
\usepackage{threeparttable}
\usepackage{makecell}
\usepackage{textcomp}
\usepackage{printlen}
\makeatletter
\AtBeginDocument{\let\LS@rot\@undefined}
\makeatother
\newcolumntype{x}[1]{>{\centering\let\newline\\\arraybackslash\hspace{0pt}}p{#1}}
\DeclareMathAlphabet{\mathbbold}{U}{bbold}{m}{n}
\newcounter{subeqn} %

\renewcommand{\Im}{\operatorname{Im}} 
\makeatletter
\@addtoreset{subeqn}{equation}
\makeatother

\renewcommand{\Im}{\operatorname{Im}}

\begin{document}
	
\title{Complex Frequency Fingerprint: Basic Concept and Theory}

\author{Juntao Huang$^1$}
\author{Kun Ding$^2$}
\author{Jiangping Hu$^{3,4,5}$}
\email[Corresponding author: ]{jphu@iphy.ac.cn}
\author{Zhesen Yang$^1$}
\email[Corresponding author: ]{yangzs@xmu.edu.cn}

\affiliation{$^1$ Department of Physics, Xiamen University, Xiamen 361005, Fujian Province, China}
\affiliation{$^2$ Department of Physics, State Key Laboratory of Surface Physics, and Key Laboratory of Micro and Nano Photonic Structures (Ministry of Education), Fudan University, Shanghai 200438, China}
\affiliation{$^3$ Beijing National Laboratory for Condensed Matter Physics and Institute of Physics, Chinese Academy of Sciences, Beijing 100190, China}
\affiliation{$^4$ School of Physical Sciences, University of Chinese Academy of Sciences, Beijing 100190, China}
\affiliation{$^5$ New Cornerstone Science Laboratory, Beijing, 100190, China}
	
\date{\today}
	
\begin{abstract}

We introduce the complex frequency fingerprint (CFF), an experimentally accessible method for detecting the complex frequency Green's function (GF). Unlike the real frequency GF, where $\omega$ is real, this complex frequency GF is shown to play a necessary role in both non-Hermitian and quantum many-body systems. For non-Hermitian systems, we will prove that our method detects complex energy spectra, eigenstates, and complex frequency GFs throughout the complex plane, providing necessary identification of the non-Hermitian skin effect. For quantum many-body systems, our method reveals quasiparticle peaks across the complex plane and intuitively illustrates interaction effects. This information is difficult to obtain with real frequency detection. Our method paves the way for exploring exotic phenomena in both non-Hermitian and quantum many-body systems, bridging theory and experiment across diverse physical areas.

\end{abstract}
	
\maketitle

{\color{red} \em Introduction.}---The Hermiticity of a Hamiltonian is a foundational assumption in quantum mechanics. 
However, when a system interacts with its external environment, an effective non-Hermitian description can naturally arise~\cite{Ashida,Rmp93015005,NH-PT,NHP-PT,EP-optics,Ozde,Dingkrev,re-nhse,NHTP-rev,Lin2023Topo,PrxUeda,Prxkawabata,FoaTorres}. 
Due to the non-orthogonality of eigenstates in a non-Hermitian Hamiltonian, a significant number of eigenstates can become localized at the boundary, giving rise to the non-Hermitian skin effect (NHSE)~\cite{Prl1210868,Prl121kunst,Prb971214,PrlYSW,Prl125126402,prlOkuma,PrlYM,PrlYZFH,llhcritical,helbig,xiao-wang,gdse,Prlborgnia,lchTR,Prl123170401,songfeiprl,Prl125186,prlgjb,prbhigherorder,secnhse,tpswitch,transferkunst,lyc,entangle-nhse,PrbKOM,gcx-cs,hym4,zhjprb,self-heal,Prl124066602,polari,Prbjh,PrbDY,Prr1023013,Prb102085151,Prbzxz}.  
This phenomenon has been experimentally studied across a variety of  physical platforms~\cite{Dingkrev,Lin2023Topo,re-nhse,Rmp93015005}, including photonic systems~\cite{scienceaaz8727,xiao-wang,NB-EP,xpanderson,Prl129113601,Prl130263801,2024NatPh,Prl133073803,Prl133070801,Prl132113802,Prl132063804,Prb110094308,Lpx1,vicencio2024}, acoustic systems~\cite{acous1,lzy,transient,zxj,yzj,Prb106134112,Prl133126601,Prb110L140305,zqc,szq-gh,zhong2024,wang2024disorder,hu2024}, cold atoms~\cite{Prl129070401,zhao2024}, electric circuits~\cite{Liu-Zhang,zoudy,helbig,Prr20232,Prr4033109,shang-tie,Prb107085426,osdbe,higherank,Prr5043034,Prb108035410,Prb107184108,jin2024} and mechanical systems~\cite{mgc1,active-elastic,ghatak,branden,cyy,mgc2,mgc3,sciadvadf7299}. 

Experimentally, most of the existing detection protocols for the NHSE can be broadly classified into two main approaches: (i) real-frequency Green's function (GF) measurements and (ii) wavefunction dynamics. 
In many experiments, nonreciprocal correlations and dynamics are observed and used as key indicators to identify the presence of NHSE.

This work addresses two fundamental questions. 
The first is: What fundamentally distinguishes the non-Hermitian skin effect from a conventional nonreciprocal system that does not exhibit the NHSE? 
Importantly, not all nonreciprocal systems have to display the non-Hermitian skin effect~\cite{FT1}. 
For instance, when subjected to uniform dissipation, a one-dimensional Hermitian system that breaks both time-reversal and inversion symmetries will belong to a dissipative nonreciprocal system without showing the NHSE~\cite{FT2}. Such systems,
as summarized in the third row of Tab.~\ref{T1} and detailed in Appendix A, also %are indistinguishable from those with NHSE when analyzed using real-frequency GF or wavefunction dynamics, i.e., both 
exhibit the nonreciprocal correlations and dynamics. Their real-frequency GF shows the same boundary insensitivity behaviors as those with the non-Hermitian skin effect~\cite{FT3}. %to the boundary conditions~\cite{FT3}.  
Therefore, detecting nonreciprocal dynamics and correlations alone is insufficient to identify the presence of the NHSE, although it allows us to conclude that such systems indeed belong to the nonreciprocal systems as shown in Tab.~\ref{T1}. 
However, the point is that the nonreciprocal systems do not necessarily exhibit the non-Hermitian skin effect.%in dissipative systems. 

This leads to our second question: Are there any intrinsic features unique to the NHSE, and if so, how can they be detected? 
Moreover, since the system is purely dissipative, identifying a steady-state response that does not decay to zero over time is also crucial. 
These questions present significant challenges for the non-Hermitian physics community.

\begin{figure}[t]
	\begin{center}
		\includegraphics[width=0.95\linewidth]{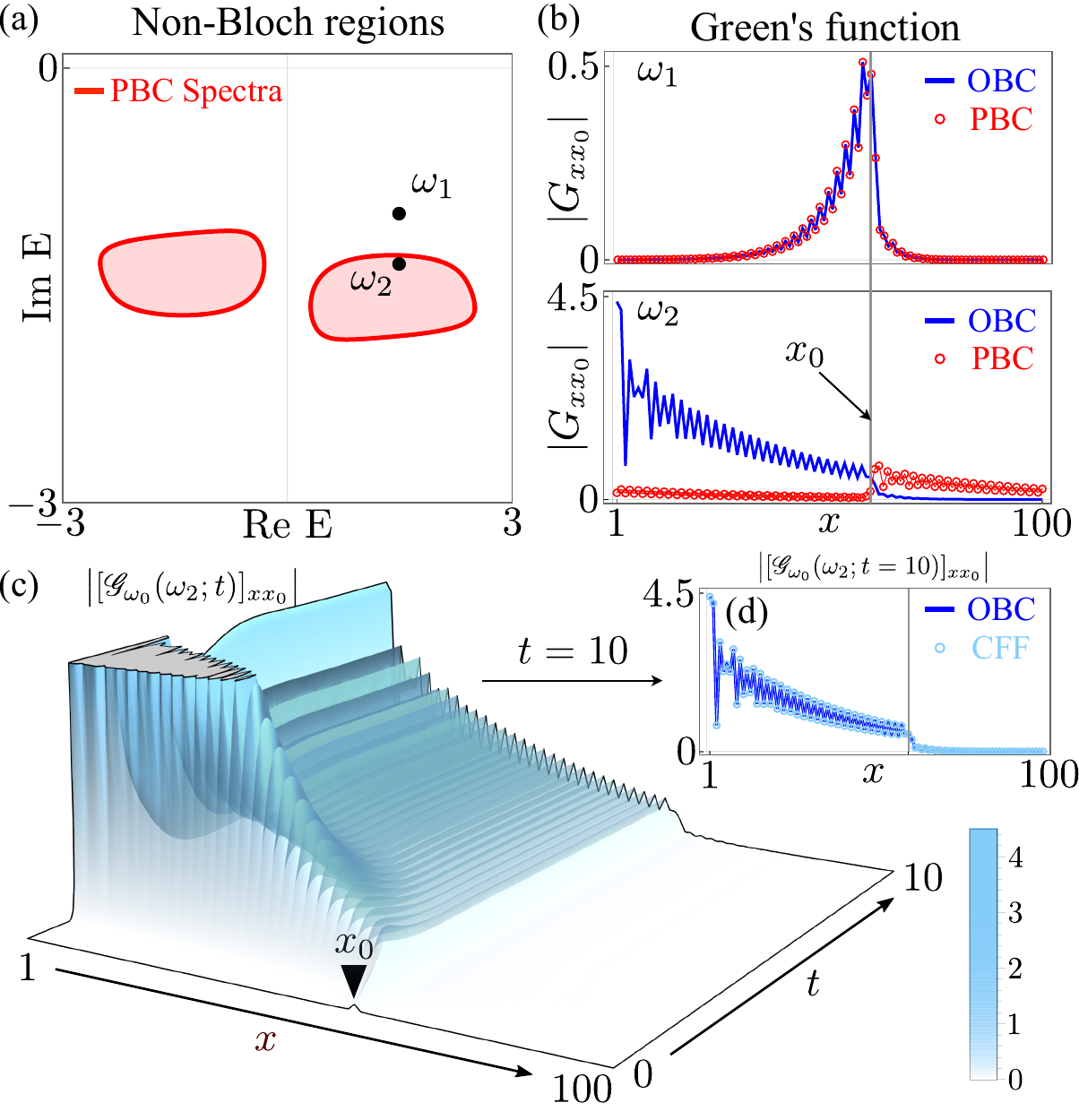}
		\par\end{center}
	\protect\caption{
		(a) Illustration of the region possessing the non-Bloch response in the complex plane, with $\omega_{1}=1.3-i$ and $\omega_{2}=1.3-1.35i$.
		(b) Comparison of $|G_{xx_{0}}(\omega_{c})|$ for frequencies with zero and nonzero spectral winding numbers. 
		(c) Time evolution of $|[\mathscr{G}_{\omega_{0}}(\omega_{c}=\omega_{2};t)]_{xx_{0}}|$. (d) Convergence of $|\mathscr{G}_{\omega_{0}}(\omega_{2};t=10)|_{xx_{0}}$ to the complex frequency GF.
		Here the model is based on Eq.~(\ref{E2}) with $N=100$, $t_{1}=1.5,t_{2}=1,\mu=0.3,\lambda=-1,\gamma_{1}=2,$ and $\gamma_{2}=1$. } 
	\label{F1}
\end{figure}

In this work, we demonstrate that complex-frequency GFs with nonzero spectral winding numbers, e.g., the red regions in Fig.~\ref{F1}(a), yield unique physical responses induced by the NHSE.
To detect these responses, we propose an innovative theoretical tool: the complex frequency fingerprint (CFF). 
This method provides a steady-state detection scheme for characterizing not only the complex-frequency GF but also the non-Hermitian spectra and the corresponding non-Hermitian skin modes. 
Therefore, the CFF enables a powerful and reliable approach to identify the presence or absence of the NHSE, further leading us to the study of the novel physical responses associated with the complex frequency domain.

\begin{table}[b]
\begin{threeparttable}
\centering
\caption{Diagnostic signatures for resolving NHSE ambiguity in dissipative systems.}
\setlength{\arrayrulewidth}{1pt}
\renewcommand{\arraystretch}{1.9} 
\setlength{\tabcolsep}{3pt}
\begin{tabular}{c|c|c|c|c}
\hline %\toprule
\multicolumn{2}{c|}{}& \makecell{Non-\\ reciprocity}&\makecell{Boundary\\ sensitivity}& \makecell{Non-Bloch\\ response} \\
\hline%\midrule
\multicolumn{2}{c|}{Reciprocal systems}&\raisebox{0.2ex}{\scalebox{0.85}[1]{$\times^{\dag}$}}&\raisebox{0.2ex}{\scalebox{0.85}[1]{$\times$}}&\raisebox{0.2ex}{\scalebox{0.85}[1]{$\times$}}\tnote{*}\\
\hline
\multirow{2.4}{*}{\makecell{Nonreciprocal\\ systems}}&\makecell{without\\ NHSE}& \raisebox{0.6ex}{\scalebox{0.7}{$\sqrt{}^{\dag}$}}&\raisebox{0.2ex}{\scalebox{0.85}[1]{$\times$}}&\raisebox{0.2ex}{\scalebox{0.85}[1]{$\times$}}\tnote{*}\\
\cline{2-5}
&\makecell{with\\ NHSE} &\raisebox{0.6ex}{\scalebox{0.7}{$\sqrt{}^{\dag}$}}&\raisebox{0.2ex}{\scalebox{0.85}[1]{$\times$}}&\raisebox{0.6ex}{\scalebox{0.7}{$\sqrt{}$}}\tnote{*}\\
\hline
\end{tabular}
\label{T1}
\begin{tablenotes}
\item[$\dag$]{denotes existence of nonreciprocal response in real frequency GF or nonreciprocal dynamics in the wavefunction.}
\item[*]{denotes existence of non-Bloch response in complex plane.}
\end{tablenotes}
\end{threeparttable}
\end{table}

Beyond non-Hermitian systems, we find that the complex frequency fingerprint is a highly universal concept applicable to general quantum many-body systems. 
As a first application, the CFF reveals the concept of quasiparticle resolution in the full complex frequency plane, exposing resonance peaks that are otherwise obscured in the real-frequency domain.
As a second application, the CFF also offers an intuitive single-particle perspective to study many-body interactions.

In summary, the CFF serves as a unique single-particle fingerprint of the quantum many-body system. Importantly, we will show its full experimental accessibility in current classical wave experiments. This framework provides a necessary tool, both theoretically and experimentally, for investigating and understanding our quantum many-body systems, as well as the dissipative non-Hermitian systems. 

{\em \color{red}Bloch and non-Bloch responses.}---In order to characterize the NHSE, we first clarify the concept of Bloch and non-Bloch responses. 
In this work, we define a physical response of an OBC Hamiltonian as a Bloch response if it is determined by the Bloch Hamiltonian. In contrast, we classify it as a non-Bloch response if it is determined by the non-Bloch Hamiltonian.
To illustrate this distinction, consider the GF as an example. 
As shown in Fig.~\ref{F1}(a), when the complex frequency, say,  $\omega_c \in \mathbb{C}$, carries a nonzero spectral winding number, the corresponding OBC Green's function $G^{\text{OBC}}(\omega_c)$~\cite{FT4} will fall into the non-Bloch response paradigm, say, red regions in Fig.~\ref{F1}(a).
Otherwise, it is classified as a Bloch response paradigm, that is, the white regions in Fig.~\ref{F1}(a). 
As shown in Fig.~\ref{F1}(b), two representative GFs are demonstrated, i.e., $\omega_{1}$ for the Bloch response and $\omega_{2}$ for the non-Bloch response.
One can summarize two hallmark features of non-Bloch responses:  
(i) a divergent behavior of the Green's function along at least one direction in real space, and  
(ii) a boundary sensitivity in the bulk, i.e., $|G_{xx_0}^{\mathrm{OBC}}(\omega_{c})|\neq |G_{xx_0}^{\mathrm{PBC}}(\omega_{c})|$. 
We will show that this divergence in the OBC GF has a one-to-one correspondence with the nontrivial point gap topology in the complex spectrum, and consequently, with the NHSE~\cite{SM}. 

In summary, as shown in Tab.~\ref{T1}, only nonreciprocal systems with NHSE possess non-Bloch response regions, making the non-Bloch response a unique and experimentally identifiable feature of the NHSE.

Based on the above discussion, we can prove the following no-go theorem:  
In any purely dissipative system, the real-frequency OBC GF must always belong to the Bloch response region.
This follows from the fact that, in dissipative systems, all poles of the OBC Green's function lie below the real axis in the complex plane, as exemplified in Fig.~\ref{F1}(a).
As a result, the spectral winding number $\nu(\omega_r)$~\cite{SM} vanishes for any real frequency $\omega_r \in \mathbb{R}$, implying that the system remains in the Bloch response region. 
Consequently, to detect non-Bloch responses, it is essential to probe the GF at complex frequencies.
This motivates the development of a novel detection framework capable of accessing the complex-frequency domain.

{\color{red}\em Driven-dissipative quantum system.}---To investigate the non-Bloch response in realistic systems, we simulate our system as a driven-dissipative system, which obeys the following inhomogeneous non-Hermitian Schr\"{o}dinger equation~\cite{SM}
\begin{equation}\begin{aligned}
		i\frac{d \langle \hat{\boldsymbol{a}}(t)\rangle}{dt}&=H_{\rm nH}\langle \hat{\boldsymbol{a}}(t)\rangle +\boldsymbol{F}(t),
		\label{Speq}
\end{aligned}\end{equation}
where $H_{\rm nH}$ in the momentum space is given by~\cite{Prl125186}:
\begin{equation}\begin{aligned}
	H_{\rm nH}(k)=&(t_{1}+t_{2}\mathrm{cos}k)\sigma_{x}+t_{2}\mathrm{sin}k\sigma_{y}+(\lambda \mathrm{sin}k+\mu)\sigma_{z}\\
	&
	-i(\gamma_{0}\sigma_{0}+\gamma_z\sigma_{z})/2.
	\label{E2}
\end{aligned}\end{equation}
Here, $\gamma_1 = (\gamma_0 + \gamma_z)/2$ and $\gamma_2 = (\gamma_0 - \gamma_z)/2$ represent the on-site dissipation at the $A$ and $B$ sublattices, respectively. 
$\bm{F}(t) = \{F_1(t), \dots, F_N(t)\}^T$ denotes the external drive, and $\langle \hat{\bm a}(t) \rangle = \{\mathrm{Tr}[\hat{a}_1 \hat{\rho}(t)], \dots, \mathrm{Tr}[\hat{a}_N \hat{\rho}(t)]\}^T$ is the field operator expectation value, with $\hat{\rho}(t)$ as the density matrix. 
Here, $N$ denotes the total number of lattice sites, including any internal degrees of freedom, ordered as $(1A,1B,2A,2B,\dots).$
When $\lambda \neq 0$ and $\gamma_1 = \gamma_2$, the system is nonreciprocal but does not exhibit the NHSE.
In contrast, when $\gamma_1 \neq \gamma_2$, the NHSE emerges, as demonstrated in Fig.~\ref{F1}(a) with red lines representing its PBC spectra. 
Fig.~\ref{F1}(b) further illustrates the OBC and PBC GFs, i.e., $|G_{xx_0}^{\text{OBC}}(\omega_c)|$ and $|G_{xx_0}^{\text{PBC}}(\omega_c)|$ as $x$ evolves from $1$ to $N=100$ with $\omega_c=\omega_1,\omega_2$ and $x_0=60$ which represents the $B$ sublattice in the $30$-th unit cell. 

Notably, (i) Eq.~(\ref{Speq}) can be derived from the Lindblad quantum master equation exactly~\cite{SM};
(ii) Eq.~(\ref{Speq}) is also equivalent to the dynamic equation in classical wave systems~\cite{Haus,Huang94,Li10,Fan03,Fan04}, enabling the application of all conclusions from this work to classical wave systems;
(iii) In experiments, % an external drive $\bm{F}(t)$ is applied to the system, and the resulting response $\langle \hat{\bm a}(t) \rangle$ is measured. 
both the amplitude and phase of $\langle \hat{\bm a}(t) \rangle$ can be experimentally accessed using techniques demonstrated in various platforms~\cite{scienceaaz8727,xiao-wang,2024NatPh,xpanderson,Prl133073803,Prl133070801,Prl132113802,Prl132063804,Prl130263801,Prl129113601,Prb110094308,Lpx1,vicencio2024,acous1,transient,Prb106134112,zxj,lzy,Prl133126601,Prb110L140305,zqc,szq-gh,zhong2024,wang2024disorder,hu2024,yzj,helbig,Prr20232,Liu-Zhang,zoudy,Prr4033109,shang-tie,Prb107085426,osdbe,higherank,Prr5043034,Prb108035410,Prb107184108,jin2024,ghatak,branden,cyy,mgc1,mgc2,mgc3,sciadvadf7299}. 

{\color{red} \em Complex frequency fingerprint.}---We now introduce the central concept of this work: the CFF. 
This framework provides an experimentally accessible, steady-state method to detect the complex-frequency GF in driven-dissipative systems, enabling direct observation of non-Bloch responses and other non-Hermitian spectral or eigenstate features.
The CFF is constructed from time-domain response measurements under harmonic driving. 
The procedure involves the following steps.

{\em Step 1:} A harmonic external drive is applied at site $i=1$, defined as:
\begin{equation}
	\boldsymbol{F}(t)=\theta(t)e^{-i\omega_0 t}\{F_{0},0,\dots,0\}^T,
\end{equation}
where $F_0$ is the driving amplitude, $\omega_0\in\mathbb{R}$ is the driving frequency, and $\theta(t)$ is the step function.
The response at each site $j=1,\dots,N$ is measured as:
\begin{equation}
	\delta\langle\hat{a}_{j}(t) \rangle_{F_1}=\langle\hat{a}_{j}(t) \rangle_{F_1}-\langle\hat{a}_{j}(t) \rangle_{F_1=0}.
\end{equation} 
The first column of the response matrix is then defined as:
\begin{equation}
	[\chi_{\omega_0}(t)]_{j1}=\frac{\delta\langle\hat{a}_{j}(t) \rangle_{F_1}}{F_0e^{-i\omega_0 t}},~~~~j=1,\dots,N.
\end{equation}

{\em Step 2:} Repeat the above procedure by applying the drive sequentially to each site $i=2,\dots, N$, i.e., 
\begin{equation}
	\boldsymbol{F}(t)=\theta(t)e^{-i\omega_0 t}\{0,\dots,F_{0},\dots,0\}^T,
\end{equation}
obtaining the $i$-th column of the response matrix $\chi_{\omega_0}(t)$. After completing this process for all sites, the full response matrix is constructed, which satisfies $\delta\langle\hat{\boldsymbol{a}}(t)\rangle=\chi_{\omega_{0}}(t)\boldsymbol{F}(t)$.

{\em Step 3:} With the full matrix $\chi_{\omega_0}(t)$, we define the CFF as:
\begin{equation}
	\mathscr{G}_{\omega_{0}}(\omega_c \in \mathbb{C};t)=\frac{1}{(\omega_c-\omega_{0})+[\chi_{\omega_{0}}(t)]^{-1}},
\end{equation}
where $\omega_c \in \mathbb{C}$ is an auxiliary complex frequency parameter, and $[\chi_{\omega_{0}}(t)]^{-1}$ represents the inverse of the response matrix. 

{\color{red} \em Application to non-Hermitian systems I: non-Bloch response.}---We now demonstrate how the CFF detects the complex frequency GF under OBC in a steady-state manner, i.e., 
\begin{equation}\begin{aligned}
		\lim_{t\rightarrow\infty}\mathscr{G}_{\omega_{0}}(\omega_c \in \mathbb{C};t)&=G(\omega_c\in \mathbb{C})=\frac{1}{\omega_c-H_{\rm nH}}. 
		\label{E8}
\end{aligned}\end{equation}
To verify this numerically, we compute the time evolution of $|[\mathscr{G}_{\omega_0}(\omega_c; t)]_{xx_0}|$ for a specific complex frequency $\omega_c = \omega_2$ with a nonzero spectral winding number, driving frequency $\omega_0 = 0$, and fixed input site $x_0 = 60$. 
The results, shown in Fig.~\ref{F1}(c) and \ref{F1}(d), reveal that as time increases, the CFF converges to the complex-frequency GF %, $G(\omega_2)$, 
as confirmed by the comparison in Fig.~\ref{F1}(d). 
Importantly, since $\omega_2$ lies within a spectral region with a nonzero winding number, a divergent behavior of $|[\mathscr{G}_{\omega_0}(\omega_c; t)]_{xx_0}|$ is also observed in the short-time dynamics. 

The mechanism behind the convergence in Eq.~(\ref{E8}) can be understood as follows: it can be proved that the time-dependent response function takes the form $\chi_{\omega_0}(t)=G(\omega_{0})-G(\omega_{0})e^{-i(H_{\rm nH}-\omega_{0})t}$~\cite{SM}.
In a dissipative system, the second term vanishes as $t\rightarrow \infty$ due to the negative imaginary components of the eigenvalues of $H_{\text{nH}}$. As a result, the steady-state contribution $[G(\omega_{0})]_{ji}$ becomes dominant, ensuring  the validity of Eq.~(\ref{E8}), which is independent of $\omega_{0}$.

\begin{figure}[b]
\centering
		\includegraphics[width=0.95\linewidth]{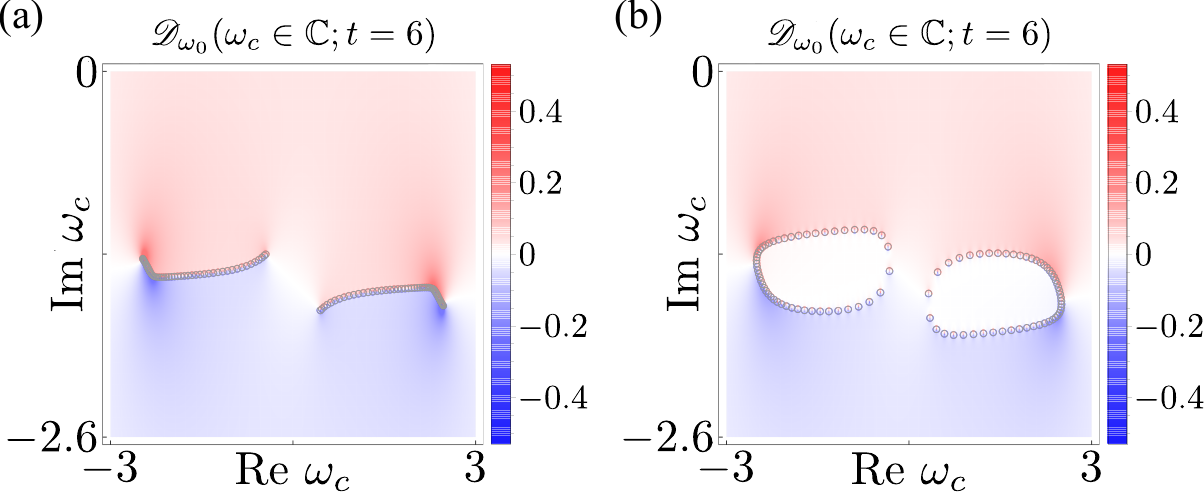}
\caption{The OBC (a) and PBC (b) complex frequency DOS of Eq.~(\ref{E2}), respectively.
The parameters are identical to those in Fig.~\ref{F1}. } 
	\label{F2}
\end{figure}

\begin{figure*}[t]
	\begin{center}
		\includegraphics[width=0.95\textwidth]{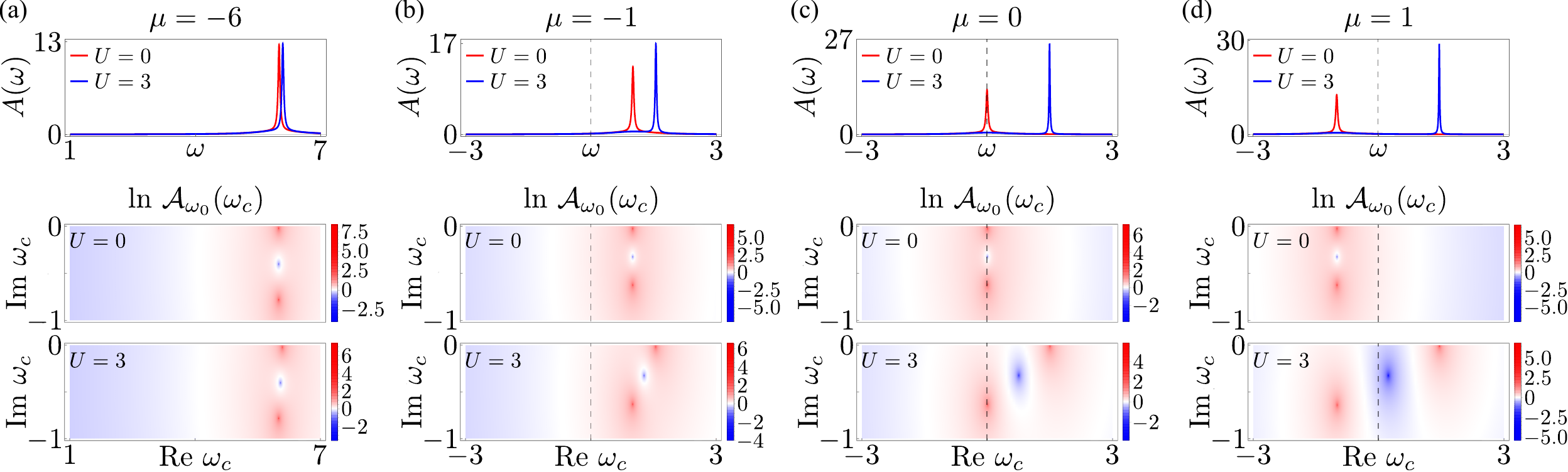}
		\par\end{center}
	\protect\caption{
		(a)-(d) Comparison of real-frequency spectral function $A(\omega\in\mathbb{R})$ with complex-frequency spectral log-modulus $\ln~\mathcal{A}_{\omega_{0}}(\omega_{c})$ as chemical potential varies. Dashed lines in (b)-(d) denote the Fermi level. Here, $t_{b}=5,V=0.1,t_{\downarrow}=\sqrt{2\pi},\eta=0.01$, and $T=0$ K. } 
	\label{F3}
\end{figure*}

{\color{red} \em Application to non-Hermitian systems II: non-Hermitian spectra and eigenstates.}---Beyond detecting non-Bloch responses, the CFF also enables direct access to the non-Hermitian spectrum and eigenstates.
We define the complex-frequency density of states (DOS) via the CFF as:
\begin{equation}\begin{aligned}
		&\mathscr{D}_{\omega_0} (\omega_c \in \mathbb{C};t)
		=-\frac{1}{N\pi}\Im {\rm Tr}~\mathscr{G}_{\omega_0}(\omega_c \in \mathbb{C};t), 
	\end{aligned}\label{DOS-Im}
\end{equation}
which converges in the long-time limit to the conventional complex-frequency DOS of the non-Hermitian Hamiltonian~\cite{Prl125186,Koziiprb,Prldmft,PrbKaneshiro,cw1,Prlshaokai}:
\begin{equation}\begin{aligned}
		\lim_{t\rightarrow\infty}\mathscr{D}_{\omega_0} (\omega_c \in \mathbb{C};t)&=D(\omega_c)=-\frac{1}{N\pi}\Im {\rm Tr}~G(\omega_c).
		\label{DOS-Im2}
\end{aligned}\end{equation}
As $\omega_c$ approaches an eigenvalue $E_n$, the complex-frequency DOS $D(\omega_c)$ diverges, allowing us to locate the eigenvalues directly in the complex plane. 
Fig.~\ref{F2} shows the comparison between the $\mathscr{D}_{\omega_0} (\omega_{c} \in \mathbb{C};t)$ under OBC and PBC with $\omega_{0}=0$. 
One can find that they can indeed reveal the spectral difference for the NHSE. 
Here, the circles in Fig.~\ref{F2}(a) and \ref{F2}(b) represent the OBC and PBC spectra, respectively. 

Furthermore, the CFF can also reveal the spatial profile of non-Hermitian eigenstates. As shown in Appendix B,
near $\omega_c \approx E_n$, matrix elements such as $[\mathscr{G}_{\omega_0}(\omega_c; t \to \infty)]_{xx_0}$ and $[\mathscr{G}_{\omega_0}(\omega_c; t \to \infty)]_{x_0x}$ are proportional to the components of the right and left eigenstates, respectively. 
This enables the experimental reconstruction of eigenmode localization, including the non-Hermitian skin modes.
In Appendix B, we will use the above method to detect the strong geometry-dependent behaviors of the two-dimensional (2D) NHSE. Additionally, this method can also be used to detect the point gap bound state~\cite{Prb108165}, as detailed in Supplementary Material~\cite{SM}.

In summary, we have proved that the CFF not only provides a necessary tool to identify the presence of non-Bloch response but also can be applied to detect the spectra and eigenstates of the non-Hermitian systems. Remarkably, this persists even when the corresponding non-Hermitian eigenvalues are complex numbers located far away from the real axis.

{\color{red} \em CFF in the quantum many-body system.}---In general (Hermitian) quantum many-body systems, the single particle retarded GF under OBC takes the following form: 
\begin{equation}
	G^R_S(\omega)=\frac{1}{(\omega+i\eta)-H_S-\Sigma_S(\omega+i\eta)}.\label{E11}
\end{equation}
Here, $H_S$ is the non-interacting Hamiltonian of the system written in the first quantized form, $\eta=0^+$, and $\Sigma_S(\omega)$ is the self-energy arising from the coupling to the external bath or many-body interaction~\cite{Prl125186,Koziiprb,Prldmft,PrbKaneshiro,cw1,Prlshaokai}. 
Notably, Eq.~(\ref{E11}) is defined at equilibrium and is exact without approximations, although the self-energy is hard to calculate for interacting systems. 

From the above retarded GF, one can define the following frequency-dependent non-Hermitian Hamiltonian, 
\begin{equation}
	H_S^{\text{eff}}(\omega)=(H_S-i\eta)+\Sigma_S(\omega+i\eta). 
\end{equation}
In this case, the CFF can be applied to detect the following double frequency GF (the detection procedure is discussed in Supplementary Material~\cite{SM}), 
\begin{equation}
	G^{\rm CFF}_S(\omega_c\in \mathbb{C},\omega_0)=\frac{1}{\omega_c-H_S^{\text{eff}}(\omega_0)}. 
		\label{E13}
\end{equation}
Note that for any given $\omega_0\in\mathbb{R}$, $H_S^{\text{eff}}(\omega_0)$ is in general non-Hermitian and $\omega_0$-dependent. Therefore, Eq.~(\ref{E13}) is nothing but the complex frequency Green's function of the non-Hermitian Hamiltonian $H_S^{\text{eff}}(\omega_{0})$.
Since Eq.~(\ref{E11}) is exact even for interacting systems, the $H_S^{\text{eff}}(\omega)$ contains all the information of the single particle excitation information.
As a result, the double frequency GF or CFF in Eq.~(\ref{E13}) provides a unique fingerprint for the quantum many-body system containing complete quasi-particle information from a single-particle perspective, and surpassing the real-frequency measurements.

Notably, unlike conventional pole descriptions of the GF, our CFF approach remains applicable even in cases where a quasiparticle description is absent, such as when the quasiparticle weight $Z\rightarrow 0$. 
Furthermore, as will be discussed in the following content, the CFF surpasses the real-frequency paradigm by encompassing all information in the entire complex frequency plane. This makes it an indispensable tool for understanding quantum many-body systems. 
Since the double frequency GF in Eq.~(\ref{E13}) is a physical observable, we will explore its properties in the following example.

{\color{red} \em Application to quantum many-body systems: quasiparticle resolution and interacting analysis.}---To showcase the powerful role of CFF in quantum many-body systems, we consider a two-level fermionic interacting system coupled to a thermal bath: $	\hat{H}_{sys}=\hat{H}_{S}+\hat{H}_{B}+\hat{H}_{S\text{-}B}$, where
\begin{equation}\begin{aligned}
		\hat{H}_S=&\begin{pmatrix}
			\hat{d}^\dag_{\uparrow} &\hat{d}^\dag_{\downarrow}
		\end{pmatrix}\begin{pmatrix}
			-\mu&V\\
			V&-\mu
		\end{pmatrix}
		\begin{pmatrix}
			\hat{d}_{\uparrow}\\
			\hat{d}_{\downarrow}
		\end{pmatrix}
		+U\hat{n}_{\uparrow}\hat{n}_{\downarrow},
\end{aligned}\end{equation}
describes the two-level system with chemical potential $\mu$, inter-level coupling $V$, and interaction strength $U$.
The bath Hamiltonian $\hat{H}_{B}=\sum_{k\sigma}(2t_{b} \mathrm{cos}k-\mu)\hat{c}^{\dag}_{k\sigma}\hat{c}_{k\sigma}$ 
models a one-dimensional thermal reservoir with temperature $T$, while $\hat{H}_{S\text{-}B}=\frac{1}{\sqrt{N}}\sum_{k\sigma}(t_{\downarrow}\delta_{\sigma \downarrow}\hat{d}^{\dag}_{\sigma}\hat{c}_{k\sigma}+h.c.)$ represents spin-selective coupling between the subsystem and the bath. 

Under a mean-field approximation, the retarded GF of the system takes the standard form as Eq.~(\ref{E11}) with 
\begin{equation}\begin{aligned}
	&H_{S}^{\text{eff}}(\omega)=H_S-i\eta+\Sigma_S(\omega+i\eta)+H_{\text{MF}}\\
	&=\begin{pmatrix}
		-\mu&V\\
		V&-\mu
	\end{pmatrix}+
	\begin{pmatrix}
	-i\eta&0\\
	0&\Sigma_{\downarrow\downarrow}(\omega+i\eta)
	\end{pmatrix}+
	U\begin{pmatrix}
		\langle \hat{n}_{\downarrow}\rangle &0\\
		0&\langle  \hat{n}_{\uparrow}\rangle
	\end{pmatrix},
\end{aligned}\end{equation}
where the explicit form of $\Sigma_{\downarrow\downarrow}(\omega)$ and the numerical procedure for calculating thermal average $\langle \hat{n}_{\downarrow}\rangle$ and $\langle \hat{n}_{\uparrow}\rangle$ are detailed in Supplementary Material~\cite{SM}.
In this model, the mean-field term $H_{\mathrm{MF}}$ only induces a real energy shift for the spin-up and down particles, while the self-energy mainly determines the finite lifetime of the spin-down particle. Notably, since $[H_S,\Sigma_S(\omega)]\neq0$ and $[H_S,H_{\rm MF}]\neq0$, it is nontrivial to understand the role of many-body interaction and the external bath in a single particle picture. 

Here we first focus on the real frequency spectral function, defined as $A(\omega\in\mathbb{R})=-(1/\pi)\mathrm{Im}~\mathrm{Tr}~1/(\omega-H^{\mathrm{eff}}_{S}(\omega))$. As illustrated in Fig.~\ref{F3}, for $U=0$, a single resonance peak is observed, shifting systematically with variations in the chemical potential $\mu$ due to the presence of $-\mu\sigma_{0}$ term. When $U=3$, this peak deviates from its non-interacting counterpart due to the mean-field interaction. Importantly, this deviation is explicitly dependent on the chemical potential: at low filling, the interaction effects within the two-level system are weak, leading to a slight deviation as shown in Fig.~\ref{F3}(a), whereas at higher filling, the stronger interaction causes a more pronounced deviation as shown in Fig.~\ref{F3}(b)-\ref{F3}(d).

However, the single peak picture observed in the single-frequency spectral function does not fully capture the complete quasi-particle information. To address this, we apply the CFF framework to the two-level system and compute the complex frequency spectral log-modulus, defined as
\begin{equation}
\ln~\mathcal{A}_{\omega_{0}}(\omega_{c}\in\mathbb{C})=\ln~|\mathrm{Tr}~G^{\mathrm{CFF}}_{S}(\omega_{c},\omega_{0})|.
\end{equation}
Here we choose $\omega_{0}=0$ in our calculation, namely, only focusing on the quasiparticles near the Fermi level.
This reveals two distinct quasi-particle peaks in the entire complex frequency plane, separated by the GF zeros as shown in Fig.~\ref{F3} with blue color. 
Regardless of the presence of interactions, one quasi-particle resides near the real axis, while the other is located far from it, explaining the absence of the second peak in the real-frequency spectral function due to strong dissipation.
Besides, with the increase of the chemical potential, one can observe that the two quasiparticles in the complex energy plane are indeed repulsive, which provides an intuitive physical picture of the role of many-body interactions. 
This point is hard to demonstrate based on the real frequency paradigm.
These findings vividly exhibit the necessary role of the CFF in characterizing the quantum many-body system. 
For the calculation details of this model, please refer to the Supplementary Material~\cite{SM}.

{\color{red} \em Conclusions and discussion}---In this work, we have established a robust theoretical framework based on the CFF for probing both driven-dissipative systems and quantum many-body systems. 
The CFF enables direct access to the complex-frequency GF, thereby unveiling non-Bloch responses that are intrinsic to non-Hermitian systems exhibiting the NHSE and their associated spectral properties.

Our approach successfully distinguishes between nonreciprocal systems that exhibit the NHSE and those that do not by identifying unique steady-state signatures in non-Bloch response.
Moreover, when generalized to quantum many-body settings, the CFF framework exposes the full structure of the double-frequency Green's function defined in Eq.~(\ref{E13}). 
This generalization allows for a refined quasiparticle resolution, revealing multiple resonance features that are otherwise obscured in conventional single-frequency analyses. 
The additional degree of freedom offered by the complex frequency domain provides a more comprehensive understanding of many-body interactions and self-energy effects.

These advances position the CFF as a versatile diagnostic tool for exploring non-Hermitian physics and quantum many-body dynamics. 
In future studies, this methodology could facilitate deeper insights into the origins of frequency-dependent non-Hermitian Hamiltonians and the role of interactions in modifying spectral properties and eigenstate localizations. 
Ultimately, our framework offers a promising experimental route to address longstanding challenges in identifying and characterizing complex dynamical features in open quantum systems.
\\

J. Huang. and Z.Yang were sponsored by the National Key R$\&$D Program of China (No. 2023YFA1407500), the National Natural Science Foundation of China (No. 12322405, 12104450, 12047503). 
K. Ding was sponsored by the National Natural Science Foundation of China (No. 12174072, 2021hwyq05) and the National Key R$\&$D Program of China (No. 2022YFA1404500, 2022YFA1404701).
J. Hu was sponsored by the Ministry of Science and Technology (Grant No. 2022YFA1403901), National Natural Science Foundation of China (No. 11920101005, No. 11888101, No. 12047503, No. 12322405, No. 12104450) and the New Cornerstone Investigator Program.

\appendix
\section*{Appendix A: The nonreciprocity in dissipative systems }
\label{AppIndex A}

In this section, we will demonstrate that in dissipative systems, both systems exhibiting NHSE and those without NHSE can display nonreciprocity, i.e., nonreciprocal correlations in real-frequency GFs and nonreciprocal dynamics in the wavefunction, making it challenging to distinguish between these two systems.

\begin{figure}[b]
	\begin{center}
		\includegraphics[width=0.95\linewidth]{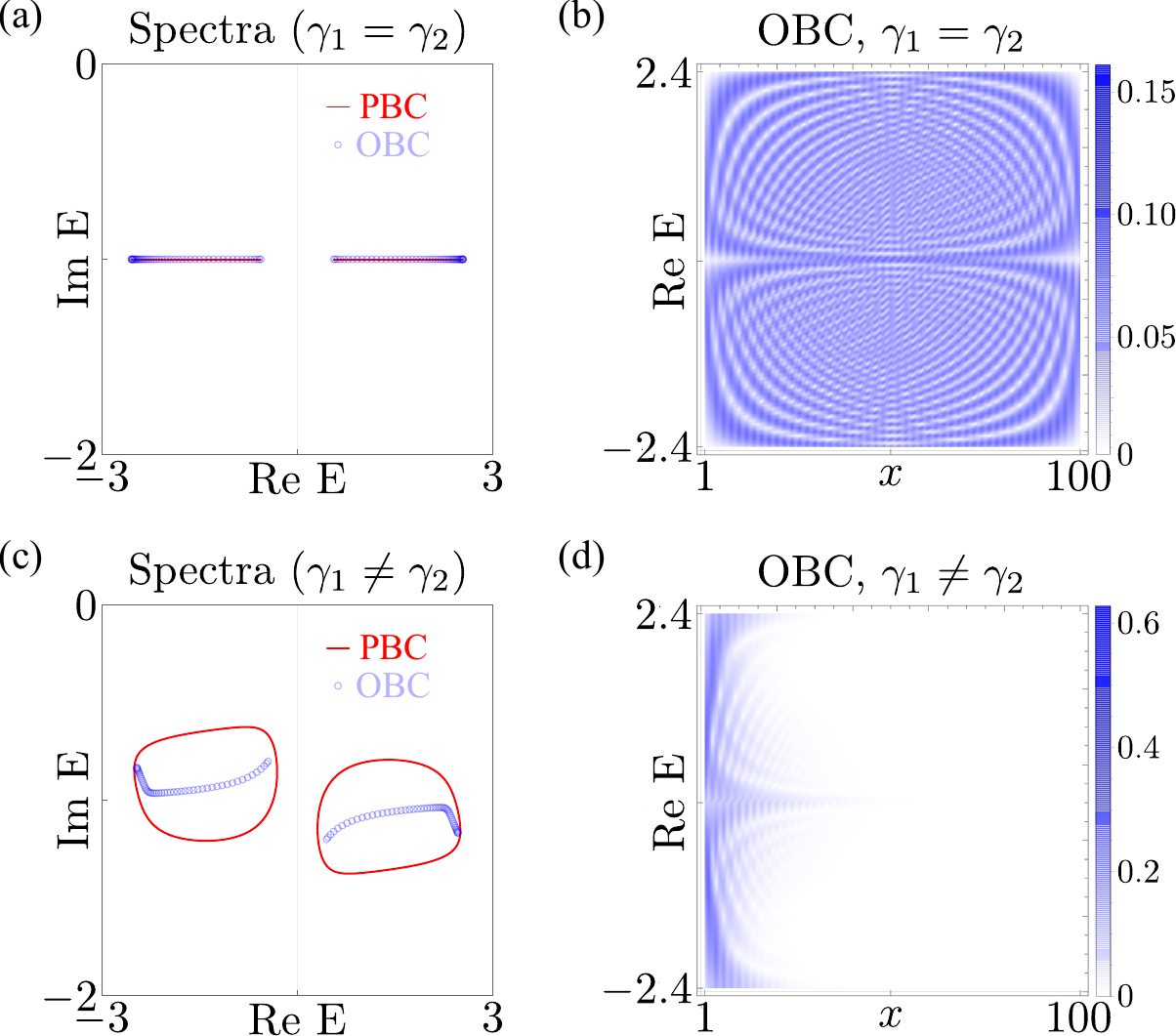}
		\par\end{center}
	\protect\caption{
		(a) Spectra for $\gamma_{1}=\gamma_{2}=1$.
		(b) The distribution of OBC eigenstates $|\psi_{x}|$ of (a).
		(c) Spectra for $\gamma_{1}=1.5,\gamma_{2}=0.5$.
       (d) The non-Hermitian skin modes of (c).
		Here, $N=100$, $t_{1}=1.5,t_{2}=1,\mu=0.3,\lambda=-1$. } 
	\label{S1}
\end{figure}

To illustrate this, we use the model of Eq.~(\ref{E2}) as an example. As discussed in the main text, this model does not exhibit NHSE when $\gamma_{1}=\gamma_{2}$ and exhibits NHSE when $\gamma_{1}\neq\gamma_{2}$. The spectra for these two cases are shown in Fig.~\ref{S1}(a) and \ref{S1}(c), respectively, where the latter exhibits a non-trivial point gap topology, indicating the presence of non-Hermitian skin effect. The corresponding eigenstates $|\psi_{x}|$ ordered by the real part of eigenvalues show extended features when $\gamma_{1}=\gamma_{2}$ and otherwise show skin modes localized at the boundary, as depicted in Fig.~\ref{S1}(b) and \ref{S1}(d).

However, despite these clear differences in spectra and eigenstates, both cases belong to the nonreciprocal system regardless of the value of $\gamma_1$ and $\gamma_2$.
This is because both time-reversal symmetry and inversion symmetry are broken when $\lambda\neq 0$ and $\mu\neq0$.

\begin{figure}[t]
	\begin{center}
		\includegraphics[width=0.95\linewidth]{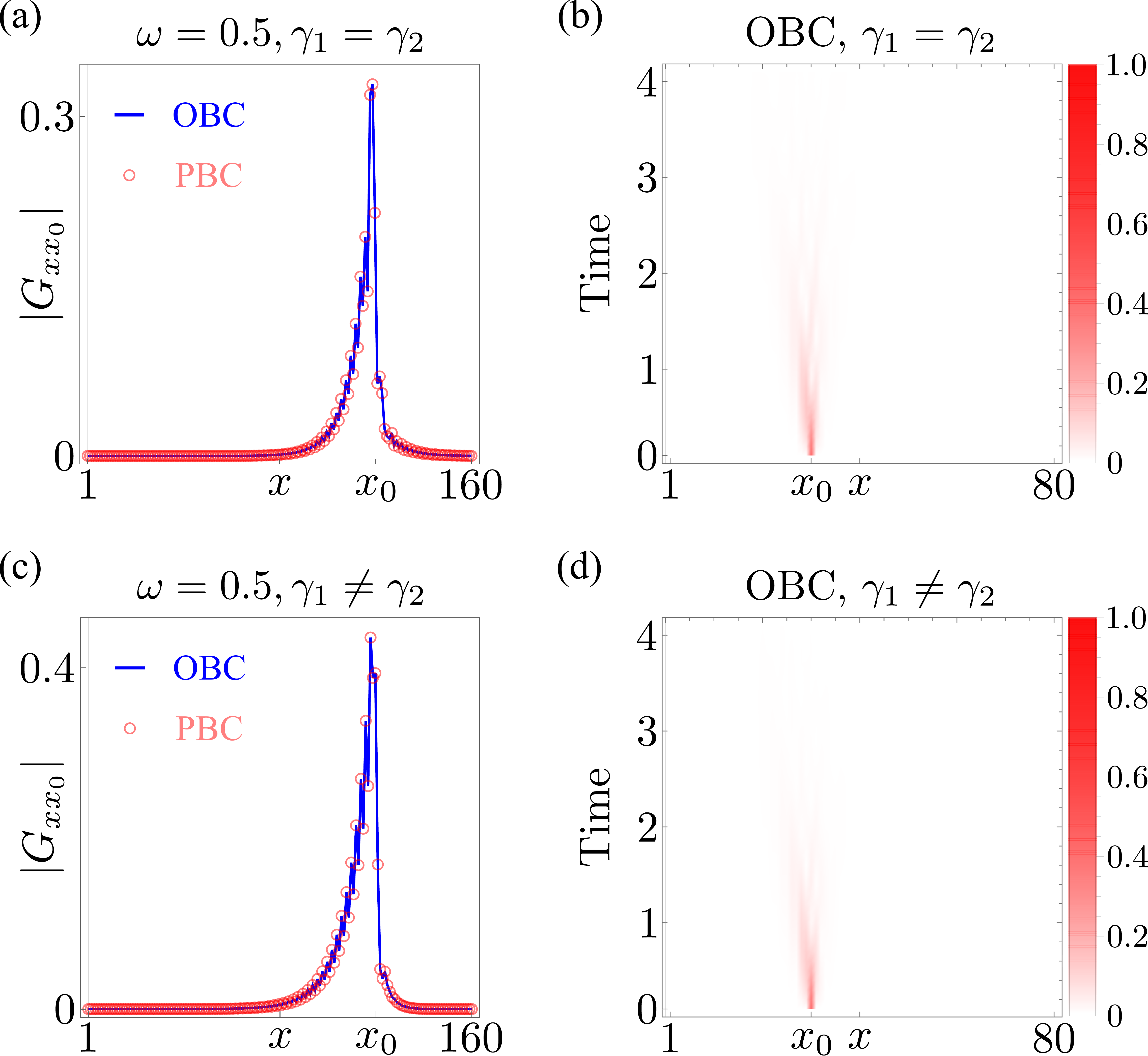}
		\par\end{center}
	\protect\caption{
		(a) The Green's function for $\gamma_{1}=\gamma_{2}=0.5$ and $\lambda=-2$ without NHSE.
		(b) The wavefunction dynamics under OBC for $\gamma_{1}=\gamma_{2}=1$ and $\lambda=-1$ without NHSE. 
		(c) The Green's function for $\gamma_{1}=1,\gamma_{2}=0$ and $\lambda=-1$ with NHSE.
       (d) The wavefunction dynamics under OBC for $\gamma_{1}=1.5,\gamma_{2}=1$ and $\lambda=-1$ with NHSE. 
		In (a)/(c) $x_{0}=120$. In (b)/(d) $x_{0}=30$, with $x\text{-}$direction plot range $[1,80]$. Here $N=160$, $t_{1}=1.5,t_{2}=1,\mu=0.3$. } 
	\label{S2}
\end{figure}

To illustrate this, we first consider the real frequency GF. As depicted in Fig.~\ref{S2}(a) and \ref{S2}(c), regardless of whether the system exhibits the NHSE, the real frequency Green's function displays nonreciprocal correlations ($\omega=0.5$ with $x_{0}=120$). Specifically, the GF favors one direction propagating along the skin localized direction (to the left). This is further evidenced by the different decay rates when $x$ deviates from $x_{0}$ to the left versus to the right. Besides, the real frequency Green's function also exhibits no boundary sensitivity, as the relation $|G^{\mathrm{PBC}}_{xx_{0}}(\omega)|=|G^{\mathrm{OBC}}_{xx_{0}}(\omega)|$ holds in the bulk region. These observations demonstrate that the real frequency GF is insufficient to identify the presence or absence of the NHSE in a dissipative system.

Second, we analyze the wavefunction dynamics. Consider an initial state $|\psi(t=0)\rangle=|x_{0}\rangle$. Under the time evolution governed by the dynamical equation,
\begin{equation}
i\frac{\partial }{\partial t}|\psi(t)\rangle=H_{\mathrm{nH}}|\psi(t)\rangle,
\end{equation}
the wavefunction evolves as $|\psi(t)\rangle=e^{-iH_{\mathrm{nH}}t}|\psi(0)\rangle$. As shown in Fig.~\ref{S2}(b) and (d), the wavefunction decays rapidly due to the system’s strong intrinsic dissipation for an initial state $x_{0}=30$. This rapid decay, although it exhibits the nonreciprocal behaviors for both cases, limits the visibility of nonreciprocal evolution to a short time window after the initial state, making it difficult to distinguish between systems with and without NHSE. 

Notably, when the detection time is extended and the wavefunction is normalized at each time step, systems with NHSE exhibit more pronounced nonreciprocity compared to those without NHSE. Specifically, the normalized wavefunction stabilizes into a state localized at the left boundary, which is a hallmark of NHSE. However, this distinction cannot be experimentally verified in strongly dissipative systems due to a low signal-to-noise ratio, which complicates normalization procedures and obscures such subtle features.

In conclusion, neither the real-frequency GF nor the wavefunction dynamics can reliably identify the presence of the NHSE in realistic experimental settings. This limitation underscores the necessity of our CFF framework as a powerful and promising tool for detecting the intrinsic non-Bloch response associated with the NHSE.

\section*{Appendix B: Detection of OBC eigenstates}\label{AppIndex B}

In this section, we demonstrate how the OBC eigenstates can be detected using the CFF method. By using the biorthogonal basis~\cite{brody,Prl121kunst}, the OBC eigenstates are related to the Green's function through the expression
\begin{equation}
G(\omega_{c})=\sum\limits_{m}\frac{|\psi^{R}_{m}\rangle \langle \psi^{L}_{m}|}{\omega_{c}-E_{m}},
\end{equation}
where $|\psi^{R/L}_{m}\rangle$ represents the $m$-th right/left eigenstate. Thus, as $\omega_{c}$ approaches an eigenvalue of a non-Hermitian Hamiltonian $H_{\mathrm{nH}}$, such as $E_{n}$, the corresponding correlation can precisely reveal the eigenstate as shown by
\begin{equation}
G_{ij}(\omega_{c}\rightarrow E_{n})\sim \frac{\langle i|\psi^{R}_{n}\rangle \langle \psi^{L}_{n}|j\rangle}{\omega_{c}-E_{n}}.
\end{equation}
Furthermore, when fixing a site $j=j_{0}$, then the responses $G_{ij_{0}}(\omega_{c}\rightarrow E_{n})$ and $G_{j_{0}i}(\omega_{c}\rightarrow E_{n})$ are proportional to the $i$-th component of the right and left eigenstates, respectively. Consequently, since $\mathscr{G}_{\omega_{0}}(E_{n};t\rightarrow\infty)=G(\omega_{c}\rightarrow E_{n})$, the eigenstates can be detected by using the CFF method:
\begin{equation}
	\begin{split}
		\lim\limits_{t\rightarrow \infty}\mathscr{G}_{ij}(E_{n};t)&\sim \langle i|\psi^{R}_{n}\rangle \langle \psi^{L}_{n}|j\rangle,\\
	\end{split}
\end{equation}
where the subscript $\omega_{0}$ is omitted for simplicity, as this section focuses on the application of CFF to a frequency-independent non-Hermitian Hamiltonian, rendering the choice of $\omega_{0}$ irrelevant to the outcome. Notably, the extension to a frequency-dependent $H^{\mathrm{eff}}_{S}(\omega_{0})$ is also straightforward. In conclusion, we can detect the eigenstates through $\mathscr{G}_{ij_{0}}(E_{n};t\rightarrow \infty)\propto \langle i|\psi^{R}_{n}\rangle$ and $\mathscr{G}_{j_{0}i}(E_{n};t\rightarrow \infty)\propto \langle \psi^{L}_{n}|i\rangle$. 

\begin{figure}[t]
	\begin{center}
		\includegraphics[width=0.95\linewidth]{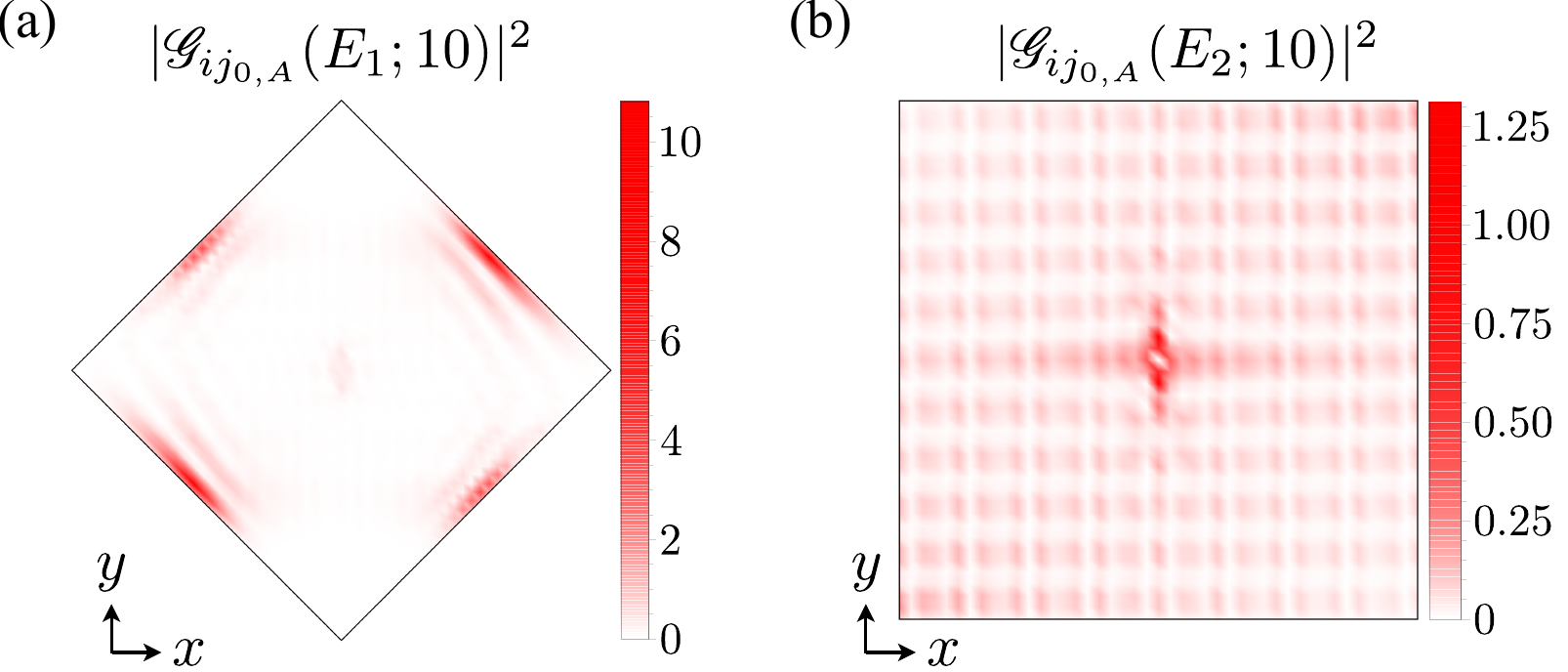}
		\par\end{center}
	\protect\caption{
		(a) $|\mathscr{G}_{ij_{0,A}}(\omega_{c}=E_{1};10)|^{2}$ for the diamond geometry of the Hamiltonian Eq.~(\ref{2d}). 
		The system's size is $L_{x}=L_{y}=63$. 
		(b) $|\mathscr{G}_{ij_{0,A}}(\omega_{c}=E_{2};10)|^{2}$ for the square geometry. 
		The system's size is $L_{x}=L_{y}=45$. 
		The parameters are selected as $t_{1}=t_{2}=0.5,\gamma_{0}=1.5,\gamma_{z}=-0.5, \omega_{0}=0$ and $\mu_{z}=0$. } 
	\label{S3}
\end{figure}

To exemplify the detection, the chosen Hamiltonian is 
\begin{equation}
H^{\mathrm{2D}}_{\mathrm{nH}}(\boldsymbol{k})=\boldsymbol{d}(\boldsymbol{k})\cdot \boldsymbol{\sigma}-i(\gamma_{0}\sigma_{0}+\gamma_{z}\sigma_{z})/2,\label{2d}
\end{equation}
where $\boldsymbol{\sigma}=(\sigma_{x},\sigma_{y},\sigma_{z})$ and $\boldsymbol{d}(\boldsymbol{k})=(t_{1}\mathrm{sin}k_{x},2t_{1}\mathrm{sin}k_{y},\mu_{z}+t_{2}\mathrm{cos}k_{x}+2t_{2}\mathrm{cos}k_{y})$.
In this model, the emergence of the NHSE depends on the choice of OBC geometry~\cite{gdse,lzy,hu2024,Fang-Ding,mgc3,yzj,PrA109023317}. 
Focusing on our study, we analyze the diamond and square geometries.
To detect the eigenstates corresponding to specific eigenvalues, we introduce $|\mathscr{G}_{ij_A}(\omega_{c};t)|^2=|\mathscr{G}_{i_{A}j_{A}}(\omega_{c};t)|^{2}+|\mathscr{G}_{i_{B}j_{A}}(\omega_{c};t)|^{2}$ for the two-band model, where $i$ specifically denotes the $i$-th unit cell. The results of $|\mathscr{G}_{ij_{0,A}}(\omega_{c}=E_{1},E_{2};10)|^2$ are presented in Fig.~\ref{S3}(a) and \ref{S3}(b), respectively, where $E_{1}=1.009-0.603i,E_{2}=1.001-0.605i$ are eigenvalues approximated to three decimal places, corresponding to the diamond and square geometries. In both cases, $i_{0}$ is chosen as the central site.
From Fig.~\ref{S3}(a) and \ref{S3}(b), it can be observed that as $\omega_{c}$ approaches the OBC eigenvalues, i.e., $E_{1}$ for (a) and $E_{2}$ for (b), the corresponding eigenstate exhibits the skin mode for the diamond geometry and the Bloch mode for the square geometry.

\bibliography{ref}

\begin{thebibliography}{116}%
\makeatletter
\providecommand \@ifxundefined [1]{%
 \@ifx{#1\undefined}
}%
\providecommand \@ifnum [1]{%
 \ifnum #1\expandafter \@firstoftwo
 \else \expandafter \@secondoftwo
 \fi
}%
\providecommand \@ifx [1]{%
 \ifx #1\expandafter \@firstoftwo
 \else \expandafter \@secondoftwo
 \fi
}%
\providecommand \natexlab [1]{#1}%
\providecommand \enquote  [1]{``#1''}%
\providecommand \bibnamefont  [1]{#1}%
\providecommand \bibfnamefont [1]{#1}%
\providecommand \citenamefont [1]{#1}%
\providecommand \href@noop [0]{\@secondoftwo}%
\providecommand \href [0]{\begingroup \@sanitize@url \@href}%
\providecommand \@href[1]{\@@startlink{#1}\@@href}%
\providecommand \@@href[1]{\endgroup#1\@@endlink}%
\providecommand \@sanitize@url [0]{\catcode `\\12\catcode `\$12\catcode
  `\&12\catcode `\#12\catcode `\^12\catcode `\_12\catcode `\%12\relax}%
\providecommand \@@startlink[1]{}%
\providecommand \@@endlink[0]{}%
\providecommand \url  [0]{\begingroup\@sanitize@url \@url }%
\providecommand \@url [1]{\endgroup\@href {#1}{\urlprefix }}%
\providecommand \urlprefix  [0]{URL }%
\providecommand \Eprint [0]{\href }%
\providecommand \doibase [0]{http://dx.doi.org/}%
\providecommand \selectlanguage [0]{\@gobble}%
\providecommand \bibinfo  [0]{\@secondoftwo}%
\providecommand \bibfield  [0]{\@secondoftwo}%
\providecommand \translation [1]{[#1]}%
\providecommand \BibitemOpen [0]{}%
\providecommand \bibitemStop [0]{}%
\providecommand \bibitemNoStop [0]{.\EOS\space}%
\providecommand \EOS [0]{\spacefactor3000\relax}%
\providecommand \BibitemShut  [1]{\csname bibitem#1\endcsname}%
\let\auto@bib@innerbib\@empty
%</preamble>
\bibitem [{\citenamefont {Ashida}\ \emph {et~al.}(2020)\citenamefont {Ashida},
  \citenamefont {Gong},\ and\ \citenamefont {Ueda}}]{Ashida}%
  \BibitemOpen
  \bibfield  {author} {\bibinfo {author} {\bibfnamefont {Y.}~\bibnamefont
  {Ashida}}, \bibinfo {author} {\bibfnamefont {Z.}~\bibnamefont {Gong}}, \ and\
  \bibinfo {author} {\bibfnamefont {M.}~\bibnamefont {Ueda}},\ }\href
  {https://doi.org/10.1080/00018732.2021.1876991} {\bibfield  {journal}
  {\bibinfo  {journal} {Adv. Phys.}\ }\textbf {\bibinfo {volume} {69}},\
  \bibinfo {pages} {249} (\bibinfo {year} {2020})}\BibitemShut {NoStop}%
\bibitem [{\citenamefont {Bergholtz}\ \emph {et~al.}(2021)\citenamefont
  {Bergholtz}, \citenamefont {Budich},\ and\ \citenamefont
  {Kunst}}]{Rmp93015005}%
  \BibitemOpen
  \bibfield  {author} {\bibinfo {author} {\bibfnamefont {E.~J.}\ \bibnamefont
  {Bergholtz}}, \bibinfo {author} {\bibfnamefont {J.~C.}\ \bibnamefont
  {Budich}}, \ and\ \bibinfo {author} {\bibfnamefont {F.~K.}\ \bibnamefont
  {Kunst}},\ }\href {\doibase 10.1103/RevModPhys.93.015005} {\bibfield
  {journal} {\bibinfo  {journal} {Rev. Mod. Phys.}\ }\textbf {\bibinfo {volume}
  {93}},\ \bibinfo {pages} {015005} (\bibinfo {year} {2021})}\BibitemShut
  {NoStop}%
\bibitem [{\citenamefont {Elganainy}\ \emph {et~al.}(2018)\citenamefont
  {Elganainy}, \citenamefont {Makris}, \citenamefont {Khajavikhan},
  \citenamefont {Musslimani}, \citenamefont {Rotter},\ and\ \citenamefont
  {Christodoulides}}]{NH-PT}%
  \BibitemOpen
  \bibfield  {author} {\bibinfo {author} {\bibfnamefont {R.}~\bibnamefont
  {Elganainy}}, \bibinfo {author} {\bibfnamefont {K.}~\bibnamefont {Makris}},
  \bibinfo {author} {\bibfnamefont {M.}~\bibnamefont {Khajavikhan}}, \bibinfo
  {author} {\bibfnamefont {Z.}~\bibnamefont {Musslimani}}, \bibinfo {author}
  {\bibfnamefont {S.}~\bibnamefont {Rotter}}, \ and\ \bibinfo {author}
  {\bibfnamefont {D.}~\bibnamefont {Christodoulides}},\ }\href {\doibase
  https://doi.org/10.1038/nphys4323} {\bibfield  {journal} {\bibinfo  {journal}
  {Nat. Phys.}\ }\textbf {\bibinfo {volume} {14}},\ \bibinfo {pages} {11}
  (\bibinfo {year} {2018})}\BibitemShut {NoStop}%
\bibitem [{\citenamefont {Feng}\ \emph {et~al.}(2017)\citenamefont {Feng},
  \citenamefont {Elganainy},\ and\ \citenamefont {Ge}}]{NHP-PT}%
  \BibitemOpen
  \bibfield  {author} {\bibinfo {author} {\bibfnamefont {L.}~\bibnamefont
  {Feng}}, \bibinfo {author} {\bibfnamefont {R.}~\bibnamefont {Elganainy}}, \
  and\ \bibinfo {author} {\bibfnamefont {L.}~\bibnamefont {Ge}},\ }\href
  {\doibase https://doi.org/10.1038/s41566-017-0031-1} {\bibfield  {journal}
  {\bibinfo  {journal} {Nat. Photonics}\ }\textbf {\bibinfo {volume} {11}},\
  \bibinfo {pages} {752} (\bibinfo {year} {2017})}\BibitemShut {NoStop}%
\bibitem [{\citenamefont {Miri}\ and\ \citenamefont {Alù}(2019)}]{EP-optics}%
  \BibitemOpen
  \bibfield  {author} {\bibinfo {author} {\bibfnamefont {M.-A.}\ \bibnamefont
  {Miri}}\ and\ \bibinfo {author} {\bibfnamefont {A.}~\bibnamefont {Alù}},\
  }\href {\doibase 10.1126/science.aar7709} {\bibfield  {journal} {\bibinfo
  {journal} {Science}\ }\textbf {\bibinfo {volume} {363}},\ \bibinfo {pages}
  {eaar7709} (\bibinfo {year} {2019})}\BibitemShut {NoStop}%
\bibitem [{\citenamefont {Ozdemir}\ \emph {et~al.}(2019)\citenamefont
  {Ozdemir}, \citenamefont {Rotter}, \citenamefont {Nori},\ and\ \citenamefont
  {Yang}}]{Ozde}%
  \BibitemOpen
  \bibfield  {author} {\bibinfo {author} {\bibfnamefont {S.}~\bibnamefont
  {Ozdemir}}, \bibinfo {author} {\bibfnamefont {S.}~\bibnamefont {Rotter}},
  \bibinfo {author} {\bibfnamefont {F.}~\bibnamefont {Nori}}, \ and\ \bibinfo
  {author} {\bibfnamefont {L.}~\bibnamefont {Yang}},\ }\href {\doibase
  https://doi.org/10.1038/s41563-019-0304-9} {\bibfield  {journal} {\bibinfo
  {journal} {Nat. Mater.}\ }\textbf {\bibinfo {volume} {18}},\ \bibinfo {pages}
  {783} (\bibinfo {year} {2019})}\BibitemShut {NoStop}%
\bibitem [{\citenamefont {Ding}\ \emph {et~al.}(2022)\citenamefont {Ding},
  \citenamefont {Fang},\ and\ \citenamefont {Ma}}]{Dingkrev}%
  \BibitemOpen
  \bibfield  {author} {\bibinfo {author} {\bibfnamefont {K.}~\bibnamefont
  {Ding}}, \bibinfo {author} {\bibfnamefont {C.}~\bibnamefont {Fang}}, \ and\
  \bibinfo {author} {\bibfnamefont {G.}~\bibnamefont {Ma}},\ }\href
  {https://doi.org/10.1038/s42254-022-00516-5} {\bibfield  {journal} {\bibinfo
  {journal} {Nat. Rev. Phys.}\ }\textbf {\bibinfo {volume} {4}},\ \bibinfo
  {pages} {745} (\bibinfo {year} {2022})}\BibitemShut {NoStop}%
\bibitem [{\citenamefont {Zhang}\ \emph
  {et~al.}(2022{\natexlab{a}})\citenamefont {Zhang}, \citenamefont {Zhang},
  \citenamefont {Lu},\ and\ \citenamefont {Chen}}]{re-nhse}%
  \BibitemOpen
  \bibfield  {author} {\bibinfo {author} {\bibfnamefont {X.}~\bibnamefont
  {Zhang}}, \bibinfo {author} {\bibfnamefont {T.}~\bibnamefont {Zhang}},
  \bibinfo {author} {\bibfnamefont {M.-H.}\ \bibnamefont {Lu}}, \ and\ \bibinfo
  {author} {\bibfnamefont {Y.-F.}\ \bibnamefont {Chen}},\ }\href
  {https://doi.org/10.1080/23746149.2022.2109431} {\bibfield  {journal}
  {\bibinfo  {journal} {Adv. Phys.: X.}\ }\textbf {\bibinfo {volume} {7}},\
  \bibinfo {pages} {2109431} (\bibinfo {year}
  {2022}{\natexlab{a}})}\BibitemShut {NoStop}%
\bibitem [{\citenamefont {Okuma}\ and\ \citenamefont {Sato}(2022)}]{NHTP-rev}%
  \BibitemOpen
  \bibfield  {author} {\bibinfo {author} {\bibfnamefont {N.}~\bibnamefont
  {Okuma}}\ and\ \bibinfo {author} {\bibfnamefont {M.}~\bibnamefont {Sato}},\
  }\href {\doibase https://doi.org/10.1146/annurev-conmatphys-040521-033133}
  {\bibfield  {journal} {\bibinfo  {journal} {Annu. Rev. Condens. Matter
  Phys.}\ }\textbf {\bibinfo {volume} {14}},\ \bibinfo {pages} {83} (\bibinfo
  {year} {2022})}\BibitemShut {NoStop}%
\bibitem [{\citenamefont {Lin}\ \emph {et~al.}(2023)\citenamefont {Lin},
  \citenamefont {Tai}, \citenamefont {Li},\ and\ \citenamefont
  {Lee}}]{Lin2023Topo}%
  \BibitemOpen
  \bibfield  {author} {\bibinfo {author} {\bibfnamefont {R.}~\bibnamefont
  {Lin}}, \bibinfo {author} {\bibfnamefont {T.}~\bibnamefont {Tai}}, \bibinfo
  {author} {\bibfnamefont {L.}~\bibnamefont {Li}}, \ and\ \bibinfo {author}
  {\bibfnamefont {C.~H.}\ \bibnamefont {Lee}},\ }\href {\doibase
  https://doi.org/10.1007/s11467-023-1309-z} {\bibfield  {journal} {\bibinfo
  {journal} {Front. Phys.}\ }\textbf {\bibinfo {volume} {18}},\ \bibinfo
  {pages} {53605} (\bibinfo {year} {2023})}\BibitemShut {NoStop}%
\bibitem [{\citenamefont {Gong}\ \emph {et~al.}(2018)\citenamefont {Gong},
  \citenamefont {Ashida}, \citenamefont {Kawabata}, \citenamefont {Takasan},
  \citenamefont {Higashikawa},\ and\ \citenamefont {Ueda}}]{PrxUeda}%
  \BibitemOpen
  \bibfield  {author} {\bibinfo {author} {\bibfnamefont {Z.}~\bibnamefont
  {Gong}}, \bibinfo {author} {\bibfnamefont {Y.}~\bibnamefont {Ashida}},
  \bibinfo {author} {\bibfnamefont {K.}~\bibnamefont {Kawabata}}, \bibinfo
  {author} {\bibfnamefont {K.}~\bibnamefont {Takasan}}, \bibinfo {author}
  {\bibfnamefont {S.}~\bibnamefont {Higashikawa}}, \ and\ \bibinfo {author}
  {\bibfnamefont {M.}~\bibnamefont {Ueda}},\ }\href
  {https://link.aps.org/doi/10.1103/PhysRevX.8.031079} {\bibfield  {journal}
  {\bibinfo  {journal} {Phys. Rev. X}\ }\textbf {\bibinfo {volume} {8}},\
  \bibinfo {pages} {031079} (\bibinfo {year} {2018})}\BibitemShut {NoStop}%
\bibitem [{\citenamefont {Kawabata}\ \emph {et~al.}(2019)\citenamefont
  {Kawabata}, \citenamefont {Shiozaki}, \citenamefont {Ueda},\ and\
  \citenamefont {Sato}}]{Prxkawabata}%
  \BibitemOpen
  \bibfield  {author} {\bibinfo {author} {\bibfnamefont {K.}~\bibnamefont
  {Kawabata}}, \bibinfo {author} {\bibfnamefont {K.}~\bibnamefont {Shiozaki}},
  \bibinfo {author} {\bibfnamefont {M.}~\bibnamefont {Ueda}}, \ and\ \bibinfo
  {author} {\bibfnamefont {M.}~\bibnamefont {Sato}},\ }\href
  {https://link.aps.org/doi/10.1103/PhysRevX.9.041015} {\bibfield  {journal}
  {\bibinfo  {journal} {Phys. Rev. X}\ }\textbf {\bibinfo {volume} {9}},\
  \bibinfo {pages} {041015} (\bibinfo {year} {2019})}\BibitemShut {NoStop}%
\bibitem [{\citenamefont {Torres}(2019)}]{FoaTorres}%
  \BibitemOpen
  \bibfield  {author} {\bibinfo {author} {\bibfnamefont {L.~E. F.~F.}\
  \bibnamefont {Torres}},\ }\href {https://dx.doi.org/10.1088/2515-7639/ab4092}
  {\bibfield  {journal} {\bibinfo  {journal} {JPhys mater.}\ }\textbf {\bibinfo
  {volume} {3}},\ \bibinfo {pages} {014002} (\bibinfo {year}
  {2019})}\BibitemShut {NoStop}%
\bibitem [{\citenamefont {Yao}\ and\ \citenamefont {Wang}(2018)}]{Prl1210868}%
  \BibitemOpen
  \bibfield  {author} {\bibinfo {author} {\bibfnamefont {S.}~\bibnamefont
  {Yao}}\ and\ \bibinfo {author} {\bibfnamefont {Z.}~\bibnamefont {Wang}},\
  }\href {\doibase 10.1103/PhysRevLett.121.086803} {\bibfield  {journal}
  {\bibinfo  {journal} {Phys. Rev. Lett.}\ }\textbf {\bibinfo {volume} {121}},\
  \bibinfo {pages} {086803} (\bibinfo {year} {2018})}\BibitemShut {NoStop}%
\bibitem [{\citenamefont {Kunst}\ \emph {et~al.}(2018)\citenamefont {Kunst},
  \citenamefont {Edvardsson}, \citenamefont {Budich},\ and\ \citenamefont
  {Bergholtz}}]{Prl121kunst}%
  \BibitemOpen
  \bibfield  {author} {\bibinfo {author} {\bibfnamefont {F.~K.}\ \bibnamefont
  {Kunst}}, \bibinfo {author} {\bibfnamefont {E.}~\bibnamefont {Edvardsson}},
  \bibinfo {author} {\bibfnamefont {J.~C.}\ \bibnamefont {Budich}}, \ and\
  \bibinfo {author} {\bibfnamefont {E.~J.}\ \bibnamefont {Bergholtz}},\ }\href
  {https://link.aps.org/doi/10.1103/PhysRevLett.121.026808} {\bibfield
  {journal} {\bibinfo  {journal} {Phys. Rev. Lett.}\ }\textbf {\bibinfo
  {volume} {121}},\ \bibinfo {pages} {026808} (\bibinfo {year}
  {2018})}\BibitemShut {NoStop}%
\bibitem [{\citenamefont {Martinez~Alvarez}\ \emph {et~al.}(2018)\citenamefont
  {Martinez~Alvarez}, \citenamefont {Barrios~Vargas},\ and\ \citenamefont
  {Foa~Torres}}]{Prb971214}%
  \BibitemOpen
  \bibfield  {author} {\bibinfo {author} {\bibfnamefont {V.~M.}\ \bibnamefont
  {Martinez~Alvarez}}, \bibinfo {author} {\bibfnamefont {J.~E.}\ \bibnamefont
  {Barrios~Vargas}}, \ and\ \bibinfo {author} {\bibfnamefont {L.~E.~F.}\
  \bibnamefont {Foa~Torres}},\ }\href {\doibase 10.1103/PhysRevB.97.121401}
  {\bibfield  {journal} {\bibinfo  {journal} {Phys. Rev. B}\ }\textbf {\bibinfo
  {volume} {97}},\ \bibinfo {pages} {121401} (\bibinfo {year}
  {2018})}\BibitemShut {NoStop}%
\bibitem [{\citenamefont {Yao}\ \emph {et~al.}(2018)\citenamefont {Yao},
  \citenamefont {Song},\ and\ \citenamefont {Wang}}]{PrlYSW}%
  \BibitemOpen
  \bibfield  {author} {\bibinfo {author} {\bibfnamefont {S.}~\bibnamefont
  {Yao}}, \bibinfo {author} {\bibfnamefont {F.}~\bibnamefont {Song}}, \ and\
  \bibinfo {author} {\bibfnamefont {Z.}~\bibnamefont {Wang}},\ }\href
  {https://link.aps.org/doi/10.1103/PhysRevLett.121.136802} {\bibfield
  {journal} {\bibinfo  {journal} {Phys. Rev. Lett.}\ }\textbf {\bibinfo
  {volume} {121}},\ \bibinfo {pages} {136802} (\bibinfo {year}
  {2018})}\BibitemShut {NoStop}%
\bibitem [{\citenamefont {Zhang}\ \emph {et~al.}(2020)\citenamefont {Zhang},
  \citenamefont {Yang},\ and\ \citenamefont {Fang}}]{Prl125126402}%
  \BibitemOpen
  \bibfield  {author} {\bibinfo {author} {\bibfnamefont {K.}~\bibnamefont
  {Zhang}}, \bibinfo {author} {\bibfnamefont {Z.}~\bibnamefont {Yang}}, \ and\
  \bibinfo {author} {\bibfnamefont {C.}~\bibnamefont {Fang}},\ }\href
  {https://link.aps.org/doi/10.1103/PhysRevLett.125.126402} {\bibfield
  {journal} {\bibinfo  {journal} {Phys. Rev. Lett.}\ }\textbf {\bibinfo
  {volume} {125}},\ \bibinfo {pages} {126402} (\bibinfo {year}
  {2020})}\BibitemShut {NoStop}%
\bibitem [{\citenamefont {Okuma}\ \emph {et~al.}(2020)\citenamefont {Okuma},
  \citenamefont {Kawabata}, \citenamefont {Shiozaki},\ and\ \citenamefont
  {Sato}}]{prlOkuma}%
  \BibitemOpen
  \bibfield  {author} {\bibinfo {author} {\bibfnamefont {N.}~\bibnamefont
  {Okuma}}, \bibinfo {author} {\bibfnamefont {K.}~\bibnamefont {Kawabata}},
  \bibinfo {author} {\bibfnamefont {K.}~\bibnamefont {Shiozaki}}, \ and\
  \bibinfo {author} {\bibfnamefont {M.}~\bibnamefont {Sato}},\ }\href
  {https://link.aps.org/doi/10.1103/PhysRevLett.124.086801} {\bibfield
  {journal} {\bibinfo  {journal} {Phys. Rev. Lett.}\ }\textbf {\bibinfo
  {volume} {124}},\ \bibinfo {pages} {086801} (\bibinfo {year}
  {2020})}\BibitemShut {NoStop}%
\bibitem [{\citenamefont {Yokomizo}\ and\ \citenamefont
  {Murakami}(2019)}]{PrlYM}%
  \BibitemOpen
  \bibfield  {author} {\bibinfo {author} {\bibfnamefont {K.}~\bibnamefont
  {Yokomizo}}\ and\ \bibinfo {author} {\bibfnamefont {S.}~\bibnamefont
  {Murakami}},\ }\href {\doibase 10.1103/PhysRevLett.123.066404} {\bibfield
  {journal} {\bibinfo  {journal} {Phys. Rev. Lett.}\ }\textbf {\bibinfo
  {volume} {123}},\ \bibinfo {pages} {066404} (\bibinfo {year}
  {2019})}\BibitemShut {NoStop}%
\bibitem [{\citenamefont {Yang}\ \emph {et~al.}(2020)\citenamefont {Yang},
  \citenamefont {Zhang}, \citenamefont {Fang},\ and\ \citenamefont
  {Hu}}]{PrlYZFH}%
  \BibitemOpen
  \bibfield  {author} {\bibinfo {author} {\bibfnamefont {Z.}~\bibnamefont
  {Yang}}, \bibinfo {author} {\bibfnamefont {K.}~\bibnamefont {Zhang}},
  \bibinfo {author} {\bibfnamefont {C.}~\bibnamefont {Fang}}, \ and\ \bibinfo
  {author} {\bibfnamefont {J.}~\bibnamefont {Hu}},\ }\href
  {https://link.aps.org/doi/10.1103/PhysRevLett.125.226402} {\bibfield
  {journal} {\bibinfo  {journal} {Phys. Rev. Lett.}\ }\textbf {\bibinfo
  {volume} {125}},\ \bibinfo {pages} {226402} (\bibinfo {year}
  {2020})}\BibitemShut {NoStop}%
\bibitem [{\citenamefont {Li}\ \emph {et~al.}(2020{\natexlab{a}})\citenamefont
  {Li}, \citenamefont {Lee}, \citenamefont {Mu},\ and\ \citenamefont
  {Gong}}]{llhcritical}%
  \BibitemOpen
  \bibfield  {author} {\bibinfo {author} {\bibfnamefont {L.}~\bibnamefont
  {Li}}, \bibinfo {author} {\bibfnamefont {C.~H.}\ \bibnamefont {Lee}},
  \bibinfo {author} {\bibfnamefont {S.}~\bibnamefont {Mu}}, \ and\ \bibinfo
  {author} {\bibfnamefont {J.}~\bibnamefont {Gong}},\ }\href
  {https://doi.org/10.1038/s41467-020-18917-4} {\bibfield  {journal} {\bibinfo
  {journal} {Nat. Commun.}\ }\textbf {\bibinfo {volume} {11}},\ \bibinfo
  {pages} {5491} (\bibinfo {year} {2020}{\natexlab{a}})}\BibitemShut {NoStop}%
\bibitem [{\citenamefont {Helbig}\ \emph {et~al.}(2020)\citenamefont {Helbig},
  \citenamefont {Hofmann}, \citenamefont {Imhof}, \citenamefont {Abdelghany},
  \citenamefont {Kiessling}, \citenamefont {Molenkamp}, \citenamefont {Lee},
  \citenamefont {Szameit}, \citenamefont {Greiter},\ and\ \citenamefont
  {Thomale}}]{helbig}%
  \BibitemOpen
  \bibfield  {author} {\bibinfo {author} {\bibfnamefont {T.}~\bibnamefont
  {Helbig}}, \bibinfo {author} {\bibfnamefont {T.}~\bibnamefont {Hofmann}},
  \bibinfo {author} {\bibfnamefont {S.}~\bibnamefont {Imhof}}, \bibinfo
  {author} {\bibfnamefont {M.}~\bibnamefont {Abdelghany}}, \bibinfo {author}
  {\bibfnamefont {T.}~\bibnamefont {Kiessling}}, \bibinfo {author}
  {\bibfnamefont {L.}~\bibnamefont {Molenkamp}}, \bibinfo {author}
  {\bibfnamefont {C.~H.}\ \bibnamefont {Lee}}, \bibinfo {author} {\bibfnamefont
  {A.}~\bibnamefont {Szameit}}, \bibinfo {author} {\bibfnamefont
  {M.}~\bibnamefont {Greiter}}, \ and\ \bibinfo {author} {\bibfnamefont
  {R.}~\bibnamefont {Thomale}},\ }\href
  {https://doi.org/10.1038/s41567-020-0922-9} {\bibfield  {journal} {\bibinfo
  {journal} {Nat. Phys.}\ }\textbf {\bibinfo {volume} {16}},\ \bibinfo {pages}
  {747} (\bibinfo {year} {2020})}\BibitemShut {NoStop}%
\bibitem [{\citenamefont {Xiao}\ \emph {et~al.}(2020)\citenamefont {Xiao},
  \citenamefont {Deng}, \citenamefont {Wang}, \citenamefont {Zhu},
  \citenamefont {Wang}, \citenamefont {Yi},\ and\ \citenamefont
  {Xue}}]{xiao-wang}%
  \BibitemOpen
  \bibfield  {author} {\bibinfo {author} {\bibfnamefont {L.}~\bibnamefont
  {Xiao}}, \bibinfo {author} {\bibfnamefont {T.-S.}\ \bibnamefont {Deng}},
  \bibinfo {author} {\bibfnamefont {K.}~\bibnamefont {Wang}}, \bibinfo {author}
  {\bibfnamefont {G.}~\bibnamefont {Zhu}}, \bibinfo {author} {\bibfnamefont
  {Z.}~\bibnamefont {Wang}}, \bibinfo {author} {\bibfnamefont {W.}~\bibnamefont
  {Yi}}, \ and\ \bibinfo {author} {\bibfnamefont {P.}~\bibnamefont {Xue}},\
  }\href {\doibase 10.1038/s41567-020-0836-6} {\bibfield  {journal} {\bibinfo
  {journal} {Nat. Phys.}\ }\textbf {\bibinfo {volume} {16}},\ \bibinfo {pages}
  {761} (\bibinfo {year} {2020})}\BibitemShut {NoStop}%
\bibitem [{\citenamefont {Zhang}\ \emph
  {et~al.}(2022{\natexlab{b}})\citenamefont {Zhang}, \citenamefont {Yang},\
  and\ \citenamefont {Fang}}]{gdse}%
  \BibitemOpen
  \bibfield  {author} {\bibinfo {author} {\bibfnamefont {K.}~\bibnamefont
  {Zhang}}, \bibinfo {author} {\bibfnamefont {Z.}~\bibnamefont {Yang}}, \ and\
  \bibinfo {author} {\bibfnamefont {C.}~\bibnamefont {Fang}},\ }\href
  {https://doi.org/10.1038/s41467-022-30161-6} {\bibfield  {journal} {\bibinfo
  {journal} {Nat. Commun.}\ }\textbf {\bibinfo {volume} {13}},\ \bibinfo
  {pages} {2496} (\bibinfo {year} {2022}{\natexlab{b}})}\BibitemShut {NoStop}%
\bibitem [{\citenamefont {Borgnia}\ \emph {et~al.}(2020)\citenamefont
  {Borgnia}, \citenamefont {Kruchkov},\ and\ \citenamefont
  {Slager}}]{Prlborgnia}%
  \BibitemOpen
  \bibfield  {author} {\bibinfo {author} {\bibfnamefont {D.~S.}\ \bibnamefont
  {Borgnia}}, \bibinfo {author} {\bibfnamefont {A.~J.}\ \bibnamefont
  {Kruchkov}}, \ and\ \bibinfo {author} {\bibfnamefont {R.-J.}\ \bibnamefont
  {Slager}},\ }\href {https://link.aps.org/doi/10.1103/PhysRevLett.124.056802}
  {\bibfield  {journal} {\bibinfo  {journal} {Phys. Rev. Lett.}\ }\textbf
  {\bibinfo {volume} {124}},\ \bibinfo {pages} {056802} (\bibinfo {year}
  {2020})}\BibitemShut {NoStop}%
\bibitem [{\citenamefont {Lee}\ and\ \citenamefont {Thomale}(2019)}]{lchTR}%
  \BibitemOpen
  \bibfield  {author} {\bibinfo {author} {\bibfnamefont {C.~H.}\ \bibnamefont
  {Lee}}\ and\ \bibinfo {author} {\bibfnamefont {R.}~\bibnamefont {Thomale}},\
  }\href {\doibase 10.1103/PhysRevB.99.201103} {\bibfield  {journal} {\bibinfo
  {journal} {Phys. Rev. B}\ }\textbf {\bibinfo {volume} {99}},\ \bibinfo
  {pages} {201103} (\bibinfo {year} {2019})}\BibitemShut {NoStop}%
\bibitem [{\citenamefont {Song}\ \emph
  {et~al.}(2019{\natexlab{a}})\citenamefont {Song}, \citenamefont {Yao},\ and\
  \citenamefont {Wang}}]{Prl123170401}%
  \BibitemOpen
  \bibfield  {author} {\bibinfo {author} {\bibfnamefont {F.}~\bibnamefont
  {Song}}, \bibinfo {author} {\bibfnamefont {S.}~\bibnamefont {Yao}}, \ and\
  \bibinfo {author} {\bibfnamefont {Z.}~\bibnamefont {Wang}},\ }\href
  {https://link.aps.org/doi/10.1103/PhysRevLett.123.170401} {\bibfield
  {journal} {\bibinfo  {journal} {Phys. Rev. Lett.}\ }\textbf {\bibinfo
  {volume} {123}},\ \bibinfo {pages} {170401} (\bibinfo {year}
  {2019}{\natexlab{a}})}\BibitemShut {NoStop}%
\bibitem [{\citenamefont {Song}\ \emph
  {et~al.}(2019{\natexlab{b}})\citenamefont {Song}, \citenamefont {Yao},\ and\
  \citenamefont {Wang}}]{songfeiprl}%
  \BibitemOpen
  \bibfield  {author} {\bibinfo {author} {\bibfnamefont {F.}~\bibnamefont
  {Song}}, \bibinfo {author} {\bibfnamefont {S.}~\bibnamefont {Yao}}, \ and\
  \bibinfo {author} {\bibfnamefont {Z.}~\bibnamefont {Wang}},\ }\href
  {https://link.aps.org/doi/10.1103/PhysRevLett.123.246801} {\bibfield
  {journal} {\bibinfo  {journal} {Phys. Rev. Lett.}\ }\textbf {\bibinfo
  {volume} {123}},\ \bibinfo {pages} {246801} (\bibinfo {year}
  {2019}{\natexlab{b}})}\BibitemShut {NoStop}%
\bibitem [{\citenamefont {Yi}\ and\ \citenamefont {Yang}(2020)}]{Prl125186}%
  \BibitemOpen
  \bibfield  {author} {\bibinfo {author} {\bibfnamefont {Y.}~\bibnamefont
  {Yi}}\ and\ \bibinfo {author} {\bibfnamefont {Z.}~\bibnamefont {Yang}},\
  }\href {\doibase 10.1103/PhysRevLett.125.186802} {\bibfield  {journal}
  {\bibinfo  {journal} {Phys. Rev. Lett.}\ }\textbf {\bibinfo {volume} {125}},\
  \bibinfo {pages} {186802} (\bibinfo {year} {2020})}\BibitemShut {NoStop}%
\bibitem [{\citenamefont {Lee}\ \emph {et~al.}(2019)\citenamefont {Lee},
  \citenamefont {Li},\ and\ \citenamefont {Gong}}]{prlgjb}%
  \BibitemOpen
  \bibfield  {author} {\bibinfo {author} {\bibfnamefont {C.~H.}\ \bibnamefont
  {Lee}}, \bibinfo {author} {\bibfnamefont {L.}~\bibnamefont {Li}}, \ and\
  \bibinfo {author} {\bibfnamefont {J.}~\bibnamefont {Gong}},\ }\href
  {https://link.aps.org/doi/10.1103/PhysRevLett.123.016805} {\bibfield
  {journal} {\bibinfo  {journal} {Phys. Rev. Lett.}\ }\textbf {\bibinfo
  {volume} {123}},\ \bibinfo {pages} {016805} (\bibinfo {year}
  {2019})}\BibitemShut {NoStop}%
\bibitem [{\citenamefont {Kawabata}\ \emph
  {et~al.}(2020{\natexlab{a}})\citenamefont {Kawabata}, \citenamefont {Sato},\
  and\ \citenamefont {Shiozaki}}]{prbhigherorder}%
  \BibitemOpen
  \bibfield  {author} {\bibinfo {author} {\bibfnamefont {K.}~\bibnamefont
  {Kawabata}}, \bibinfo {author} {\bibfnamefont {M.}~\bibnamefont {Sato}}, \
  and\ \bibinfo {author} {\bibfnamefont {K.}~\bibnamefont {Shiozaki}},\ }\href
  {\doibase 10.1103/PhysRevB.102.205118} {\bibfield  {journal} {\bibinfo
  {journal} {Phys. Rev. B}\ }\textbf {\bibinfo {volume} {102}},\ \bibinfo
  {pages} {205118} (\bibinfo {year} {2020}{\natexlab{a}})}\BibitemShut
  {NoStop}%
\bibitem [{\citenamefont {Okugawa}\ \emph {et~al.}(2020)\citenamefont
  {Okugawa}, \citenamefont {Takahashi},\ and\ \citenamefont
  {Yokomizo}}]{secnhse}%
  \BibitemOpen
  \bibfield  {author} {\bibinfo {author} {\bibfnamefont {R.}~\bibnamefont
  {Okugawa}}, \bibinfo {author} {\bibfnamefont {R.}~\bibnamefont {Takahashi}},
  \ and\ \bibinfo {author} {\bibfnamefont {K.}~\bibnamefont {Yokomizo}},\
  }\href {https://link.aps.org/doi/10.1103/PhysRevB.102.241202} {\bibfield
  {journal} {\bibinfo  {journal} {Phys. Rev. B}\ }\textbf {\bibinfo {volume}
  {102}},\ \bibinfo {pages} {241202} (\bibinfo {year} {2020})}\BibitemShut
  {NoStop}%
\bibitem [{\citenamefont {Li}\ \emph {et~al.}(2020{\natexlab{b}})\citenamefont
  {Li}, \citenamefont {Lee},\ and\ \citenamefont {Gong}}]{tpswitch}%
  \BibitemOpen
  \bibfield  {author} {\bibinfo {author} {\bibfnamefont {L.}~\bibnamefont
  {Li}}, \bibinfo {author} {\bibfnamefont {C.~H.}\ \bibnamefont {Lee}}, \ and\
  \bibinfo {author} {\bibfnamefont {J.}~\bibnamefont {Gong}},\ }\href
  {https://link.aps.org/doi/10.1103/PhysRevLett.124.250402} {\bibfield
  {journal} {\bibinfo  {journal} {Phys. Rev. Lett.}\ }\textbf {\bibinfo
  {volume} {124}},\ \bibinfo {pages} {250402} (\bibinfo {year}
  {2020}{\natexlab{b}})}\BibitemShut {NoStop}%
\bibitem [{\citenamefont {Kunst}\ and\ \citenamefont
  {Dwivedi}(2019)}]{transferkunst}%
  \BibitemOpen
  \bibfield  {author} {\bibinfo {author} {\bibfnamefont {F.~K.}\ \bibnamefont
  {Kunst}}\ and\ \bibinfo {author} {\bibfnamefont {V.}~\bibnamefont
  {Dwivedi}},\ }\href {\doibase 10.1103/PhysRevB.99.245116} {\bibfield
  {journal} {\bibinfo  {journal} {Phys. Rev. B}\ }\textbf {\bibinfo {volume}
  {99}},\ \bibinfo {pages} {245116} (\bibinfo {year} {2019})}\BibitemShut
  {NoStop}%
\bibitem [{\citenamefont {Li}\ \emph {et~al.}(2022)\citenamefont {Li},
  \citenamefont {Liang}, \citenamefont {Wang}, \citenamefont {Lu},\ and\
  \citenamefont {Liu}}]{lyc}%
  \BibitemOpen
  \bibfield  {author} {\bibinfo {author} {\bibfnamefont {Y.}~\bibnamefont
  {Li}}, \bibinfo {author} {\bibfnamefont {C.}~\bibnamefont {Liang}}, \bibinfo
  {author} {\bibfnamefont {C.}~\bibnamefont {Wang}}, \bibinfo {author}
  {\bibfnamefont {C.}~\bibnamefont {Lu}}, \ and\ \bibinfo {author}
  {\bibfnamefont {Y.-C.}\ \bibnamefont {Liu}},\ }\href
  {https://link.aps.org/doi/10.1103/PhysRevLett.128.223903} {\bibfield
  {journal} {\bibinfo  {journal} {Phys. Rev. Lett.}\ }\textbf {\bibinfo
  {volume} {128}},\ \bibinfo {pages} {223903} (\bibinfo {year}
  {2022})}\BibitemShut {NoStop}%
\bibitem [{\citenamefont {Kawabata}\ \emph {et~al.}(2023)\citenamefont
  {Kawabata}, \citenamefont {Numasawa},\ and\ \citenamefont
  {Ryu}}]{entangle-nhse}%
  \BibitemOpen
  \bibfield  {author} {\bibinfo {author} {\bibfnamefont {K.}~\bibnamefont
  {Kawabata}}, \bibinfo {author} {\bibfnamefont {T.}~\bibnamefont {Numasawa}},
  \ and\ \bibinfo {author} {\bibfnamefont {S.}~\bibnamefont {Ryu}},\ }\href
  {https://link.aps.org/doi/10.1103/PhysRevX.13.021007} {\bibfield  {journal}
  {\bibinfo  {journal} {Phys. Rev. X}\ }\textbf {\bibinfo {volume} {13}},\
  \bibinfo {pages} {021007} (\bibinfo {year} {2023})}\BibitemShut {NoStop}%
\bibitem [{\citenamefont {Kawabata}\ \emph
  {et~al.}(2020{\natexlab{b}})\citenamefont {Kawabata}, \citenamefont {Okuma},\
  and\ \citenamefont {Sato}}]{PrbKOM}%
  \BibitemOpen
  \bibfield  {author} {\bibinfo {author} {\bibfnamefont {K.}~\bibnamefont
  {Kawabata}}, \bibinfo {author} {\bibfnamefont {N.}~\bibnamefont {Okuma}}, \
  and\ \bibinfo {author} {\bibfnamefont {M.}~\bibnamefont {Sato}},\ }\href
  {https://link.aps.org/doi/10.1103/PhysRevB.101.195147} {\bibfield  {journal}
  {\bibinfo  {journal} {Phys. Rev. B}\ }\textbf {\bibinfo {volume} {101}},\
  \bibinfo {pages} {195147} (\bibinfo {year} {2020}{\natexlab{b}})}\BibitemShut
  {NoStop}%
\bibitem [{\citenamefont {Guo}\ \emph {et~al.}(2021)\citenamefont {Guo},
  \citenamefont {Liu}, \citenamefont {Zhao}, \citenamefont {Liu},\ and\
  \citenamefont {Chen}}]{gcx-cs}%
  \BibitemOpen
  \bibfield  {author} {\bibinfo {author} {\bibfnamefont {C.-X.}\ \bibnamefont
  {Guo}}, \bibinfo {author} {\bibfnamefont {C.-H.}\ \bibnamefont {Liu}},
  \bibinfo {author} {\bibfnamefont {X.-M.}\ \bibnamefont {Zhao}}, \bibinfo
  {author} {\bibfnamefont {Y.}~\bibnamefont {Liu}}, \ and\ \bibinfo {author}
  {\bibfnamefont {S.}~\bibnamefont {Chen}},\ }\href
  {https://link.aps.org/doi/10.1103/PhysRevLett.127.116801} {\bibfield
  {journal} {\bibinfo  {journal} {Phys. Rev. Lett.}\ }\textbf {\bibinfo
  {volume} {127}},\ \bibinfo {pages} {116801} (\bibinfo {year}
  {2021})}\BibitemShut {NoStop}%
\bibitem [{\citenamefont {Xue}\ \emph {et~al.}(2022)\citenamefont {Xue},
  \citenamefont {Hu}, \citenamefont {Song},\ and\ \citenamefont {Wang}}]{hym4}%
  \BibitemOpen
  \bibfield  {author} {\bibinfo {author} {\bibfnamefont {W.-T.}\ \bibnamefont
  {Xue}}, \bibinfo {author} {\bibfnamefont {Y.-M.}\ \bibnamefont {Hu}},
  \bibinfo {author} {\bibfnamefont {F.}~\bibnamefont {Song}}, \ and\ \bibinfo
  {author} {\bibfnamefont {Z.}~\bibnamefont {Wang}},\ }\href
  {https://link.aps.org/doi/10.1103/PhysRevLett.128.120401} {\bibfield
  {journal} {\bibinfo  {journal} {Phys. Rev. Lett.}\ }\textbf {\bibinfo
  {volume} {128}},\ \bibinfo {pages} {120401} (\bibinfo {year}
  {2022})}\BibitemShut {NoStop}%
\bibitem [{\citenamefont {Wang}\ \emph {et~al.}(2019)\citenamefont {Wang},
  \citenamefont {Ruan},\ and\ \citenamefont {Zhang}}]{zhjprb}%
  \BibitemOpen
  \bibfield  {author} {\bibinfo {author} {\bibfnamefont {H.}~\bibnamefont
  {Wang}}, \bibinfo {author} {\bibfnamefont {J.}~\bibnamefont {Ruan}}, \ and\
  \bibinfo {author} {\bibfnamefont {H.}~\bibnamefont {Zhang}},\ }\href
  {https://link.aps.org/doi/10.1103/PhysRevB.99.075130} {\bibfield  {journal}
  {\bibinfo  {journal} {Phys. Rev. B}\ }\textbf {\bibinfo {volume} {99}},\
  \bibinfo {pages} {075130} (\bibinfo {year} {2019})}\BibitemShut {NoStop}%
\bibitem [{\citenamefont {Longhi}(2022)}]{self-heal}%
  \BibitemOpen
  \bibfield  {author} {\bibinfo {author} {\bibfnamefont {S.}~\bibnamefont
  {Longhi}},\ }\href {https://link.aps.org/doi/10.1103/PhysRevLett.128.157601}
  {\bibfield  {journal} {\bibinfo  {journal} {Phys. Rev. Lett.}\ }\textbf
  {\bibinfo {volume} {128}},\ \bibinfo {pages} {157601} (\bibinfo {year}
  {2022})}\BibitemShut {NoStop}%
\bibitem [{\citenamefont {Longhi}(2020)}]{Prl124066602}%
  \BibitemOpen
  \bibfield  {author} {\bibinfo {author} {\bibfnamefont {S.}~\bibnamefont
  {Longhi}},\ }\href {\doibase 10.1103/PhysRevLett.124.066602} {\bibfield
  {journal} {\bibinfo  {journal} {Phys. Rev. Lett.}\ }\textbf {\bibinfo
  {volume} {124}},\ \bibinfo {pages} {066602} (\bibinfo {year}
  {2020})}\BibitemShut {NoStop}%
\bibitem [{\citenamefont {Mandal}\ \emph {et~al.}(2020)\citenamefont {Mandal},
  \citenamefont {Banerjee}, \citenamefont {Ostrovskaya},\ and\ \citenamefont
  {Liew}}]{polari}%
  \BibitemOpen
  \bibfield  {author} {\bibinfo {author} {\bibfnamefont {S.}~\bibnamefont
  {Mandal}}, \bibinfo {author} {\bibfnamefont {R.}~\bibnamefont {Banerjee}},
  \bibinfo {author} {\bibfnamefont {E.~A.}\ \bibnamefont {Ostrovskaya}}, \ and\
  \bibinfo {author} {\bibfnamefont {T.~C.~H.}\ \bibnamefont {Liew}},\ }\href
  {https://link.aps.org/doi/10.1103/PhysRevLett.125.123902} {\bibfield
  {journal} {\bibinfo  {journal} {Phys. Rev. Lett.}\ }\textbf {\bibinfo
  {volume} {125}},\ \bibinfo {pages} {123902} (\bibinfo {year}
  {2020})}\BibitemShut {NoStop}%
\bibitem [{\citenamefont {Jiang}\ \emph {et~al.}(2019)\citenamefont {Jiang},
  \citenamefont {Lang}, \citenamefont {Yang}, \citenamefont {Zhu},\ and\
  \citenamefont {Chen}}]{Prbjh}%
  \BibitemOpen
  \bibfield  {author} {\bibinfo {author} {\bibfnamefont {H.}~\bibnamefont
  {Jiang}}, \bibinfo {author} {\bibfnamefont {L.-J.}\ \bibnamefont {Lang}},
  \bibinfo {author} {\bibfnamefont {C.}~\bibnamefont {Yang}}, \bibinfo {author}
  {\bibfnamefont {S.-L.}\ \bibnamefont {Zhu}}, \ and\ \bibinfo {author}
  {\bibfnamefont {S.}~\bibnamefont {Chen}},\ }\href
  {https://link.aps.org/doi/10.1103/PhysRevB.100.054301} {\bibfield  {journal}
  {\bibinfo  {journal} {Phys. Rev. B}\ }\textbf {\bibinfo {volume} {100}},\
  \bibinfo {pages} {054301} (\bibinfo {year} {2019})}\BibitemShut {NoStop}%
\bibitem [{\citenamefont {Deng}\ and\ \citenamefont {Yi}(2019)}]{PrbDY}%
  \BibitemOpen
  \bibfield  {author} {\bibinfo {author} {\bibfnamefont {T.-S.}\ \bibnamefont
  {Deng}}\ and\ \bibinfo {author} {\bibfnamefont {W.}~\bibnamefont {Yi}},\
  }\href {https://link.aps.org/doi/10.1103/PhysRevB.100.035102} {\bibfield
  {journal} {\bibinfo  {journal} {Phys. Rev. B}\ }\textbf {\bibinfo {volume}
  {100}},\ \bibinfo {pages} {035102} (\bibinfo {year} {2019})}\BibitemShut
  {NoStop}%
\bibitem [{\citenamefont {Longhi}(2019)}]{Prr1023013}%
  \BibitemOpen
  \bibfield  {author} {\bibinfo {author} {\bibfnamefont {S.}~\bibnamefont
  {Longhi}},\ }\href
  {https://link.aps.org/doi/10.1103/PhysRevResearch.1.023013} {\bibfield
  {journal} {\bibinfo  {journal} {Phys. Rev. Res.}\ }\textbf {\bibinfo {volume}
  {1}},\ \bibinfo {pages} {023013} (\bibinfo {year} {2019})}\BibitemShut
  {NoStop}%
\bibitem [{\citenamefont {Lee}\ \emph {et~al.}(2020)\citenamefont {Lee},
  \citenamefont {Li}, \citenamefont {Thomale},\ and\ \citenamefont
  {Gong}}]{Prb102085151}%
  \BibitemOpen
  \bibfield  {author} {\bibinfo {author} {\bibfnamefont {C.~H.}\ \bibnamefont
  {Lee}}, \bibinfo {author} {\bibfnamefont {L.}~\bibnamefont {Li}}, \bibinfo
  {author} {\bibfnamefont {R.}~\bibnamefont {Thomale}}, \ and\ \bibinfo
  {author} {\bibfnamefont {J.}~\bibnamefont {Gong}},\ }\href
  {https://link.aps.org/doi/10.1103/PhysRevB.102.085151} {\bibfield  {journal}
  {\bibinfo  {journal} {Phys. Rev. B}\ }\textbf {\bibinfo {volume} {102}},\
  \bibinfo {pages} {085151} (\bibinfo {year} {2020})}\BibitemShut {NoStop}%
\bibitem [{\citenamefont {Zhang}\ and\ \citenamefont {Gong}(2020)}]{Prbzxz}%
  \BibitemOpen
  \bibfield  {author} {\bibinfo {author} {\bibfnamefont {X.}~\bibnamefont
  {Zhang}}\ and\ \bibinfo {author} {\bibfnamefont {J.}~\bibnamefont {Gong}},\
  }\href {\doibase 10.1103/PhysRevB.101.045415} {\bibfield  {journal} {\bibinfo
   {journal} {Phys. Rev. B}\ }\textbf {\bibinfo {volume} {101}},\ \bibinfo
  {pages} {045415} (\bibinfo {year} {2020})}\BibitemShut {NoStop}%
\bibitem [{\citenamefont {Weidemann}\ \emph {et~al.}(2020)\citenamefont
  {Weidemann}, \citenamefont {Kremer}, \citenamefont {Helbig}, \citenamefont
  {Hofmann}, \citenamefont {Stegmaier}, \citenamefont {Greiter}, \citenamefont
  {Thomale},\ and\ \citenamefont {Szameit}}]{scienceaaz8727}%
  \BibitemOpen
  \bibfield  {author} {\bibinfo {author} {\bibfnamefont {S.}~\bibnamefont
  {Weidemann}}, \bibinfo {author} {\bibfnamefont {M.}~\bibnamefont {Kremer}},
  \bibinfo {author} {\bibfnamefont {T.}~\bibnamefont {Helbig}}, \bibinfo
  {author} {\bibfnamefont {T.}~\bibnamefont {Hofmann}}, \bibinfo {author}
  {\bibfnamefont {A.}~\bibnamefont {Stegmaier}}, \bibinfo {author}
  {\bibfnamefont {M.}~\bibnamefont {Greiter}}, \bibinfo {author} {\bibfnamefont
  {R.}~\bibnamefont {Thomale}}, \ and\ \bibinfo {author} {\bibfnamefont
  {A.}~\bibnamefont {Szameit}},\ }\href
  {https://www.science.org/doi/abs/10.1126/science.aaz8727} {\bibfield
  {journal} {\bibinfo  {journal} {Science}\ }\textbf {\bibinfo {volume}
  {368}},\ \bibinfo {pages} {311} (\bibinfo {year} {2020})}\BibitemShut
  {NoStop}%
\bibitem [{\citenamefont {Xiao}\ \emph {et~al.}(2021)\citenamefont {Xiao},
  \citenamefont {Deng}, \citenamefont {Wang}, \citenamefont {Wang},
  \citenamefont {Yi},\ and\ \citenamefont {Xue}}]{NB-EP}%
  \BibitemOpen
  \bibfield  {author} {\bibinfo {author} {\bibfnamefont {L.}~\bibnamefont
  {Xiao}}, \bibinfo {author} {\bibfnamefont {T.}~\bibnamefont {Deng}}, \bibinfo
  {author} {\bibfnamefont {K.}~\bibnamefont {Wang}}, \bibinfo {author}
  {\bibfnamefont {Z.}~\bibnamefont {Wang}}, \bibinfo {author} {\bibfnamefont
  {W.}~\bibnamefont {Yi}}, \ and\ \bibinfo {author} {\bibfnamefont
  {P.}~\bibnamefont {Xue}},\ }\href
  {https://link.aps.org/doi/10.1103/PhysRevLett.126.230402} {\bibfield
  {journal} {\bibinfo  {journal} {Phys. Rev. Lett.}\ }\textbf {\bibinfo
  {volume} {126}},\ \bibinfo {pages} {230402} (\bibinfo {year}
  {2021})}\BibitemShut {NoStop}%
\bibitem [{\citenamefont {Lin}\ \emph {et~al.}(2022{\natexlab{a}})\citenamefont
  {Lin}, \citenamefont {Li}, \citenamefont {Xiao}, \citenamefont {Wang},
  \citenamefont {Yi},\ and\ \citenamefont {Xue}}]{xpanderson}%
  \BibitemOpen
  \bibfield  {author} {\bibinfo {author} {\bibfnamefont {Q.}~\bibnamefont
  {Lin}}, \bibinfo {author} {\bibfnamefont {T.}~\bibnamefont {Li}}, \bibinfo
  {author} {\bibfnamefont {L.}~\bibnamefont {Xiao}}, \bibinfo {author}
  {\bibfnamefont {K.}~\bibnamefont {Wang}}, \bibinfo {author} {\bibfnamefont
  {W.}~\bibnamefont {Yi}}, \ and\ \bibinfo {author} {\bibfnamefont
  {P.}~\bibnamefont {Xue}},\ }\href
  {https://doi.org/10.1038/s41467-022-30938-9} {\bibfield  {journal} {\bibinfo
  {journal} {Nat. Commun.}\ }\textbf {\bibinfo {volume} {13}},\ \bibinfo
  {pages} {3229} (\bibinfo {year} {2022}{\natexlab{a}})}\BibitemShut {NoStop}%
\bibitem [{\citenamefont {Lin}\ \emph {et~al.}(2022{\natexlab{b}})\citenamefont
  {Lin}, \citenamefont {Li}, \citenamefont {Xiao}, \citenamefont {Wang},
  \citenamefont {Yi},\ and\ \citenamefont {Xue}}]{Prl129113601}%
  \BibitemOpen
  \bibfield  {author} {\bibinfo {author} {\bibfnamefont {Q.}~\bibnamefont
  {Lin}}, \bibinfo {author} {\bibfnamefont {T.}~\bibnamefont {Li}}, \bibinfo
  {author} {\bibfnamefont {L.}~\bibnamefont {Xiao}}, \bibinfo {author}
  {\bibfnamefont {K.}~\bibnamefont {Wang}}, \bibinfo {author} {\bibfnamefont
  {W.}~\bibnamefont {Yi}}, \ and\ \bibinfo {author} {\bibfnamefont
  {P.}~\bibnamefont {Xue}},\ }\href
  {https://link.aps.org/doi/10.1103/PhysRevLett.129.113601} {\bibfield
  {journal} {\bibinfo  {journal} {Phys. Rev. Lett.}\ }\textbf {\bibinfo
  {volume} {129}},\ \bibinfo {pages} {113601} (\bibinfo {year}
  {2022}{\natexlab{b}})}\BibitemShut {NoStop}%
\bibitem [{\citenamefont {Gao}\ \emph {et~al.}(2023)\citenamefont {Gao},
  \citenamefont {Qiao}, \citenamefont {Pan}, \citenamefont {Wu}, \citenamefont
  {Yim}, \citenamefont {Chen}, \citenamefont {Midya}, \citenamefont {Ge},\ and\
  \citenamefont {Feng}}]{Prl130263801}%
  \BibitemOpen
  \bibfield  {author} {\bibinfo {author} {\bibfnamefont {Z.}~\bibnamefont
  {Gao}}, \bibinfo {author} {\bibfnamefont {X.}~\bibnamefont {Qiao}}, \bibinfo
  {author} {\bibfnamefont {M.}~\bibnamefont {Pan}}, \bibinfo {author}
  {\bibfnamefont {S.}~\bibnamefont {Wu}}, \bibinfo {author} {\bibfnamefont
  {J.}~\bibnamefont {Yim}}, \bibinfo {author} {\bibfnamefont {K.}~\bibnamefont
  {Chen}}, \bibinfo {author} {\bibfnamefont {B.}~\bibnamefont {Midya}},
  \bibinfo {author} {\bibfnamefont {L.}~\bibnamefont {Ge}}, \ and\ \bibinfo
  {author} {\bibfnamefont {L.}~\bibnamefont {Feng}},\ }\href
  {https://link.aps.org/doi/10.1103/PhysRevLett.130.263801} {\bibfield
  {journal} {\bibinfo  {journal} {Phys. Rev. Lett.}\ }\textbf {\bibinfo
  {volume} {130}},\ \bibinfo {pages} {263801} (\bibinfo {year}
  {2023})}\BibitemShut {NoStop}%
\bibitem [{\citenamefont {{Leefmans}}\ \emph {et~al.}(2024)\citenamefont
  {{Leefmans}}, \citenamefont {{Parto}}, \citenamefont {{Williams}},
  \citenamefont {{Li}}, \citenamefont {{Dutt}}, \citenamefont {{Nori}},\ and\
  \citenamefont {{Marandi}}}]{2024NatPh}%
  \BibitemOpen
  \bibfield  {author} {\bibinfo {author} {\bibfnamefont {C.~R.}\ \bibnamefont
  {{Leefmans}}}, \bibinfo {author} {\bibfnamefont {M.}~\bibnamefont {{Parto}}},
  \bibinfo {author} {\bibfnamefont {J.}~\bibnamefont {{Williams}}}, \bibinfo
  {author} {\bibfnamefont {G.~H.~Y.}\ \bibnamefont {{Li}}}, \bibinfo {author}
  {\bibfnamefont {A.}~\bibnamefont {{Dutt}}}, \bibinfo {author} {\bibfnamefont
  {F.}~\bibnamefont {{Nori}}}, \ and\ \bibinfo {author} {\bibfnamefont
  {A.}~\bibnamefont {{Marandi}}},\ }\href
  {https://doi.org/10.1038/s41567-024-02420-4} {\bibfield  {journal} {\bibinfo
  {journal} {Nat. Phys.}\ }\textbf {\bibinfo {volume} {20}},\ \bibinfo {pages}
  {852} (\bibinfo {year} {2024})}\BibitemShut {NoStop}%
\bibitem [{\citenamefont {Lin}\ \emph {et~al.}(2024)\citenamefont {Lin},
  \citenamefont {Song}, \citenamefont {Wang}, \citenamefont {Xin},
  \citenamefont {Sun}, \citenamefont {Wu}, \citenamefont {Huang}, \citenamefont
  {Zhu}, \citenamefont {Jiang},\ and\ \citenamefont {Li}}]{Prl133073803}%
  \BibitemOpen
  \bibfield  {author} {\bibinfo {author} {\bibfnamefont {Z.}~\bibnamefont
  {Lin}}, \bibinfo {author} {\bibfnamefont {W.}~\bibnamefont {Song}}, \bibinfo
  {author} {\bibfnamefont {L.-W.}\ \bibnamefont {Wang}}, \bibinfo {author}
  {\bibfnamefont {H.}~\bibnamefont {Xin}}, \bibinfo {author} {\bibfnamefont
  {J.}~\bibnamefont {Sun}}, \bibinfo {author} {\bibfnamefont {S.}~\bibnamefont
  {Wu}}, \bibinfo {author} {\bibfnamefont {C.}~\bibnamefont {Huang}}, \bibinfo
  {author} {\bibfnamefont {S.}~\bibnamefont {Zhu}}, \bibinfo {author}
  {\bibfnamefont {J.-H.}\ \bibnamefont {Jiang}}, \ and\ \bibinfo {author}
  {\bibfnamefont {T.}~\bibnamefont {Li}},\ }\href
  {https://link.aps.org/doi/10.1103/PhysRevLett.133.073803} {\bibfield
  {journal} {\bibinfo  {journal} {Phys. Rev. Lett.}\ }\textbf {\bibinfo
  {volume} {133}},\ \bibinfo {pages} {073803} (\bibinfo {year}
  {2024})}\BibitemShut {NoStop}%
\bibitem [{\citenamefont {Xiao}\ \emph {et~al.}(2024)\citenamefont {Xiao},
  \citenamefont {Xue}, \citenamefont {Song}, \citenamefont {Hu}, \citenamefont
  {Yi}, \citenamefont {Wang},\ and\ \citenamefont {Xue}}]{Prl133070801}%
  \BibitemOpen
  \bibfield  {author} {\bibinfo {author} {\bibfnamefont {L.}~\bibnamefont
  {Xiao}}, \bibinfo {author} {\bibfnamefont {W.-T.}\ \bibnamefont {Xue}},
  \bibinfo {author} {\bibfnamefont {F.}~\bibnamefont {Song}}, \bibinfo {author}
  {\bibfnamefont {Y.-M.}\ \bibnamefont {Hu}}, \bibinfo {author} {\bibfnamefont
  {W.}~\bibnamefont {Yi}}, \bibinfo {author} {\bibfnamefont {Z.}~\bibnamefont
  {Wang}}, \ and\ \bibinfo {author} {\bibfnamefont {P.}~\bibnamefont {Xue}},\
  }\href {https://link.aps.org/doi/10.1103/PhysRevLett.133.070801} {\bibfield
  {journal} {\bibinfo  {journal} {Phys. Rev. Lett.}\ }\textbf {\bibinfo
  {volume} {133}},\ \bibinfo {pages} {070801} (\bibinfo {year}
  {2024})}\BibitemShut {NoStop}%
\bibitem [{\citenamefont {Liu}\ \emph {et~al.}(2024)\citenamefont {Liu},
  \citenamefont {Mandal}, \citenamefont {Zhou}, \citenamefont {Xi},
  \citenamefont {Banerjee}, \citenamefont {Hu}, \citenamefont {Wei},
  \citenamefont {Wang}, \citenamefont {Wang}, \citenamefont {Gao},
  \citenamefont {Chen}, \citenamefont {Yang}, \citenamefont {Chong},\ and\
  \citenamefont {Zhang}}]{Prl132113802}%
  \BibitemOpen
  \bibfield  {author} {\bibinfo {author} {\bibfnamefont {G.-G.}\ \bibnamefont
  {Liu}}, \bibinfo {author} {\bibfnamefont {S.}~\bibnamefont {Mandal}},
  \bibinfo {author} {\bibfnamefont {P.}~\bibnamefont {Zhou}}, \bibinfo {author}
  {\bibfnamefont {X.}~\bibnamefont {Xi}}, \bibinfo {author} {\bibfnamefont
  {R.}~\bibnamefont {Banerjee}}, \bibinfo {author} {\bibfnamefont {Y.-H.}\
  \bibnamefont {Hu}}, \bibinfo {author} {\bibfnamefont {M.}~\bibnamefont
  {Wei}}, \bibinfo {author} {\bibfnamefont {M.}~\bibnamefont {Wang}}, \bibinfo
  {author} {\bibfnamefont {Q.}~\bibnamefont {Wang}}, \bibinfo {author}
  {\bibfnamefont {Z.}~\bibnamefont {Gao}}, \bibinfo {author} {\bibfnamefont
  {H.}~\bibnamefont {Chen}}, \bibinfo {author} {\bibfnamefont {Y.}~\bibnamefont
  {Yang}}, \bibinfo {author} {\bibfnamefont {Y.}~\bibnamefont {Chong}}, \ and\
  \bibinfo {author} {\bibfnamefont {B.}~\bibnamefont {Zhang}},\ }\href
  {https://link.aps.org/doi/10.1103/PhysRevLett.132.113802} {\bibfield
  {journal} {\bibinfo  {journal} {Phys. Rev. Lett.}\ }\textbf {\bibinfo
  {volume} {132}},\ \bibinfo {pages} {113802} (\bibinfo {year}
  {2024})}\BibitemShut {NoStop}%
\bibitem [{\citenamefont {Sun}\ \emph {et~al.}(2024)\citenamefont {Sun},
  \citenamefont {Hou}, \citenamefont {Wan}, \citenamefont {Wang}, \citenamefont
  {Zhu}, \citenamefont {Ruan},\ and\ \citenamefont {Yang}}]{Prl132063804}%
  \BibitemOpen
  \bibfield  {author} {\bibinfo {author} {\bibfnamefont {Y.}~\bibnamefont
  {Sun}}, \bibinfo {author} {\bibfnamefont {X.}~\bibnamefont {Hou}}, \bibinfo
  {author} {\bibfnamefont {T.}~\bibnamefont {Wan}}, \bibinfo {author}
  {\bibfnamefont {F.}~\bibnamefont {Wang}}, \bibinfo {author} {\bibfnamefont
  {S.}~\bibnamefont {Zhu}}, \bibinfo {author} {\bibfnamefont {Z.}~\bibnamefont
  {Ruan}}, \ and\ \bibinfo {author} {\bibfnamefont {Z.}~\bibnamefont {Yang}},\
  }\href {https://link.aps.org/doi/10.1103/PhysRevLett.132.063804} {\bibfield
  {journal} {\bibinfo  {journal} {Phys. Rev. Lett.}\ }\textbf {\bibinfo
  {volume} {132}},\ \bibinfo {pages} {063804} (\bibinfo {year}
  {2024})}\BibitemShut {NoStop}%
\bibitem [{\citenamefont {Gao}\ \emph {et~al.}(2024{\natexlab{a}})\citenamefont
  {Gao}, \citenamefont {Sheng}, \citenamefont {Zhao}, \citenamefont {He},
  \citenamefont {Lu}, \citenamefont {Chen}, \citenamefont {Ding}, \citenamefont
  {Zhu},\ and\ \citenamefont {Liu}}]{Prb110094308}%
  \BibitemOpen
  \bibfield  {author} {\bibinfo {author} {\bibfnamefont {M.}~\bibnamefont
  {Gao}}, \bibinfo {author} {\bibfnamefont {C.}~\bibnamefont {Sheng}}, \bibinfo
  {author} {\bibfnamefont {Y.}~\bibnamefont {Zhao}}, \bibinfo {author}
  {\bibfnamefont {R.}~\bibnamefont {He}}, \bibinfo {author} {\bibfnamefont
  {L.}~\bibnamefont {Lu}}, \bibinfo {author} {\bibfnamefont {W.}~\bibnamefont
  {Chen}}, \bibinfo {author} {\bibfnamefont {K.}~\bibnamefont {Ding}}, \bibinfo
  {author} {\bibfnamefont {S.}~\bibnamefont {Zhu}}, \ and\ \bibinfo {author}
  {\bibfnamefont {H.}~\bibnamefont {Liu}},\ }\href
  {https://link.aps.org/doi/10.1103/PhysRevB.110.094308} {\bibfield  {journal}
  {\bibinfo  {journal} {Phys. Rev. B}\ }\textbf {\bibinfo {volume} {110}},\
  \bibinfo {pages} {094308} (\bibinfo {year} {2024}{\natexlab{a}})}\BibitemShut
  {NoStop}%
\bibitem [{\citenamefont {Wang}\ \emph {et~al.}()\citenamefont {Wang},
  \citenamefont {Wang}, \citenamefont {Liu}, \citenamefont {Qin}, \citenamefont
  {Zhao}, \citenamefont {Liu}, \citenamefont {Longhi},\ and\ \citenamefont
  {Lu}}]{Lpx1}%
  \BibitemOpen
  \bibfield  {author} {\bibinfo {author} {\bibfnamefont {S.}~\bibnamefont
  {Wang}}, \bibinfo {author} {\bibfnamefont {B.}~\bibnamefont {Wang}}, \bibinfo
  {author} {\bibfnamefont {C.}~\bibnamefont {Liu}}, \bibinfo {author}
  {\bibfnamefont {C.}~\bibnamefont {Qin}}, \bibinfo {author} {\bibfnamefont
  {L.}~\bibnamefont {Zhao}}, \bibinfo {author} {\bibfnamefont {W.}~\bibnamefont
  {Liu}}, \bibinfo {author} {\bibfnamefont {S.}~\bibnamefont {Longhi}}, \ and\
  \bibinfo {author} {\bibfnamefont {P.}~\bibnamefont {Lu}},\ }\href
  {https://arxiv.org/abs/2409.19693} {\ }\Eprint
  {http://arxiv.org/abs/2409.19693} {arXiv:2409.19693} \BibitemShut {NoStop}%
\bibitem [{\citenamefont {Vicencio}\ \emph {et~al.}()\citenamefont {Vicencio},
  \citenamefont {Román-Cortés}, \citenamefont {Rubio-Saldías}, \citenamefont
  {Vildoso},\ and\ \citenamefont {Torres}}]{vicencio2024}%
  \BibitemOpen
  \bibfield  {author} {\bibinfo {author} {\bibfnamefont {R.~A.}\ \bibnamefont
  {Vicencio}}, \bibinfo {author} {\bibfnamefont {D.}~\bibnamefont
  {Román-Cortés}}, \bibinfo {author} {\bibfnamefont {M.}~\bibnamefont
  {Rubio-Saldías}}, \bibinfo {author} {\bibfnamefont {P.}~\bibnamefont
  {Vildoso}}, \ and\ \bibinfo {author} {\bibfnamefont {L.~E. F.~F.}\
  \bibnamefont {Torres}},\ }\href {https://arxiv.org/abs/2407.18174} {\
  }\Eprint {http://arxiv.org/abs/2407.18174} {arXiv:2407.18174} \BibitemShut
  {NoStop}%
\bibitem [{\citenamefont {Zhang}\ \emph
  {et~al.}(2021{\natexlab{a}})\citenamefont {Zhang}, \citenamefont {Yang},
  \citenamefont {Yong}, \citenamefont {Jun}, \citenamefont {Chen},
  \citenamefont {Yan}, \citenamefont {Chen}, \citenamefont {Xi}, \citenamefont
  {Li}, \citenamefont {Jia}, \citenamefont {Yuan}, \citenamefont {Sun},
  \citenamefont {Chen},\ and\ \citenamefont {Zhang}}]{acous1}%
  \BibitemOpen
  \bibfield  {author} {\bibinfo {author} {\bibfnamefont {L.}~\bibnamefont
  {Zhang}}, \bibinfo {author} {\bibfnamefont {Y.}~\bibnamefont {Yang}},
  \bibinfo {author} {\bibfnamefont {g.}~\bibnamefont {Yong}}, \bibinfo {author}
  {\bibfnamefont {G.}~\bibnamefont {Jun}}, \bibinfo {author} {\bibfnamefont
  {Q.}~\bibnamefont {Chen}}, \bibinfo {author} {\bibfnamefont {Q.}~\bibnamefont
  {Yan}}, \bibinfo {author} {\bibfnamefont {F.}~\bibnamefont {Chen}}, \bibinfo
  {author} {\bibfnamefont {R.}~\bibnamefont {Xi}}, \bibinfo {author}
  {\bibfnamefont {Y.}~\bibnamefont {Li}}, \bibinfo {author} {\bibfnamefont
  {D.}~\bibnamefont {Jia}}, \bibinfo {author} {\bibfnamefont {S.-Q.}\
  \bibnamefont {Yuan}}, \bibinfo {author} {\bibfnamefont {H.-x.}\ \bibnamefont
  {Sun}}, \bibinfo {author} {\bibfnamefont {H.}~\bibnamefont {Chen}}, \ and\
  \bibinfo {author} {\bibfnamefont {B.}~\bibnamefont {Zhang}},\ }\href
  {\doibase 10.1038/s41467-021-26619-8} {\bibfield  {journal} {\bibinfo
  {journal} {Nat. Commun.}\ }\textbf {\bibinfo {volume} {12}},\ \bibinfo
  {pages} {6297} (\bibinfo {year} {2021}{\natexlab{a}})}\BibitemShut {NoStop}%
\bibitem [{\citenamefont {Zhou}\ \emph {et~al.}(2023)\citenamefont {Zhou},
  \citenamefont {Wu}, \citenamefont {Pu}, \citenamefont {Lu}, \citenamefont
  {Huang}, \citenamefont {Deng}, \citenamefont {Ke},\ and\ \citenamefont
  {Liu}}]{lzy}%
  \BibitemOpen
  \bibfield  {author} {\bibinfo {author} {\bibfnamefont {Q.}~\bibnamefont
  {Zhou}}, \bibinfo {author} {\bibfnamefont {J.}~\bibnamefont {Wu}}, \bibinfo
  {author} {\bibfnamefont {Z.}~\bibnamefont {Pu}}, \bibinfo {author}
  {\bibfnamefont {J.}~\bibnamefont {Lu}}, \bibinfo {author} {\bibfnamefont
  {X.}~\bibnamefont {Huang}}, \bibinfo {author} {\bibfnamefont
  {W.}~\bibnamefont {Deng}}, \bibinfo {author} {\bibfnamefont {M.}~\bibnamefont
  {Ke}}, \ and\ \bibinfo {author} {\bibfnamefont {Z.}~\bibnamefont {Liu}},\
  }\href {https://doi.org/10.1038/s41467-023-40236-7} {\bibfield  {journal}
  {\bibinfo  {journal} {Nat. Commun.}\ }\textbf {\bibinfo {volume} {14}},\
  \bibinfo {pages} {4569} (\bibinfo {year} {2023})}\BibitemShut {NoStop}%
\bibitem [{\citenamefont {Gu}\ \emph {et~al.}(2022)\citenamefont {Gu},
  \citenamefont {Gao}, \citenamefont {Xue}, \citenamefont {Li}, \citenamefont
  {Su},\ and\ \citenamefont {Zhu}}]{transient}%
  \BibitemOpen
  \bibfield  {author} {\bibinfo {author} {\bibfnamefont {Z.}~\bibnamefont
  {Gu}}, \bibinfo {author} {\bibfnamefont {H.}~\bibnamefont {Gao}}, \bibinfo
  {author} {\bibfnamefont {H.}~\bibnamefont {Xue}}, \bibinfo {author}
  {\bibfnamefont {J.}~\bibnamefont {Li}}, \bibinfo {author} {\bibfnamefont
  {Z.}~\bibnamefont {Su}}, \ and\ \bibinfo {author} {\bibfnamefont
  {J.}~\bibnamefont {Zhu}},\ }\href {\doibase 10.1038/s41467-022-35448-2}
  {\bibfield  {journal} {\bibinfo  {journal} {Nat. Commun.}\ }\textbf {\bibinfo
  {volume} {13}},\ \bibinfo {pages} {7668} (\bibinfo {year}
  {2022})}\BibitemShut {NoStop}%
\bibitem [{\citenamefont {Zhang}\ \emph
  {et~al.}(2021{\natexlab{b}})\citenamefont {Zhang}, \citenamefont {Tian},
  \citenamefont {Jiang}, \citenamefont {Lu},\ and\ \citenamefont {Chen}}]{zxj}%
  \BibitemOpen
  \bibfield  {author} {\bibinfo {author} {\bibfnamefont {X.}~\bibnamefont
  {Zhang}}, \bibinfo {author} {\bibfnamefont {Y.}~\bibnamefont {Tian}},
  \bibinfo {author} {\bibfnamefont {J.-H.}\ \bibnamefont {Jiang}}, \bibinfo
  {author} {\bibfnamefont {M.-H.}\ \bibnamefont {Lu}}, \ and\ \bibinfo {author}
  {\bibfnamefont {Y.-F.}\ \bibnamefont {Chen}},\ }\href
  {https://doi.org/10.1038/s41467-021-25716-y} {\bibfield  {journal} {\bibinfo
  {journal} {Nat. Commun.}\ }\textbf {\bibinfo {volume} {12}},\ \bibinfo
  {pages} {5377} (\bibinfo {year} {2021}{\natexlab{b}})}\BibitemShut {NoStop}%
\bibitem [{\citenamefont {Wan}\ \emph {et~al.}(2023)\citenamefont {Wan},
  \citenamefont {Zhang}, \citenamefont {Li}, \citenamefont {Yang},\ and\
  \citenamefont {Yang}}]{yzj}%
  \BibitemOpen
  \bibfield  {author} {\bibinfo {author} {\bibfnamefont {T.}~\bibnamefont
  {Wan}}, \bibinfo {author} {\bibfnamefont {K.}~\bibnamefont {Zhang}}, \bibinfo
  {author} {\bibfnamefont {J.}~\bibnamefont {Li}}, \bibinfo {author}
  {\bibfnamefont {Z.}~\bibnamefont {Yang}}, \ and\ \bibinfo {author}
  {\bibfnamefont {Z.}~\bibnamefont {Yang}},\ }\href
  {https://doi.org/10.1016/j.scib.2023.09.013} {\bibfield  {journal} {\bibinfo
  {journal} {Sci. Bull.}\ }\textbf {\bibinfo {volume} {68}},\ \bibinfo {pages}
  {2330} (\bibinfo {year} {2023})}\BibitemShut {NoStop}%
\bibitem [{\citenamefont {Gao}\ \emph {et~al.}(2022)\citenamefont {Gao},
  \citenamefont {Xue}, \citenamefont {Gu}, \citenamefont {Li}, \citenamefont
  {Zhu}, \citenamefont {Su}, \citenamefont {Zhu}, \citenamefont {Zhang},\ and\
  \citenamefont {Chong}}]{Prb106134112}%
  \BibitemOpen
  \bibfield  {author} {\bibinfo {author} {\bibfnamefont {H.}~\bibnamefont
  {Gao}}, \bibinfo {author} {\bibfnamefont {H.}~\bibnamefont {Xue}}, \bibinfo
  {author} {\bibfnamefont {Z.}~\bibnamefont {Gu}}, \bibinfo {author}
  {\bibfnamefont {L.}~\bibnamefont {Li}}, \bibinfo {author} {\bibfnamefont
  {W.}~\bibnamefont {Zhu}}, \bibinfo {author} {\bibfnamefont {Z.}~\bibnamefont
  {Su}}, \bibinfo {author} {\bibfnamefont {J.}~\bibnamefont {Zhu}}, \bibinfo
  {author} {\bibfnamefont {B.}~\bibnamefont {Zhang}}, \ and\ \bibinfo {author}
  {\bibfnamefont {Y.~D.}\ \bibnamefont {Chong}},\ }\href
  {https://link.aps.org/doi/10.1103/PhysRevB.106.134112} {\bibfield  {journal}
  {\bibinfo  {journal} {Phys. Rev. B}\ }\textbf {\bibinfo {volume} {106}},\
  \bibinfo {pages} {134112} (\bibinfo {year} {2022})}\BibitemShut {NoStop}%
\bibitem [{\citenamefont {Wu}\ \emph {et~al.}(2024)\citenamefont {Wu},
  \citenamefont {Zheng}, \citenamefont {Liang}, \citenamefont {Ke},
  \citenamefont {Lu}, \citenamefont {Deng}, \citenamefont {Huang},\ and\
  \citenamefont {Liu}}]{Prl133126601}%
  \BibitemOpen
  \bibfield  {author} {\bibinfo {author} {\bibfnamefont {J.}~\bibnamefont
  {Wu}}, \bibinfo {author} {\bibfnamefont {R.}~\bibnamefont {Zheng}}, \bibinfo
  {author} {\bibfnamefont {J.}~\bibnamefont {Liang}}, \bibinfo {author}
  {\bibfnamefont {M.}~\bibnamefont {Ke}}, \bibinfo {author} {\bibfnamefont
  {J.}~\bibnamefont {Lu}}, \bibinfo {author} {\bibfnamefont {W.}~\bibnamefont
  {Deng}}, \bibinfo {author} {\bibfnamefont {X.}~\bibnamefont {Huang}}, \ and\
  \bibinfo {author} {\bibfnamefont {Z.}~\bibnamefont {Liu}},\ }\href
  {https://link.aps.org/doi/10.1103/PhysRevLett.133.126601} {\bibfield
  {journal} {\bibinfo  {journal} {Phys. Rev. Lett.}\ }\textbf {\bibinfo
  {volume} {133}},\ \bibinfo {pages} {126601} (\bibinfo {year}
  {2024})}\BibitemShut {NoStop}%
\bibitem [{\citenamefont {Xiong}\ \emph {et~al.}(2024)\citenamefont {Xiong},
  \citenamefont {Zhang}, \citenamefont {Feng}, \citenamefont {Leng},
  \citenamefont {Pi}, \citenamefont {Tong},\ and\ \citenamefont
  {Qiu}}]{Prb110L140305}%
  \BibitemOpen
  \bibfield  {author} {\bibinfo {author} {\bibfnamefont {L.}~\bibnamefont
  {Xiong}}, \bibinfo {author} {\bibfnamefont {Q.}~\bibnamefont {Zhang}},
  \bibinfo {author} {\bibfnamefont {X.}~\bibnamefont {Feng}}, \bibinfo {author}
  {\bibfnamefont {Y.}~\bibnamefont {Leng}}, \bibinfo {author} {\bibfnamefont
  {M.}~\bibnamefont {Pi}}, \bibinfo {author} {\bibfnamefont {S.}~\bibnamefont
  {Tong}}, \ and\ \bibinfo {author} {\bibfnamefont {C.}~\bibnamefont {Qiu}},\
  }\href {\doibase 10.1103/PhysRevB.110.L140305} {\bibfield  {journal}
  {\bibinfo  {journal} {Phys. Rev. B}\ }\textbf {\bibinfo {volume} {110}},\
  \bibinfo {pages} {L140305} (\bibinfo {year} {2024})}\BibitemShut {NoStop}%
\bibitem [{\citenamefont {Zhang}\ \emph {et~al.}(2024)\citenamefont {Zhang},
  \citenamefont {Leng}, \citenamefont {Xiong}, \citenamefont {Li},
  \citenamefont {Zhang}, \citenamefont {Qi},\ and\ \citenamefont {Qiu}}]{zqc}%
  \BibitemOpen
  \bibfield  {author} {\bibinfo {author} {\bibfnamefont {Q.}~\bibnamefont
  {Zhang}}, \bibinfo {author} {\bibfnamefont {Y.}~\bibnamefont {Leng}},
  \bibinfo {author} {\bibfnamefont {L.}~\bibnamefont {Xiong}}, \bibinfo
  {author} {\bibfnamefont {Y.}~\bibnamefont {Li}}, \bibinfo {author}
  {\bibfnamefont {K.}~\bibnamefont {Zhang}}, \bibinfo {author} {\bibfnamefont
  {L.}~\bibnamefont {Qi}}, \ and\ \bibinfo {author} {\bibfnamefont
  {C.}~\bibnamefont {Qiu}},\ }\href {https://doi.org/10.1002/adma.202403108}
  {\bibfield  {journal} {\bibinfo  {journal} {Adv. Mater.}\ }\textbf {\bibinfo
  {volume} {36}},\ \bibinfo {pages} {2403108} (\bibinfo {year}
  {2024})}\BibitemShut {NoStop}%
\bibitem [{\citenamefont {Gao}\ \emph {et~al.}(2024{\natexlab{b}})\citenamefont
  {Gao}, \citenamefont {Zhu}, \citenamefont {Xue}, \citenamefont {Ma},\ and\
  \citenamefont {Su}}]{szq-gh}%
  \BibitemOpen
  \bibfield  {author} {\bibinfo {author} {\bibfnamefont {H.}~\bibnamefont
  {Gao}}, \bibinfo {author} {\bibfnamefont {W.}~\bibnamefont {Zhu}}, \bibinfo
  {author} {\bibfnamefont {H.}~\bibnamefont {Xue}}, \bibinfo {author}
  {\bibfnamefont {G.}~\bibnamefont {Ma}}, \ and\ \bibinfo {author}
  {\bibfnamefont {Z.}~\bibnamefont {Su}},\ }\href
  {https://doi.org/10.1063/5.0213867} {\bibfield  {journal} {\bibinfo
  {journal} {Appl. Phys. Rev.}\ }\textbf {\bibinfo {volume} {11}},\ \bibinfo
  {pages} {031410} (\bibinfo {year} {2024}{\natexlab{b}})}\BibitemShut
  {NoStop}%
\bibitem [{\citenamefont {Zhong}\ \emph {et~al.}(2025)\citenamefont {Zhong},
  \citenamefont {de~Castro}, \citenamefont {Lu}, \citenamefont {Kim},
  \citenamefont {Oudich}, \citenamefont {Ji}, \citenamefont {Shi},
  \citenamefont {Chen}, \citenamefont {Lu}, \citenamefont {Jing},\ and\
  \citenamefont {Benalcazar}}]{zhong2024}%
  \BibitemOpen
  \bibfield  {author} {\bibinfo {author} {\bibfnamefont {J.-X.}\ \bibnamefont
  {Zhong}}, \bibinfo {author} {\bibfnamefont {P.~F.}\ \bibnamefont
  {de~Castro}}, \bibinfo {author} {\bibfnamefont {T.}~\bibnamefont {Lu}},
  \bibinfo {author} {\bibfnamefont {J.}~\bibnamefont {Kim}}, \bibinfo {author}
  {\bibfnamefont {M.}~\bibnamefont {Oudich}}, \bibinfo {author} {\bibfnamefont
  {J.}~\bibnamefont {Ji}}, \bibinfo {author} {\bibfnamefont {L.}~\bibnamefont
  {Shi}}, \bibinfo {author} {\bibfnamefont {K.}~\bibnamefont {Chen}}, \bibinfo
  {author} {\bibfnamefont {J.}~\bibnamefont {Lu}}, \bibinfo {author}
  {\bibfnamefont {Y.}~\bibnamefont {Jing}}, \ and\ \bibinfo {author}
  {\bibfnamefont {W.~A.}\ \bibnamefont {Benalcazar}},\ }\href
  {https://link.aps.org/doi/10.1103/PhysRevB.111.014314} {\bibfield  {journal}
  {\bibinfo  {journal} {Phys. Rev. B}\ }\textbf {\bibinfo {volume} {111}},\
  \bibinfo {pages} {014314} (\bibinfo {year} {2025})}\BibitemShut {NoStop}%
\bibitem [{\citenamefont {bing Wang}\ \emph {et~al.}()\citenamefont {bing
  Wang}, \citenamefont {Cheng}, \citenamefont {yu~Zou}, \citenamefont {Ge},
  \citenamefont {qi~Zhao}, \citenamefont {rui Si}, \citenamefont {qi~Yuan},
  \citenamefont {xiang Sun}, \citenamefont {Xue},\ and\ \citenamefont
  {Zhang}}]{wang2024disorder}%
  \BibitemOpen
  \bibfield  {author} {\bibinfo {author} {\bibfnamefont {B.}~\bibnamefont {bing
  Wang}}, \bibinfo {author} {\bibfnamefont {Z.}~\bibnamefont {Cheng}}, \bibinfo
  {author} {\bibfnamefont {H.}~\bibnamefont {yu~Zou}}, \bibinfo {author}
  {\bibfnamefont {Y.}~\bibnamefont {Ge}}, \bibinfo {author} {\bibfnamefont
  {K.}~\bibnamefont {qi~Zhao}}, \bibinfo {author} {\bibfnamefont
  {Q.}~\bibnamefont {rui Si}}, \bibinfo {author} {\bibfnamefont
  {S.}~\bibnamefont {qi~Yuan}}, \bibinfo {author} {\bibfnamefont
  {H.}~\bibnamefont {xiang Sun}}, \bibinfo {author} {\bibfnamefont
  {H.}~\bibnamefont {Xue}}, \ and\ \bibinfo {author} {\bibfnamefont
  {B.}~\bibnamefont {Zhang}},\ }\href {https://arxiv.org/abs/2402.10989} {\
  }\Eprint {http://arxiv.org/abs/2402.10989} {arXiv:2402.10989} \BibitemShut
  {NoStop}%
\bibitem [{\citenamefont {Hu}\ \emph {et~al.}()\citenamefont {Hu},
  \citenamefont {Wu}, \citenamefont {Ye}, \citenamefont {Deng}, \citenamefont
  {Lu}, \citenamefont {Huang}, \citenamefont {Wang}, \citenamefont {Ke},\ and\
  \citenamefont {Liu}}]{hu2024}%
  \BibitemOpen
  \bibfield  {author} {\bibinfo {author} {\bibfnamefont {Y.}~\bibnamefont
  {Hu}}, \bibinfo {author} {\bibfnamefont {J.}~\bibnamefont {Wu}}, \bibinfo
  {author} {\bibfnamefont {P.}~\bibnamefont {Ye}}, \bibinfo {author}
  {\bibfnamefont {W.}~\bibnamefont {Deng}}, \bibinfo {author} {\bibfnamefont
  {J.}~\bibnamefont {Lu}}, \bibinfo {author} {\bibfnamefont {X.}~\bibnamefont
  {Huang}}, \bibinfo {author} {\bibfnamefont {Z.}~\bibnamefont {Wang}},
  \bibinfo {author} {\bibfnamefont {M.}~\bibnamefont {Ke}}, \ and\ \bibinfo
  {author} {\bibfnamefont {Z.}~\bibnamefont {Liu}},\ }\href
  {https://arxiv.org/abs/2407.03868} {\ }\Eprint
  {http://arxiv.org/abs/2407.03868} {arXiv:2407.03868} \BibitemShut {NoStop}%
\bibitem [{\citenamefont {Liang}\ \emph {et~al.}(2022)\citenamefont {Liang},
  \citenamefont {Xie}, \citenamefont {Dong}, \citenamefont {Li}, \citenamefont
  {Li}, \citenamefont {Gadway}, \citenamefont {Yi},\ and\ \citenamefont
  {Yan}}]{Prl129070401}%
  \BibitemOpen
  \bibfield  {author} {\bibinfo {author} {\bibfnamefont {Q.}~\bibnamefont
  {Liang}}, \bibinfo {author} {\bibfnamefont {D.}~\bibnamefont {Xie}}, \bibinfo
  {author} {\bibfnamefont {Z.}~\bibnamefont {Dong}}, \bibinfo {author}
  {\bibfnamefont {H.}~\bibnamefont {Li}}, \bibinfo {author} {\bibfnamefont
  {H.}~\bibnamefont {Li}}, \bibinfo {author} {\bibfnamefont {B.}~\bibnamefont
  {Gadway}}, \bibinfo {author} {\bibfnamefont {W.}~\bibnamefont {Yi}}, \ and\
  \bibinfo {author} {\bibfnamefont {B.}~\bibnamefont {Yan}},\ }\href
  {https://link.aps.org/doi/10.1103/PhysRevLett.129.070401} {\bibfield
  {journal} {\bibinfo  {journal} {Phys. Rev. Lett.}\ }\textbf {\bibinfo
  {volume} {129}},\ \bibinfo {pages} {070401} (\bibinfo {year}
  {2022})}\BibitemShut {NoStop}%
\bibitem [{\citenamefont {Zhao}\ \emph {et~al.}(2025)\citenamefont {Zhao},
  \citenamefont {Wang}, \citenamefont {He}, \citenamefont {Poon}, \citenamefont
  {Pak}, \citenamefont {Liu}, \citenamefont {Ren}, \citenamefont {Liu},\ and\
  \citenamefont {Jo}}]{zhao2024}%
  \BibitemOpen
  \bibfield  {author} {\bibinfo {author} {\bibfnamefont {E.}~\bibnamefont
  {Zhao}}, \bibinfo {author} {\bibfnamefont {Z.}~\bibnamefont {Wang}}, \bibinfo
  {author} {\bibfnamefont {C.}~\bibnamefont {He}}, \bibinfo {author}
  {\bibfnamefont {T.}~\bibnamefont {Poon}}, \bibinfo {author} {\bibfnamefont
  {K.}~\bibnamefont {Pak}}, \bibinfo {author} {\bibfnamefont {Y.-J.}\
  \bibnamefont {Liu}}, \bibinfo {author} {\bibfnamefont {P.}~\bibnamefont
  {Ren}}, \bibinfo {author} {\bibfnamefont {X.-J.}\ \bibnamefont {Liu}}, \ and\
  \bibinfo {author} {\bibfnamefont {G.-B.}\ \bibnamefont {Jo}},\ }\href
  {\doibase https://doi.org/10.1038/s41586-024-08347-3} {\bibfield  {journal}
  {\bibinfo  {journal} {Nature}\ }\textbf {\bibinfo {volume} {637}},\ \bibinfo
  {pages} {565} (\bibinfo {year} {2025})}\BibitemShut {NoStop}%
\bibitem [{\citenamefont {Liu}\ \emph {et~al.}(2021)\citenamefont {Liu},
  \citenamefont {Shao}, \citenamefont {Ma}, \citenamefont {Zhang},
  \citenamefont {You}, \citenamefont {Wu}, \citenamefont {Xiang}, \citenamefont
  {Cui},\ and\ \citenamefont {Zhang}}]{Liu-Zhang}%
  \BibitemOpen
  \bibfield  {author} {\bibinfo {author} {\bibfnamefont {S.}~\bibnamefont
  {Liu}}, \bibinfo {author} {\bibfnamefont {R.}~\bibnamefont {Shao}}, \bibinfo
  {author} {\bibfnamefont {S.}~\bibnamefont {Ma}}, \bibinfo {author}
  {\bibfnamefont {L.}~\bibnamefont {Zhang}}, \bibinfo {author} {\bibfnamefont
  {O.}~\bibnamefont {You}}, \bibinfo {author} {\bibfnamefont {H.}~\bibnamefont
  {Wu}}, \bibinfo {author} {\bibfnamefont {Y.~J.}\ \bibnamefont {Xiang}},
  \bibinfo {author} {\bibfnamefont {T.~J.}\ \bibnamefont {Cui}}, \ and\
  \bibinfo {author} {\bibfnamefont {S.}~\bibnamefont {Zhang}},\ }\href
  {https://spj.science.org/doi/abs/10.34133/2021/5608038} {\bibfield  {journal}
  {\bibinfo  {journal} {Research}\ }\textbf {\bibinfo {volume} {2021}},\
  \bibinfo {pages} {5608038} (\bibinfo {year} {2021})}\BibitemShut {NoStop}%
\bibitem [{\citenamefont {Zou}\ \emph {et~al.}(2021)\citenamefont {Zou},
  \citenamefont {Chen}, \citenamefont {He}, \citenamefont {Bao}, \citenamefont
  {Lee}, \citenamefont {Sun},\ and\ \citenamefont {Zhang}}]{zoudy}%
  \BibitemOpen
  \bibfield  {author} {\bibinfo {author} {\bibfnamefont {D.}~\bibnamefont
  {Zou}}, \bibinfo {author} {\bibfnamefont {T.}~\bibnamefont {Chen}}, \bibinfo
  {author} {\bibfnamefont {W.}~\bibnamefont {He}}, \bibinfo {author}
  {\bibfnamefont {J.}~\bibnamefont {Bao}}, \bibinfo {author} {\bibfnamefont
  {C.~H.}\ \bibnamefont {Lee}}, \bibinfo {author} {\bibfnamefont
  {H.}~\bibnamefont {Sun}}, \ and\ \bibinfo {author} {\bibfnamefont
  {X.}~\bibnamefont {Zhang}},\ }\href {\doibase 10.1038/s41467-021-26414-5}
  {\bibfield  {journal} {\bibinfo  {journal} {Nat. Commun.}\ }\textbf {\bibinfo
  {volume} {12}},\ \bibinfo {pages} {7201} (\bibinfo {year}
  {2021})}\BibitemShut {NoStop}%
\bibitem [{\citenamefont {Hofmann}\ \emph {et~al.}(2020)\citenamefont
  {Hofmann}, \citenamefont {Helbig}, \citenamefont {Schindler}, \citenamefont
  {Salgo}, \citenamefont {Brzezi\ifmmode~\acute{n}\else \'{n}\fi{}ska},
  \citenamefont {Greiter}, \citenamefont {Kiessling}, \citenamefont {Wolf},
  \citenamefont {Vollhardt}, \citenamefont {Kaba\ifmmode~\check{s}\else
  \v{s}\fi{}i}, \citenamefont {Lee}, \citenamefont {Bilu\ifmmode \check{s}\else
  \v{s}\fi{}i\ifmmode~\acute{c}\else \'{c}\fi{}}, \citenamefont {Thomale},\
  and\ \citenamefont {Neupert}}]{Prr20232}%
  \BibitemOpen
  \bibfield  {author} {\bibinfo {author} {\bibfnamefont {T.}~\bibnamefont
  {Hofmann}}, \bibinfo {author} {\bibfnamefont {T.}~\bibnamefont {Helbig}},
  \bibinfo {author} {\bibfnamefont {F.}~\bibnamefont {Schindler}}, \bibinfo
  {author} {\bibfnamefont {N.}~\bibnamefont {Salgo}}, \bibinfo {author}
  {\bibfnamefont {M.}~\bibnamefont {Brzezi\ifmmode~\acute{n}\else
  \'{n}\fi{}ska}}, \bibinfo {author} {\bibfnamefont {M.}~\bibnamefont
  {Greiter}}, \bibinfo {author} {\bibfnamefont {T.}~\bibnamefont {Kiessling}},
  \bibinfo {author} {\bibfnamefont {D.}~\bibnamefont {Wolf}}, \bibinfo {author}
  {\bibfnamefont {A.}~\bibnamefont {Vollhardt}}, \bibinfo {author}
  {\bibfnamefont {A.}~\bibnamefont {Kaba\ifmmode~\check{s}\else \v{s}\fi{}i}},
  \bibinfo {author} {\bibfnamefont {C.~H.}\ \bibnamefont {Lee}}, \bibinfo
  {author} {\bibfnamefont {A.}~\bibnamefont {Bilu\ifmmode \check{s}\else
  \v{s}\fi{}i\ifmmode~\acute{c}\else \'{c}\fi{}}}, \bibinfo {author}
  {\bibfnamefont {R.}~\bibnamefont {Thomale}}, \ and\ \bibinfo {author}
  {\bibfnamefont {T.}~\bibnamefont {Neupert}},\ }\href
  {https://link.aps.org/doi/10.1103/PhysRevResearch.2.023265} {\bibfield
  {journal} {\bibinfo  {journal} {Phys. Rev. Res.}\ }\textbf {\bibinfo {volume}
  {2}},\ \bibinfo {pages} {023265} (\bibinfo {year} {2020})}\BibitemShut
  {NoStop}%
\bibitem [{\citenamefont {Deng}\ \emph {et~al.}(2022)\citenamefont {Deng},
  \citenamefont {Chen},\ and\ \citenamefont {Zhang}}]{Prr4033109}%
  \BibitemOpen
  \bibfield  {author} {\bibinfo {author} {\bibfnamefont {W.}~\bibnamefont
  {Deng}}, \bibinfo {author} {\bibfnamefont {T.}~\bibnamefont {Chen}}, \ and\
  \bibinfo {author} {\bibfnamefont {X.}~\bibnamefont {Zhang}},\ }\href
  {https://link.aps.org/doi/10.1103/PhysRevResearch.4.033109} {\bibfield
  {journal} {\bibinfo  {journal} {Phys. Rev. Res.}\ }\textbf {\bibinfo {volume}
  {4}},\ \bibinfo {pages} {033109} (\bibinfo {year} {2022})}\BibitemShut
  {NoStop}%
\bibitem [{\citenamefont {Shang}\ \emph {et~al.}(2022)\citenamefont {Shang},
  \citenamefont {Liu}, \citenamefont {Shao}, \citenamefont {Han}, \citenamefont
  {Zang}, \citenamefont {Zhang}, \citenamefont {Salama}, \citenamefont {Gao},
  \citenamefont {Lee}, \citenamefont {Thomale}, \citenamefont {Manchon},
  \citenamefont {Zhang}, \citenamefont {Cui},\ and\ \citenamefont
  {Schwingenschlögl}}]{shang-tie}%
  \BibitemOpen
  \bibfield  {author} {\bibinfo {author} {\bibfnamefont {C.}~\bibnamefont
  {Shang}}, \bibinfo {author} {\bibfnamefont {S.}~\bibnamefont {Liu}}, \bibinfo
  {author} {\bibfnamefont {R.}~\bibnamefont {Shao}}, \bibinfo {author}
  {\bibfnamefont {P.}~\bibnamefont {Han}}, \bibinfo {author} {\bibfnamefont
  {X.}~\bibnamefont {Zang}}, \bibinfo {author} {\bibfnamefont {X.}~\bibnamefont
  {Zhang}}, \bibinfo {author} {\bibfnamefont {K.}~\bibnamefont {Salama}},
  \bibinfo {author} {\bibfnamefont {W.}~\bibnamefont {Gao}}, \bibinfo {author}
  {\bibfnamefont {C.~H.}\ \bibnamefont {Lee}}, \bibinfo {author} {\bibfnamefont
  {R.}~\bibnamefont {Thomale}}, \bibinfo {author} {\bibfnamefont
  {A.}~\bibnamefont {Manchon}}, \bibinfo {author} {\bibfnamefont
  {S.}~\bibnamefont {Zhang}}, \bibinfo {author} {\bibfnamefont
  {T.}~\bibnamefont {Cui}}, \ and\ \bibinfo {author} {\bibfnamefont
  {U.}~\bibnamefont {Schwingenschlögl}},\ }\href
  {https://doi.org/10.1002/advs.202202922} {\bibfield  {journal} {\bibinfo
  {journal} {Adv. Sci.}\ }\textbf {\bibinfo {volume} {9}},\ \bibinfo {pages}
  {2202922} (\bibinfo {year} {2022})}\BibitemShut {NoStop}%
\bibitem [{\citenamefont {Zhang}\ \emph {et~al.}(2023)\citenamefont {Zhang},
  \citenamefont {Chen}, \citenamefont {Li}, \citenamefont {Lee},\ and\
  \citenamefont {Zhang}}]{Prb107085426}%
  \BibitemOpen
  \bibfield  {author} {\bibinfo {author} {\bibfnamefont {H.}~\bibnamefont
  {Zhang}}, \bibinfo {author} {\bibfnamefont {T.}~\bibnamefont {Chen}},
  \bibinfo {author} {\bibfnamefont {L.}~\bibnamefont {Li}}, \bibinfo {author}
  {\bibfnamefont {C.~H.}\ \bibnamefont {Lee}}, \ and\ \bibinfo {author}
  {\bibfnamefont {X.}~\bibnamefont {Zhang}},\ }\href
  {https://link.aps.org/doi/10.1103/PhysRevB.107.085426} {\bibfield  {journal}
  {\bibinfo  {journal} {Phys. Rev. B}\ }\textbf {\bibinfo {volume} {107}},\
  \bibinfo {pages} {085426} (\bibinfo {year} {2023})}\BibitemShut {NoStop}%
\bibitem [{\citenamefont {Su}\ \emph {et~al.}(2023{\natexlab{a}})\citenamefont
  {Su}, \citenamefont {Guo}, \citenamefont {Wang}, \citenamefont {Li},
  \citenamefont {Ruan}, \citenamefont {Du}, \citenamefont {Chen},\ and\
  \citenamefont {Zheng}}]{osdbe}%
  \BibitemOpen
  \bibfield  {author} {\bibinfo {author} {\bibfnamefont {L.}~\bibnamefont
  {Su}}, \bibinfo {author} {\bibfnamefont {C.-X.}\ \bibnamefont {Guo}},
  \bibinfo {author} {\bibfnamefont {Y.}~\bibnamefont {Wang}}, \bibinfo {author}
  {\bibfnamefont {L.}~\bibnamefont {Li}}, \bibinfo {author} {\bibfnamefont
  {X.}~\bibnamefont {Ruan}}, \bibinfo {author} {\bibfnamefont {Y.}~\bibnamefont
  {Du}}, \bibinfo {author} {\bibfnamefont {S.}~\bibnamefont {Chen}}, \ and\
  \bibinfo {author} {\bibfnamefont {D.}~\bibnamefont {Zheng}},\ }\href
  {\doibase 10.1088/1674-1056/aca9c4} {\bibfield  {journal} {\bibinfo
  {journal} {Chinese Phys. B}\ }\textbf {\bibinfo {volume} {32}},\ \bibinfo
  {pages} {038401} (\bibinfo {year} {2023}{\natexlab{a}})}\BibitemShut
  {NoStop}%
\bibitem [{\citenamefont {Zhu}\ \emph {et~al.}(2023)\citenamefont {Zhu},
  \citenamefont {Sun}, \citenamefont {Hughes},\ and\ \citenamefont
  {Bahl}}]{higherank}%
  \BibitemOpen
  \bibfield  {author} {\bibinfo {author} {\bibfnamefont {P.}~\bibnamefont
  {Zhu}}, \bibinfo {author} {\bibfnamefont {X.-Q.}\ \bibnamefont {Sun}},
  \bibinfo {author} {\bibfnamefont {T.}~\bibnamefont {Hughes}}, \ and\ \bibinfo
  {author} {\bibfnamefont {G.}~\bibnamefont {Bahl}},\ }\href
  {https://doi.org/10.1038/s41467-023-36130-x} {\bibfield  {journal} {\bibinfo
  {journal} {Nat. Commun.}\ }\textbf {\bibinfo {volume} {14}},\ \bibinfo
  {pages} {720} (\bibinfo {year} {2023})}\BibitemShut {NoStop}%
\bibitem [{\citenamefont {Liu}\ \emph {et~al.}(2023)\citenamefont {Liu},
  \citenamefont {Li}, \citenamefont {Yang}, \citenamefont {Shen}, \citenamefont
  {Yang}, \citenamefont {Hang},\ and\ \citenamefont {Ezawa}}]{Prr5043034}%
  \BibitemOpen
  \bibfield  {author} {\bibinfo {author} {\bibfnamefont {B.}~\bibnamefont
  {Liu}}, \bibinfo {author} {\bibfnamefont {Y.}~\bibnamefont {Li}}, \bibinfo
  {author} {\bibfnamefont {B.}~\bibnamefont {Yang}}, \bibinfo {author}
  {\bibfnamefont {X.}~\bibnamefont {Shen}}, \bibinfo {author} {\bibfnamefont
  {Y.}~\bibnamefont {Yang}}, \bibinfo {author} {\bibfnamefont {Z.~H.}\
  \bibnamefont {Hang}}, \ and\ \bibinfo {author} {\bibfnamefont
  {M.}~\bibnamefont {Ezawa}},\ }\href
  {https://link.aps.org/doi/10.1103/PhysRevResearch.5.043034} {\bibfield
  {journal} {\bibinfo  {journal} {Phys. Rev. Res.}\ }\textbf {\bibinfo {volume}
  {5}},\ \bibinfo {pages} {043034} (\bibinfo {year} {2023})}\BibitemShut
  {NoStop}%
\bibitem [{\citenamefont {Tang}\ \emph {et~al.}(2023)\citenamefont {Tang},
  \citenamefont {Yang}, \citenamefont {Song}, \citenamefont {Yao},
  \citenamefont {Yan},\ and\ \citenamefont {Cao}}]{Prb108035410}%
  \BibitemOpen
  \bibfield  {author} {\bibinfo {author} {\bibfnamefont {C.}~\bibnamefont
  {Tang}}, \bibinfo {author} {\bibfnamefont {H.}~\bibnamefont {Yang}}, \bibinfo
  {author} {\bibfnamefont {L.}~\bibnamefont {Song}}, \bibinfo {author}
  {\bibfnamefont {X.}~\bibnamefont {Yao}}, \bibinfo {author} {\bibfnamefont
  {P.}~\bibnamefont {Yan}}, \ and\ \bibinfo {author} {\bibfnamefont
  {Y.}~\bibnamefont {Cao}},\ }\href
  {https://link.aps.org/doi/10.1103/PhysRevB.108.035410} {\bibfield  {journal}
  {\bibinfo  {journal} {Phys. Rev. B}\ }\textbf {\bibinfo {volume} {108}},\
  \bibinfo {pages} {035410} (\bibinfo {year} {2023})}\BibitemShut {NoStop}%
\bibitem [{\citenamefont {Su}\ \emph {et~al.}(2023{\natexlab{b}})\citenamefont
  {Su}, \citenamefont {Jiang}, \citenamefont {Wang}, \citenamefont {Chen},\
  and\ \citenamefont {Zheng}}]{Prb107184108}%
  \BibitemOpen
  \bibfield  {author} {\bibinfo {author} {\bibfnamefont {L.}~\bibnamefont
  {Su}}, \bibinfo {author} {\bibfnamefont {H.}~\bibnamefont {Jiang}}, \bibinfo
  {author} {\bibfnamefont {Z.}~\bibnamefont {Wang}}, \bibinfo {author}
  {\bibfnamefont {S.}~\bibnamefont {Chen}}, \ and\ \bibinfo {author}
  {\bibfnamefont {D.}~\bibnamefont {Zheng}},\ }\href
  {https://link.aps.org/doi/10.1103/PhysRevB.107.184108} {\bibfield  {journal}
  {\bibinfo  {journal} {Phys. Rev. B}\ }\textbf {\bibinfo {volume} {107}},\
  \bibinfo {pages} {184108} (\bibinfo {year} {2023}{\natexlab{b}})}\BibitemShut
  {NoStop}%
\bibitem [{\citenamefont {Jin}\ \emph {et~al.}()\citenamefont {Jin},
  \citenamefont {Liu}, \citenamefont {Wang}, \citenamefont {Zhang},
  \citenamefont {Huang}, \citenamefont {Wei}, \citenamefont {Ju}, \citenamefont
  {Liu}, \citenamefont {Yang},\ and\ \citenamefont {Nori}}]{jin2024}%
  \BibitemOpen
  \bibfield  {author} {\bibinfo {author} {\bibfnamefont {W.-W.}\ \bibnamefont
  {Jin}}, \bibinfo {author} {\bibfnamefont {J.}~\bibnamefont {Liu}}, \bibinfo
  {author} {\bibfnamefont {X.}~\bibnamefont {Wang}}, \bibinfo {author}
  {\bibfnamefont {Y.-R.}\ \bibnamefont {Zhang}}, \bibinfo {author}
  {\bibfnamefont {X.}~\bibnamefont {Huang}}, \bibinfo {author} {\bibfnamefont
  {X.}~\bibnamefont {Wei}}, \bibinfo {author} {\bibfnamefont {W.}~\bibnamefont
  {Ju}}, \bibinfo {author} {\bibfnamefont {T.}~\bibnamefont {Liu}}, \bibinfo
  {author} {\bibfnamefont {Z.}~\bibnamefont {Yang}}, \ and\ \bibinfo {author}
  {\bibfnamefont {F.}~\bibnamefont {Nori}},\ }\href
  {https://arxiv.org/abs/2311.03777} {\ }\Eprint
  {http://arxiv.org/abs/2311.03777} {arXiv:2311.03777} \BibitemShut {NoStop}%
\bibitem [{\citenamefont {Wang}\ \emph {et~al.}(2022)\citenamefont {Wang},
  \citenamefont {Wang},\ and\ \citenamefont {Ma}}]{mgc1}%
  \BibitemOpen
  \bibfield  {author} {\bibinfo {author} {\bibfnamefont {W.}~\bibnamefont
  {Wang}}, \bibinfo {author} {\bibfnamefont {X.}~\bibnamefont {Wang}}, \ and\
  \bibinfo {author} {\bibfnamefont {G.}~\bibnamefont {Ma}},\ }\href
  {https://doi.org/10.1038/s41586-022-04929-1} {\bibfield  {journal} {\bibinfo
  {journal} {Nature}\ }\textbf {\bibinfo {volume} {608}},\ \bibinfo {pages}
  {50} (\bibinfo {year} {2022})}\BibitemShut {NoStop}%
\bibitem [{\citenamefont {Scheibner}\ \emph {et~al.}(2020)\citenamefont
  {Scheibner}, \citenamefont {Irvine},\ and\ \citenamefont
  {Vitelli}}]{active-elastic}%
  \BibitemOpen
  \bibfield  {author} {\bibinfo {author} {\bibfnamefont {C.}~\bibnamefont
  {Scheibner}}, \bibinfo {author} {\bibfnamefont {W.~T.~M.}\ \bibnamefont
  {Irvine}}, \ and\ \bibinfo {author} {\bibfnamefont {V.}~\bibnamefont
  {Vitelli}},\ }\href {https://link.aps.org/doi/10.1103/PhysRevLett.125.118001}
  {\bibfield  {journal} {\bibinfo  {journal} {Phys. Rev. Lett.}\ }\textbf
  {\bibinfo {volume} {125}},\ \bibinfo {pages} {118001} (\bibinfo {year}
  {2020})}\BibitemShut {NoStop}%
\bibitem [{\citenamefont {Ghatak}\ \emph {et~al.}(2020)\citenamefont {Ghatak},
  \citenamefont {Brandenbourger}, \citenamefont {Wezel},\ and\ \citenamefont
  {Coulais}}]{ghatak}%
  \BibitemOpen
  \bibfield  {author} {\bibinfo {author} {\bibfnamefont {A.}~\bibnamefont
  {Ghatak}}, \bibinfo {author} {\bibfnamefont {M.}~\bibnamefont
  {Brandenbourger}}, \bibinfo {author} {\bibfnamefont {J.}~\bibnamefont
  {Wezel}}, \ and\ \bibinfo {author} {\bibfnamefont {C.}~\bibnamefont
  {Coulais}},\ }\href {\doibase 10.1073/pnas.2010580117} {\bibfield  {journal}
  {\bibinfo  {journal} {Proc. Natl. Acad. Sci. U.S.A.}\ }\textbf {\bibinfo
  {volume} {117}},\ \bibinfo {pages} {29561} (\bibinfo {year}
  {2020})}\BibitemShut {NoStop}%
\bibitem [{\citenamefont {Brandenbourger}\ \emph {et~al.}(2019)\citenamefont
  {Brandenbourger}, \citenamefont {Locsin}, \citenamefont {Lerner},\ and\
  \citenamefont {Coulais}}]{branden}%
  \BibitemOpen
  \bibfield  {author} {\bibinfo {author} {\bibfnamefont {M.}~\bibnamefont
  {Brandenbourger}}, \bibinfo {author} {\bibfnamefont {X.}~\bibnamefont
  {Locsin}}, \bibinfo {author} {\bibfnamefont {E.}~\bibnamefont {Lerner}}, \
  and\ \bibinfo {author} {\bibfnamefont {C.}~\bibnamefont {Coulais}},\ }\href
  {\doibase 10.1038/s41467-019-12599-3} {\bibfield  {journal} {\bibinfo
  {journal} {Nat. Commun.}\ }\textbf {\bibinfo {volume} {10}},\ \bibinfo
  {pages} {1} (\bibinfo {year} {2019})}\BibitemShut {NoStop}%
\bibitem [{\citenamefont {Chen}\ \emph {et~al.}(2021)\citenamefont {Chen},
  \citenamefont {Li}, \citenamefont {Scheibner}, \citenamefont {Vitelli},\ and\
  \citenamefont {Huang}}]{cyy}%
  \BibitemOpen
  \bibfield  {author} {\bibinfo {author} {\bibfnamefont {Y.}~\bibnamefont
  {Chen}}, \bibinfo {author} {\bibfnamefont {X.}~\bibnamefont {Li}}, \bibinfo
  {author} {\bibfnamefont {C.}~\bibnamefont {Scheibner}}, \bibinfo {author}
  {\bibfnamefont {V.}~\bibnamefont {Vitelli}}, \ and\ \bibinfo {author}
  {\bibfnamefont {G.}~\bibnamefont {Huang}},\ }\href
  {https://doi.org/10.1038/s41467-021-26034-z} {\bibfield  {journal} {\bibinfo
  {journal} {Nat. Commun.}\ }\textbf {\bibinfo {volume} {12}},\ \bibinfo
  {pages} {5935} (\bibinfo {year} {2021})}\BibitemShut {NoStop}%
\bibitem [{\citenamefont {Li}\ \emph {et~al.}(2024)\citenamefont {Li},
  \citenamefont {Wang}, \citenamefont {Wang}, \citenamefont {Lin},
  \citenamefont {Ma},\ and\ \citenamefont {Jiang}}]{mgc2}%
  \BibitemOpen
  \bibfield  {author} {\bibinfo {author} {\bibfnamefont {Z.}~\bibnamefont
  {Li}}, \bibinfo {author} {\bibfnamefont {L.-W.}\ \bibnamefont {Wang}},
  \bibinfo {author} {\bibfnamefont {X.}~\bibnamefont {Wang}}, \bibinfo {author}
  {\bibfnamefont {Z.-K.}\ \bibnamefont {Lin}}, \bibinfo {author} {\bibfnamefont
  {G.}~\bibnamefont {Ma}}, \ and\ \bibinfo {author} {\bibfnamefont {J.-H.}\
  \bibnamefont {Jiang}},\ }\href {https://doi.org/10.1038/s41467-024-50776-1}
  {\bibfield  {journal} {\bibinfo  {journal} {Nat. Commun.}\ }\textbf {\bibinfo
  {volume} {15}},\ \bibinfo {pages} {6544} (\bibinfo {year}
  {2024})}\BibitemShut {NoStop}%
\bibitem [{\citenamefont {Wang}\ \emph
  {et~al.}(2023{\natexlab{a}})\citenamefont {Wang}, \citenamefont {Hu},
  \citenamefont {Wang}, \citenamefont {Ma},\ and\ \citenamefont {Ding}}]{mgc3}%
  \BibitemOpen
  \bibfield  {author} {\bibinfo {author} {\bibfnamefont {W.}~\bibnamefont
  {Wang}}, \bibinfo {author} {\bibfnamefont {M.}~\bibnamefont {Hu}}, \bibinfo
  {author} {\bibfnamefont {X.}~\bibnamefont {Wang}}, \bibinfo {author}
  {\bibfnamefont {G.}~\bibnamefont {Ma}}, \ and\ \bibinfo {author}
  {\bibfnamefont {K.}~\bibnamefont {Ding}},\ }\href
  {https://link.aps.org/doi/10.1103/PhysRevLett.131.207201} {\bibfield
  {journal} {\bibinfo  {journal} {Phys. Rev. Lett.}\ }\textbf {\bibinfo
  {volume} {131}},\ \bibinfo {pages} {207201} (\bibinfo {year}
  {2023}{\natexlab{a}})}\BibitemShut {NoStop}%
\bibitem [{\citenamefont {Wang}\ \emph
  {et~al.}(2023{\natexlab{b}})\citenamefont {Wang}, \citenamefont {Meng},\ and\
  \citenamefont {Chen}}]{sciadvadf7299}%
  \BibitemOpen
  \bibfield  {author} {\bibinfo {author} {\bibfnamefont {A.}~\bibnamefont
  {Wang}}, \bibinfo {author} {\bibfnamefont {Z.}~\bibnamefont {Meng}}, \ and\
  \bibinfo {author} {\bibfnamefont {C.~Q.}\ \bibnamefont {Chen}},\ }\href
  {https://www.science.org/doi/abs/10.1126/sciadv.adf7299} {\bibfield
  {journal} {\bibinfo  {journal} {Sci. Adv.}\ }\textbf {\bibinfo {volume}
  {9}},\ \bibinfo {pages} {eadf7299} (\bibinfo {year}
  {2023}{\natexlab{b}})}\BibitemShut {NoStop}%
\bibitem [{FT1()}]{FT1}%
  \BibitemOpen
  \href@noop {} {}\bibinfo {note} {A concrete model will be introduced in the
  following contents, i.e., Eq.~(\ref{E2}) with
  $\gamma_1=\gamma_2$.}\BibitemShut {Stop}%
\bibitem [{FT2()}]{FT2}%
  \BibitemOpen
  \href@noop {} {}\bibinfo {note} {The reason is that since the non-Hermitian
  term is an identical matrix, it does not exhibit NHSE. On the other hand,
  since the Hamiltonian breaks the time-reversal symmetry and inversion
  symmetry, this system will also exhibit nonreciprocal correlations or
  dynamics. For a more detailed discussion, please refer to Appendix
  A.}\BibitemShut {Stop}%
\bibitem [{FT3()}]{FT3}%
  \BibitemOpen
  \href@noop {} {}\bibinfo {note} {For a more detailed discussion, please refer
  to Appendix A.}\BibitemShut {Stop}%
\bibitem [{FT4()}]{FT4}%
  \BibitemOpen
  \href@noop {} {}\bibinfo {note} {Here the complex frequency Green's function
  is defined as $G(\omega_{c})=1/(\omega_{c}-H_{\mathrm{nH}})$, where
  $H_{\mathrm{nH}}$ is given by Eq.~(\ref{E2}).}\BibitemShut {Stop}%
\bibitem [{SM()}]{SM}%
  \BibitemOpen
  \href@noop {} {}\bibinfo {note} {See Supplementary Material.}\BibitemShut
  {Stop}%
\bibitem [{\citenamefont {Haus}\ and\ \citenamefont {Huang}(1991)}]{Haus}%
  \BibitemOpen
  \bibfield  {author} {\bibinfo {author} {\bibfnamefont {H.}~\bibnamefont
  {Haus}}\ and\ \bibinfo {author} {\bibfnamefont {W.}~\bibnamefont {Huang}},\
  }\href {\doibase 10.1109/5.104225} {\bibfield  {journal} {\bibinfo  {journal}
  {Proc. IEEE.}\ }\textbf {\bibinfo {volume} {79}},\ \bibinfo {pages} {1505}
  (\bibinfo {year} {1991})}\BibitemShut {NoStop}%
\bibitem [{\citenamefont {Huang}(1994)}]{Huang94}%
  \BibitemOpen
  \bibfield  {author} {\bibinfo {author} {\bibfnamefont {W.-P.}\ \bibnamefont
  {Huang}},\ }\href {\doibase 10.1364/JOSAA.11.000963} {\bibfield  {journal}
  {\bibinfo  {journal} {J. Opt. Soc. Am. A}\ }\textbf {\bibinfo {volume}
  {11}},\ \bibinfo {pages} {963} (\bibinfo {year} {1994})}\BibitemShut
  {NoStop}%
\bibitem [{\citenamefont {Li}\ \emph {et~al.}(2010)\citenamefont {Li},
  \citenamefont {Wang}, \citenamefont {Su}, \citenamefont {Yan},\ and\
  \citenamefont {Qiu}}]{Li10}%
  \BibitemOpen
  \bibfield  {author} {\bibinfo {author} {\bibfnamefont {Q.}~\bibnamefont
  {Li}}, \bibinfo {author} {\bibfnamefont {T.}~\bibnamefont {Wang}}, \bibinfo
  {author} {\bibfnamefont {Y.}~\bibnamefont {Su}}, \bibinfo {author}
  {\bibfnamefont {M.}~\bibnamefont {Yan}}, \ and\ \bibinfo {author}
  {\bibfnamefont {M.}~\bibnamefont {Qiu}},\ }\href
  {https://opg.optica.org/oe/abstract.cfm?URI=oe-18-8-8367} {\bibfield
  {journal} {\bibinfo  {journal} {Opt. Express}\ }\textbf {\bibinfo {volume}
  {18}},\ \bibinfo {pages} {8367} (\bibinfo {year} {2010})}\BibitemShut
  {NoStop}%
\bibitem [{\citenamefont {Fan}\ \emph {et~al.}(2003)\citenamefont {Fan},
  \citenamefont {Suh},\ and\ \citenamefont {Joannopoulos}}]{Fan03}%
  \BibitemOpen
  \bibfield  {author} {\bibinfo {author} {\bibfnamefont {S.}~\bibnamefont
  {Fan}}, \bibinfo {author} {\bibfnamefont {W.}~\bibnamefont {Suh}}, \ and\
  \bibinfo {author} {\bibfnamefont {J.~D.}\ \bibnamefont {Joannopoulos}},\
  }\href {\doibase 10.1364/JOSAA.20.000569} {\bibfield  {journal} {\bibinfo
  {journal} {J. Opt. Soc. Am. A}\ }\textbf {\bibinfo {volume} {20}},\ \bibinfo
  {pages} {569} (\bibinfo {year} {2003})}\BibitemShut {NoStop}%
\bibitem [{\citenamefont {Suh}\ \emph {et~al.}(2004)\citenamefont {Suh},
  \citenamefont {Wang},\ and\ \citenamefont {Fan}}]{Fan04}%
  \BibitemOpen
  \bibfield  {author} {\bibinfo {author} {\bibfnamefont {W.}~\bibnamefont
  {Suh}}, \bibinfo {author} {\bibfnamefont {Z.}~\bibnamefont {Wang}}, \ and\
  \bibinfo {author} {\bibfnamefont {S.}~\bibnamefont {Fan}},\ }\href
  {https://api.semanticscholar.org/CorpusID:43775509} {\bibfield  {journal}
  {\bibinfo  {journal} {IEEE J. Quantum Electron.}\ }\textbf {\bibinfo {volume}
  {40}},\ \bibinfo {pages} {1511} (\bibinfo {year} {2004})}\BibitemShut
  {NoStop}%
\bibitem [{\citenamefont {Kozii}\ and\ \citenamefont {Fu}(2024)}]{Koziiprb}%
  \BibitemOpen
  \bibfield  {author} {\bibinfo {author} {\bibfnamefont {V.}~\bibnamefont
  {Kozii}}\ and\ \bibinfo {author} {\bibfnamefont {L.}~\bibnamefont {Fu}},\
  }\href {https://link.aps.org/doi/10.1103/PhysRevB.109.235139} {\bibfield
  {journal} {\bibinfo  {journal} {Phys. Rev. B}\ }\textbf {\bibinfo {volume}
  {109}},\ \bibinfo {pages} {235139} (\bibinfo {year} {2024})}\BibitemShut
  {NoStop}%
\bibitem [{\citenamefont {Nagai}\ \emph {et~al.}(2020)\citenamefont {Nagai},
  \citenamefont {Qi}, \citenamefont {Isobe}, \citenamefont {Kozii},\ and\
  \citenamefont {Fu}}]{Prldmft}%
  \BibitemOpen
  \bibfield  {author} {\bibinfo {author} {\bibfnamefont {Y.}~\bibnamefont
  {Nagai}}, \bibinfo {author} {\bibfnamefont {Y.}~\bibnamefont {Qi}}, \bibinfo
  {author} {\bibfnamefont {H.}~\bibnamefont {Isobe}}, \bibinfo {author}
  {\bibfnamefont {V.}~\bibnamefont {Kozii}}, \ and\ \bibinfo {author}
  {\bibfnamefont {L.}~\bibnamefont {Fu}},\ }\href
  {https://link.aps.org/doi/10.1103/PhysRevLett.125.227204} {\bibfield
  {journal} {\bibinfo  {journal} {Phys. Rev. Lett.}\ }\textbf {\bibinfo
  {volume} {125}},\ \bibinfo {pages} {227204} (\bibinfo {year}
  {2020})}\BibitemShut {NoStop}%
\bibitem [{\citenamefont {Kaneshiro}\ \emph {et~al.}(2023)\citenamefont
  {Kaneshiro}, \citenamefont {Yoshida},\ and\ \citenamefont
  {Peters}}]{PrbKaneshiro}%
  \BibitemOpen
  \bibfield  {author} {\bibinfo {author} {\bibfnamefont {S.}~\bibnamefont
  {Kaneshiro}}, \bibinfo {author} {\bibfnamefont {T.}~\bibnamefont {Yoshida}},
  \ and\ \bibinfo {author} {\bibfnamefont {R.}~\bibnamefont {Peters}},\ }\href
  {https://link.aps.org/doi/10.1103/PhysRevB.107.195149} {\bibfield  {journal}
  {\bibinfo  {journal} {Phys. Rev. B}\ }\textbf {\bibinfo {volume} {107}},\
  \bibinfo {pages} {195149} (\bibinfo {year} {2023})}\BibitemShut {NoStop}%
\bibitem [{\citenamefont {Geng}\ \emph {et~al.}(2023)\citenamefont {Geng},
  \citenamefont {Wei}, \citenamefont {Zou}, \citenamefont {Sheng},
  \citenamefont {Chen},\ and\ \citenamefont {Xing}}]{cw1}%
  \BibitemOpen
  \bibfield  {author} {\bibinfo {author} {\bibfnamefont {H.}~\bibnamefont
  {Geng}}, \bibinfo {author} {\bibfnamefont {J.~Y.}\ \bibnamefont {Wei}},
  \bibinfo {author} {\bibfnamefont {M.~H.}\ \bibnamefont {Zou}}, \bibinfo
  {author} {\bibfnamefont {L.}~\bibnamefont {Sheng}}, \bibinfo {author}
  {\bibfnamefont {W.}~\bibnamefont {Chen}}, \ and\ \bibinfo {author}
  {\bibfnamefont {D.~Y.}\ \bibnamefont {Xing}},\ }\href
  {https://link.aps.org/doi/10.1103/PhysRevB.107.035306} {\bibfield  {journal}
  {\bibinfo  {journal} {Phys. Rev. B}\ }\textbf {\bibinfo {volume} {107}},\
  \bibinfo {pages} {035306} (\bibinfo {year} {2023})}\BibitemShut {NoStop}%
\bibitem [{\citenamefont {Shao}\ \emph {et~al.}(2024)\citenamefont {Shao},
  \citenamefont {Geng}, \citenamefont {Liu}, \citenamefont {Lado},
  \citenamefont {Chen},\ and\ \citenamefont {Xing}}]{Prlshaokai}%
  \BibitemOpen
  \bibfield  {author} {\bibinfo {author} {\bibfnamefont {K.}~\bibnamefont
  {Shao}}, \bibinfo {author} {\bibfnamefont {H.}~\bibnamefont {Geng}}, \bibinfo
  {author} {\bibfnamefont {E.}~\bibnamefont {Liu}}, \bibinfo {author}
  {\bibfnamefont {J.~L.}\ \bibnamefont {Lado}}, \bibinfo {author}
  {\bibfnamefont {W.}~\bibnamefont {Chen}}, \ and\ \bibinfo {author}
  {\bibfnamefont {D.~Y.}\ \bibnamefont {Xing}},\ }\href
  {https://link.aps.org/doi/10.1103/PhysRevLett.132.156301} {\bibfield
  {journal} {\bibinfo  {journal} {Phys. Rev. Lett.}\ }\textbf {\bibinfo
  {volume} {132}},\ \bibinfo {pages} {156301} (\bibinfo {year}
  {2024})}\BibitemShut {NoStop}%
\bibitem [{\citenamefont {Fang}\ \emph {et~al.}(2023)\citenamefont {Fang},
  \citenamefont {Fang},\ and\ \citenamefont {Zhang}}]{Prb108165}%
  \BibitemOpen
  \bibfield  {author} {\bibinfo {author} {\bibfnamefont {Z.}~\bibnamefont
  {Fang}}, \bibinfo {author} {\bibfnamefont {C.}~\bibnamefont {Fang}}, \ and\
  \bibinfo {author} {\bibfnamefont {K.}~\bibnamefont {Zhang}},\ }\href
  {\doibase 10.1103/PhysRevB.108.165132} {\bibfield  {journal} {\bibinfo
  {journal} {Phys. Rev. B}\ }\textbf {\bibinfo {volume} {108}},\ \bibinfo
  {pages} {165132} (\bibinfo {year} {2023})}\BibitemShut {NoStop}%
\bibitem [{\citenamefont {Brody}(2013)}]{brody}%
  \BibitemOpen
  \bibfield  {author} {\bibinfo {author} {\bibfnamefont {D.}~\bibnamefont
  {Brody}},\ }\href {\doibase 10.1088/1751-8113/47/3/035305} {\bibfield
  {journal} {\bibinfo  {journal} {J. Phys. A: Math. Theor.}\ }\textbf {\bibinfo
  {volume} {47}},\ \bibinfo {pages} {035305} (\bibinfo {year}
  {2013})}\BibitemShut {NoStop}%
\bibitem [{\citenamefont {Fang}\ \emph {et~al.}(2022)\citenamefont {Fang},
  \citenamefont {Hu}, \citenamefont {Zhou},\ and\ \citenamefont
  {Ding}}]{Fang-Ding}%
  \BibitemOpen
  \bibfield  {author} {\bibinfo {author} {\bibfnamefont {Z.}~\bibnamefont
  {Fang}}, \bibinfo {author} {\bibfnamefont {M.}~\bibnamefont {Hu}}, \bibinfo
  {author} {\bibfnamefont {L.}~\bibnamefont {Zhou}}, \ and\ \bibinfo {author}
  {\bibfnamefont {K.}~\bibnamefont {Ding}},\ }\href
  {https://doi.org/10.1515/nanoph-2022-0211} {\bibfield  {journal} {\bibinfo
  {journal} {Nanophotonics}\ }\textbf {\bibinfo {volume} {11}},\ \bibinfo
  {pages} {3447} (\bibinfo {year} {2022})}\BibitemShut {NoStop}%
\bibitem [{\citenamefont {Qin}\ \emph {et~al.}(2024)\citenamefont {Qin},
  \citenamefont {Zhang},\ and\ \citenamefont {Li}}]{PrA109023317}%
  \BibitemOpen
  \bibfield  {author} {\bibinfo {author} {\bibfnamefont {Y.}~\bibnamefont
  {Qin}}, \bibinfo {author} {\bibfnamefont {K.}~\bibnamefont {Zhang}}, \ and\
  \bibinfo {author} {\bibfnamefont {L.}~\bibnamefont {Li}},\ }\href
  {https://link.aps.org/doi/10.1103/PhysRevA.109.023317} {\bibfield  {journal}
  {\bibinfo  {journal} {Phys. Rev. A}\ }\textbf {\bibinfo {volume} {109}},\
  \bibinfo {pages} {023317} (\bibinfo {year} {2024})}\BibitemShut {NoStop}%
\end{thebibliography}%


\begin{thebibliography}{4}%
\makeatletter
\providecommand \@ifxundefined [1]{%
 \@ifx{#1\undefined}
}%
\providecommand \@ifnum [1]{%
 \ifnum #1\expandafter \@firstoftwo
 \else \expandafter \@secondoftwo
 \fi
}%
\providecommand \@ifx [1]{%
 \ifx #1\expandafter \@firstoftwo
 \else \expandafter \@secondoftwo
 \fi
}%
\providecommand \natexlab [1]{#1}%
\providecommand \enquote  [1]{``#1''}%
\providecommand \bibnamefont  [1]{#1}%
\providecommand \bibfnamefont [1]{#1}%
\providecommand \citenamefont [1]{#1}%
\providecommand \href@noop [0]{\@secondoftwo}%
\providecommand \href [0]{\begingroup \@sanitize@url \@href}%
\providecommand \@href[1]{\@@startlink{#1}\@@href}%
\providecommand \@@href[1]{\endgroup#1\@@endlink}%
\providecommand \@sanitize@url [0]{\catcode `\\12\catcode `\$12\catcode
  `\&12\catcode `\#12\catcode `\^12\catcode `\_12\catcode `\%12\relax}%
\providecommand \@@startlink[1]{}%
\providecommand \@@endlink[0]{}%
\providecommand \url  [0]{\begingroup\@sanitize@url \@url }%
\providecommand \@url [1]{\endgroup\@href {#1}{\urlprefix }}%
\providecommand \urlprefix  [0]{URL }%
\providecommand \Eprint [0]{\href }%
\providecommand \doibase [0]{http://dx.doi.org/}%
\providecommand \selectlanguage [0]{\@gobble}%
\providecommand \bibinfo  [0]{\@secondoftwo}%
\providecommand \bibfield  [0]{\@secondoftwo}%
\providecommand \translation [1]{[#1]}%
\providecommand \BibitemOpen [0]{}%
\providecommand \bibitemStop [0]{}%
\providecommand \bibitemNoStop [0]{.\EOS\space}%
\providecommand \EOS [0]{\spacefactor3000\relax}%
\providecommand \BibitemShut  [1]{\csname bibitem#1\endcsname}%
\let\auto@bib@innerbib\@empty
%</preamble>
\bibitem [{\citenamefont {Xue}\ \emph {et~al.}(2021)\citenamefont {Xue},
  \citenamefont {Li}, \citenamefont {Hu}, \citenamefont {Song},\ and\
  \citenamefont {Wang}}]{GFwz1}%
  \BibitemOpen
  \bibfield  {author} {\bibinfo {author} {\bibfnamefont {W.-T.}\ \bibnamefont
  {Xue}}, \bibinfo {author} {\bibfnamefont {M.-R.}\ \bibnamefont {Li}},
  \bibinfo {author} {\bibfnamefont {Y.-M.}\ \bibnamefont {Hu}}, \bibinfo
  {author} {\bibfnamefont {F.}~\bibnamefont {Song}}, \ and\ \bibinfo {author}
  {\bibfnamefont {Z.}~\bibnamefont {Wang}},\ }\href
  {https://link.aps.org/doi/10.1103/PhysRevB.103.L241408} {\bibfield  {journal}
  {\bibinfo  {journal} {Phys. Rev. B}\ }\textbf {\bibinfo {volume} {103}},\
  \bibinfo {pages} {L241408} (\bibinfo {year} {2021})}\BibitemShut {NoStop}%
\bibitem [{\citenamefont {Hu}\ and\ \citenamefont {Wang}(2023)}]{GFwz2}%
  \BibitemOpen
  \bibfield  {author} {\bibinfo {author} {\bibfnamefont {Y.-M.}\ \bibnamefont
  {Hu}}\ and\ \bibinfo {author} {\bibfnamefont {Z.}~\bibnamefont {Wang}},\
  }\href {https://link.aps.org/doi/10.1103/PhysRevResearch.5.043073} {\bibfield
   {journal} {\bibinfo  {journal} {Phys. Rev. Res.}\ }\textbf {\bibinfo
  {volume} {5}},\ \bibinfo {pages} {043073} (\bibinfo {year}
  {2023})}\BibitemShut {NoStop}%
\bibitem [{\citenamefont {Yi}\ and\ \citenamefont {Yang}(2020)}]{Prl125186}%
  \BibitemOpen
  \bibfield  {author} {\bibinfo {author} {\bibfnamefont {Y.}~\bibnamefont
  {Yi}}\ and\ \bibinfo {author} {\bibfnamefont {Z.}~\bibnamefont {Yang}},\
  }\href {\doibase 10.1103/PhysRevLett.125.186802} {\bibfield  {journal}
  {\bibinfo  {journal} {Phys. Rev. Lett.}\ }\textbf {\bibinfo {volume} {125}},\
  \bibinfo {pages} {186802} (\bibinfo {year} {2020})}\BibitemShut {NoStop}%
\bibitem [{\citenamefont {Fang}\ \emph {et~al.}(2023)\citenamefont {Fang},
  \citenamefont {Fang},\ and\ \citenamefont {Zhang}}]{Prb108165}%
  \BibitemOpen
  \bibfield  {author} {\bibinfo {author} {\bibfnamefont {Z.}~\bibnamefont
  {Fang}}, \bibinfo {author} {\bibfnamefont {C.}~\bibnamefont {Fang}}, \ and\
  \bibinfo {author} {\bibfnamefont {K.}~\bibnamefont {Zhang}},\ }\href
  {\doibase 10.1103/PhysRevB.108.165132} {\bibfield  {journal} {\bibinfo
  {journal} {Phys. Rev. B}\ }\textbf {\bibinfo {volume} {108}},\ \bibinfo
  {pages} {165132} (\bibinfo {year} {2023})}\BibitemShut {NoStop}%
\end{thebibliography}%
\bibliographystyle{apsrev4-1}
	
\end{document}

\title{Supplementary Information for ``Complex Frequency Fingerprint: Basic Concept and Theory"}

\author{Juntao Huang$^1$}
\author{Kun Ding$^2$}
\author{Jiangping Hu$^{3,4,5}$}
\email[Corresponding author: ]{jphu@iphy.ac.cn}
\author{Zhesen Yang$^1$}
\email[Corresponding author: ]{yangzs@xmu.edu.cn}

\affiliation{$^1$ Department of Physics, Xiamen University, Xiamen 361005, Fujian Province, China}
\affiliation{$^2$ Department of Physics, State Key Laboratory of Surface Physics, and Key Laboratory of Micro and Nano Photonic Structures (Ministry of Education), Fudan University, Shanghai 200438, China}
\affiliation{$^3$ Beijing National Laboratory for Condensed Matter Physics and Institute of Physics, Chinese Academy of Sciences, Beijing 100190, China}
\affiliation{$^4$ School of Physical Sciences, University of Chinese Academy of Sciences, Beijing 100190, China}
\affiliation{$^5$ New Cornerstone Science Laboratory, Beijing, 100190, China}

%\date{\today}
%\centering

\maketitle

%\titlecontents

\tableofcontents

\newpage

\section{The non-Bloch response of Green's functions}

This section provides a concise verification that a nonzero winding number will lead to the non-Bloch response of the Green's function, aligning with the conclusions presented in Fig. 1. 

%\begin{figure}[b]
%	\begin{center}
%		\includegraphics[width=0.95\linewidth]{appendix-1.pdf}
%		\par\end{center}
%	\protect\caption{(a) $|G_{i_{B}i_{0,B}}(\omega)|$ for $\nu(\omega)=0$, satisfying $G_{\mathrm{OBC}}\simeq G_{\mathrm{PBC}}$. Here $\omega=1.3-i$ (b) $|G_{i_{B}i_{0,B}}(\omega)|$ for $\nu(\omega)=1$, satisfying $G_{\mathrm{OBC}}\neq G_{\mathrm{PBC}}$. Here $\omega=1.3-1.35i$. In (a) and (b) $i_{0}=37$, all other parameters are identical to those used in Fig.~\ref{F2}.} 
%	\label{F5}
%\end{figure}

\begin{figure}[b]
	\begin{center}
		\includegraphics[width=0.55\linewidth]{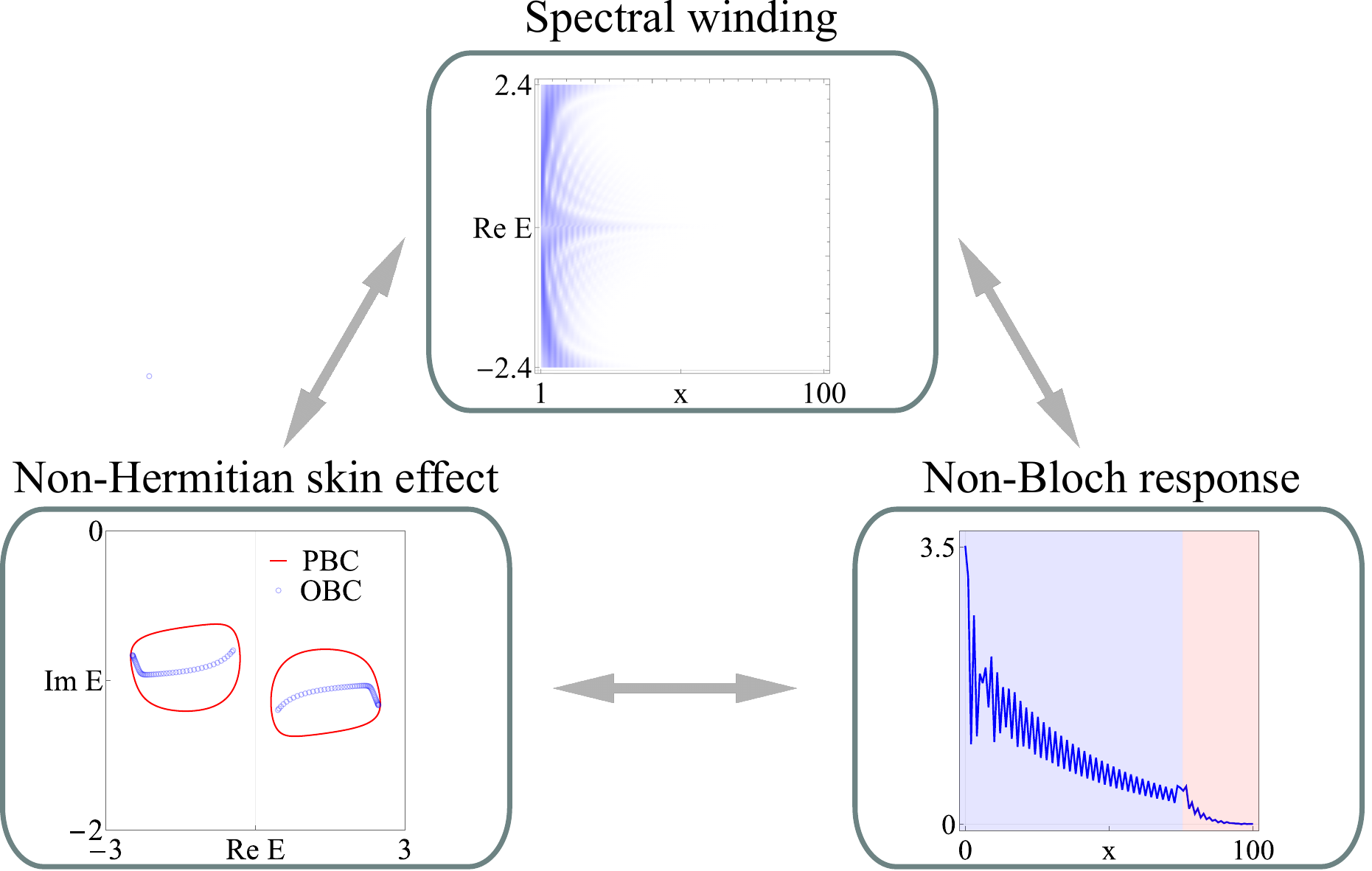}
		\par\end{center}
	\protect\caption{The schematic illustration of one-to-one correspondence between spectral winding, NHSE, and the non-Bloch response.} 
	\label{SMS1}
\end{figure}

To examine the Green's function, we introduce $\beta=e^{ik}$ and label $\beta_{n=1,\dots,2M}$ as the roots of $\mathrm{det}[\omega_{c}-H(\beta)]=\frac{P(\beta,\omega_{c})}{\beta^{M}}=0$, with $P(\beta,\omega_{c})$ denoting an algebraic polynomial for $\beta$ and $M=2$ in our one-dimensional two-band model. For a given complex frequency $\omega_{c}$, the roots can be ordered as $|\beta_{1}(\omega_{c})|\leq |\beta_{2}(\omega_{c})|\leq |\beta_{3}(\omega_{c})|\leq |\beta_{4}(\omega_{c})|$, where $\beta_{1},\beta_{2}$ are enclosed by the generalized Brillouin zone~\cite{GFwz1,GFwz2}. The scaling behavior of the Green's function under OBC is described as~\cite{GFwz1,GFwz2}
\begin{equation}
	[G^{\mathrm{OBC}}(\omega_{c})]_{i_{\alpha},i_{0,\beta}}\sim \begin{cases}
		\beta^{-(i_{0}-i)}_{3}, & i<i_{0}\\
		\beta^{i-i_{0}}_{2}, & i> i_{0}
	\end{cases}
\end{equation}
for large $|i-i_{0}| $, where $\alpha$ and $\beta$ represent orbital indices. Besides, the roots can be alternatively ordered as $|\beta_{1}(\omega_{c})|\leq\dots\leq |\beta_{a}(\omega_{c})|\leq 1\leq |\beta_{4}(\omega_{c})|$, then the result under PBC can be qualitatively characterized as~\cite{GFwz1,GFwz2}

\begin{equation}
	[G^{\mathrm{PBC}}(\omega_{c})]_{i_{\alpha},i_{0,\beta}}\sim \begin{cases}
		\beta^{-(i_{0}-i)}_{a+1}, & i<i_{0}\\
		\beta^{i-i_{0}}_{a}, & i> i_{0}
	\end{cases}\label{PBCscale}
\end{equation}
for large $|i-i_{0}|$, assuming $|i-i_{0}|$ is away from the balanced point~\cite{GFwz1}. Here, $a$ depends on the specific value of $\omega_{c}$.

The numerical result has been presented in Fig. 1, where the winding number is given by~\cite{Prl125186}
\begin{equation}
\nu(\omega_{c})=\frac{1}{2\pi i}\int^{2\pi}_{0}dk \partial_{k}\mathrm{ln}~\mathrm{det}[H_{\mathrm{nH}}(k)-\omega_{c}].
\end{equation}

When $\omega_{c}$ is outside the point gap, the roots satisfy $|\beta_{1}(\omega_{c})|\leq |\beta_{2}(\omega_{c})|< 1 < |\beta_{3}(\omega_{c})|\leq |\beta_{4}(\omega_{c})|$~\cite{GFwz1,GFwz2}. Hence, the inequalities $|\beta_{2}(\omega_{c})|<1$ and $|\beta_{3}(\omega_{c})|>1$ hold for all $\nu(\omega_{c})=0$. In this case, we conclude that the scaling behavior of OBC Green's function will exponentially decay as $i$ moves away from $i_{0}$. It is evident that the scaling behavior of $[G^{\mathrm{PBC}}(\omega_{c})]_{i_{\alpha}i_{0,\beta}}$ is identical to that of $G^{\mathrm{OBC}}(\omega_{c})$ since $a=2$ as shown in Eq.~\ref{PBCscale}. Specifically, they coalesce, i.e. $|G^{\mathrm{OBC}}(\omega_{c})|\simeq |G^{\mathrm{PBC}}(\omega_{c})|$ in the bulk as depicted in Fig. 1 (b) for $\omega_{c}=\omega_{1}$, corresponding to the Bloch response. 

When $\omega_{c}$ is inside the point gap, i.e., $\nu(\omega_{c})=1$, we can deduce that $|\beta_{3}(\omega_{c})|<1$, indicating that the OBC Green's function exhibits exponential increase as $i$ moves away from $i_{0}$ to the left, showing divergent behavior. Conversely, the PBC Green's function still exhibits decaying behavior as $i$ moves away from $i_{0}$ since now $a=3$ as shown in Eq.~\ref{PBCscale}, with $|\beta_{4}(\omega_{c})|>1$. This distinction is exemplified as shown in Fig. 1 (b) for $\omega_{c}=\omega_{2}$, i.e., $|G^{\mathrm{OBC}}(\omega_{c})|\neq |G^{\mathrm{PBC}}(\omega_{c})|$, corresponding to the non-Bloch response.

Therefore, based on the above result and the discussion in Appendix A, we can develop the one-to-one correspondence between the non-Bloch response, i.e., the divergent behavior of OBC GF, the NHSE, and the spectral winding, as illustrated in Fig.~\ref{SMS1}.

\section{Derivation of the inhomogeneous non-Hermitian Schr\"{o}dinger equation}
In this section, we provide a detailed derivation of Eq. 1 in the main text. We begin with the following quantum master equation in a driven-dissipative system:
\begin{equation}
\frac{d\hat{\rho}(t)}{dt}=-i[\hat{H}(t),\hat{\rho}(t)]+\sum\limits^{N}_{m=1}\kappa_{m}\hat{\mathcal{L}}_{m}[\hat{\rho}(t)],\label{SE4}
\end{equation}
where $\hat{H}(t)=\sum_{mn}t_{mn}\hat{a}^{\dag}_{m}\hat{a}_{n}+\sum_{m}(\hat{a}^{\dag}_{m}F_{m}(t)+F^{*}_{m}(t)\hat{a}_{m})$ is the Hermitian driven Hamiltonian with $t_{mn}=t^{*}_{nm}$,$\hat{a}_{m}^\dag$, $\hat{a}_{m}$, $\hat{\rho}(t)$, $F_m(t)$ and $N$ denoting the hopping parameter, creation and annihilation operators of the bosonic modes, density matrix, external driving field, and the total number of lattice sites including orbital degrees of freedom, respectively.
\begin{equation}
	\hat{\mathcal{L}}_{m}[\hat{\rho}(t)]=\hat{a}_m\hat{\rho}(t) \hat{a}_m^{\dag}-\frac{1}{2}\{\hat{a}_m^{\dag}\hat{a}_m,\hat{\rho}(t)\},
\end{equation}
is the dissipative superoperator, and $\kappa_{m}$ in Eq.~\ref{SE4} represents the local damping rate for each bosonic mode.
	
Following the definition of $\langle \boldsymbol{\hat{a}}(t)\rangle$ in the main text, we obtain
\begin{equation}
\begin{split}
i \frac{d\langle \hat{a}_{m}(t)\rangle}{dt}&=i\mathrm{Tr}\big[\hat{a}_{m}\frac{d\hat{\rho}(t)}{dt}\big]\\
&=i \mathrm{Tr}\big[\hat{a}_{m}\big(-i[\hat{H}(t),\hat{\rho}(t)]+\sum\limits^{N}_{j=1}\kappa_{j}\hat{\mathcal{L}}_{j}[\hat{\rho}(t)]\big)\big]\\
&=\mathrm{Tr}\big[\hat{a}_{m}[\hat{H}(t),\hat{\rho}(t)]\big]+i\mathrm{Tr}\big[\hat{a}_{m}\sum\limits_{j}\kappa_{j}\hat{\mathcal{L}}_{j}[\hat{\rho}(t)]\big].
\end{split}
\end{equation}
We then employ the identity:
\begin{equation}
\begin{split}
\mathrm{Tr}\big[\hat{a}_{m}[\hat{H}(t),\hat{\rho}(t)]\big]=\mathrm{Tr}\big[[\hat{a}_{m},\hat{H}(t)]\hat{\rho}(t)\big].
\end{split}
\end{equation}
Subsequently, utilizing the bosonic commutation relation $[\hat{a}_{i},\hat{a}^{\dag}_{j}]=\delta_{ij}$, we arrive at
\begin{equation}
[\hat{a}_{m},\hat{H}(t)]=\sum\limits_{n}t_{mn}\hat{a}_{n}+F_{m}(t).
\end{equation}
Since $\mathrm{Tr}\big[\hat{\rho}(t)\big]=1$, we deduce
\begin{equation}
\mathrm{Tr}\big[\hat{\boldsymbol{a}},[\hat{H}(t),\hat{\rho}(t)]\big]=H_{0}\langle \hat{\boldsymbol{a}}(t)\rangle+\boldsymbol{F}(t),
\end{equation}
where $[H_{0}]_{mn}=t_{mn}$ is the Hermitian Hamiltonian and $\langle \hat{\boldsymbol{a}}(t)\rangle=\{\mathrm{Tr}\big[\hat{a}_{1}\hat{\rho}(t)\big],\dots,\mathrm{Tr}\big[\hat{a}_{N}\hat{\rho}(t)\big]\}^{T}$. Then by applying the identity
\begin{equation}
\begin{split}
&\mathrm{Tr}\big[\hat{a}_{m}\big(\hat{a}_{j}\hat{\rho}(t) \hat{a}^{\dag}_{j}-\frac{1}{2}\{\hat{a}^{\dag}_{j}\hat{a}_{j},\hat{\rho}(t)\}\big)\big]\\
&=\mathrm{Tr}\big[\big(\hat{a}^{\dag}_{j}\hat{a}_{m}\hat{a}_{j}-\frac{1}{2}\{\hat{a}_{m},\hat{a}^{\dag}_{j}\hat{a}_{j}\}\big)\hat{\rho}(t)\big]\\
&=\mathrm{Tr}\big[\frac{1}{2}\big([\hat{a}^{\dag}_{j},\hat{a}_{m}]\hat{a}_{j}+\hat{a}^{\dag}_{j}[\hat{a}_{m},\hat{a}_{j}]\big)\hat{\rho}(t) \big]\\
&=-\frac{1}{2}\delta_{mj}\langle \hat{a}_{j}(t)\rangle,
\end{split}
\end{equation}
we derive
\begin{equation}
i\mathrm{Tr}\big[\hat{a}_{m}\sum\limits_{j}\kappa_{j}\hat{\mathcal{L}}_{j}[\hat{\rho}(t)]\big]=-\frac{i}{2}\kappa_{m}\langle \hat{a}_{m}(t)\rangle.
\end{equation}
Consequently, we establish
\begin{equation}
i\frac{d\langle \hat{\boldsymbol{a}}(t)\rangle}{dt}=H_{\rm nH}\langle \hat{\boldsymbol{a}}(t)\rangle +\boldsymbol{F}(t),\label{S12}
\end{equation}
with $[H_{\rm nH}]_{mn}=t_{mn}-\frac{i}{2}\kappa_{m}\delta_{mn}$ denoting the non-Hermitian Hamiltonian in the first quantized form. Notably, as written in Eq. 2 in the main text, half of the local damping rate $\kappa_{m}/2$ corresponds to the on-site dissipation terms $\gamma_{1}$ and $\gamma_{2}$.

\section{Derivation of the response function}

This section provides a brief derivation of the solution for the response function. Consider a harmonic external driving $\boldsymbol{F}_{\omega_{0}}(t)=\theta(t)e^{-i\omega_{0}t}\{F_{1}(0),\dots, F_{N}(0)\}^{T}$ with $\theta(t)$ representing the step function, then it is easy to check that the general solution of Eq.~\ref{S12} is expressed as
\begin{equation}
\langle \hat{\boldsymbol{a}}(t)\rangle=e^{-iH_{\rm nH}t}[-i\int^{t}_{0}e^{iH_{\rm nH}\tau}\boldsymbol{F}_{\omega_{0}}(\tau)d\tau+\langle \hat{\boldsymbol{a}}(0)\rangle].
\end{equation}
A straightforward integral over $\tau$ yields:
\begin{equation}
\begin{split}
\langle \hat{\boldsymbol{a}}(t)\rangle&=e^{-iH_{\rm nH}t}[-i\int^{t}_{0}e^{iH_{\rm nH}\tau}\boldsymbol{F}_{\omega_{0}}(\tau)d\tau+\langle \hat{\boldsymbol{a}}(0)\rangle]\\
&=\frac{e^{-i\omega_{0}t}-e^{-iH_{\rm nH}t}}{\omega_{0}-H_{\rm nH}}\theta(t)\{F_{1}(0),\dots,F_{N}(0)\}^{T}+e^{-iH_{\rm nH}t}\langle\hat{\boldsymbol{a}}(0)\rangle\\
&=G(\omega_{0})(1-e^{-i(H_{\rm nH}-\omega_{0})t})\boldsymbol{F}_{\omega_{0}}(t)+\langle \hat{\boldsymbol{a}}(t)\rangle_{\boldsymbol{F}=0},\\
\end{split}
\end{equation} 
where $G(\omega_{0})=\frac{1}{\omega_{0}-H_{\rm nH}}$. Subsequently, the response function is determined as:
\begin{equation}
\delta\langle \hat{\boldsymbol{a}}(t)\rangle_{\omega_{0}}=\chi_{\omega_{0}}(t)\boldsymbol{F}_{\omega_{0}}(t),\label{SE15}
\end{equation}	
with
\begin{equation}\begin{split}
	[\chi_{\omega_{0}}(t)]_{mn}&=[G(\omega_{0})]_{mn}-[G(\omega_{0})e^{-i(H_{\rm nH}-\omega_{0})t}]_{mn}. 
		\end{split}
\end{equation}

\section{Application of CFF in detecting the point gap bound state}

\begin{figure}[b]
	\begin{center}
		\includegraphics[width=0.48\linewidth]{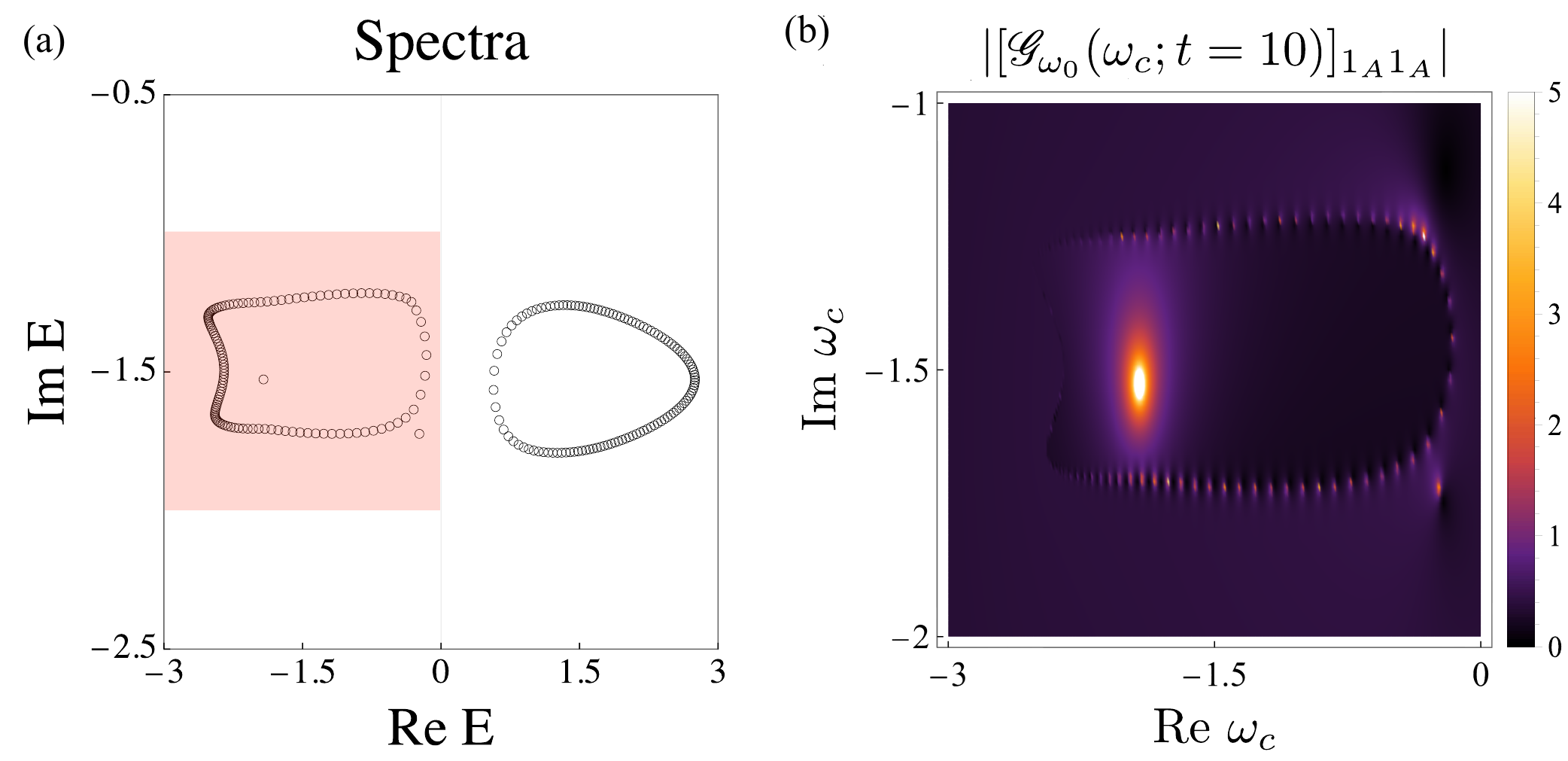}
		\par\end{center}
	\protect\caption{
		(a) The PBC spectra of Eq.~\ref{M3a} with the presence of impurities. 
		(b) The complex frequency LDOS, which is expressed as $|\mathscr{G}_{1_{A}1_{A}}(\omega;t=10)|$. 
		Here the parameters are set as $t_{1}=1,t_{2}=1.6,\mu=0.1,\gamma_{0}=3,\gamma_{z}=1,\lambda=1,t_{3}=0.5,V=4,\omega_{0}=0$ and $N=120$.} 
	\label{F4}
\end{figure}
In this section, we demonstrate that, beyond the NHSE, the CFF can also be applied to detect another observable in the non-Hermitian community, i.e., the point gap bound state~\cite{Prb108165}. This application is straightforward for a given non-Hermitian system $H_{\mathrm{nH}}$: we can detect the complex frequency local density of states (LDOS) via the CFF as follows,
\begin{equation}
	\lim_{t\rightarrow\infty}|[\mathscr{G}_{\omega_{0}}(\omega_{c} \in \mathbb{C};t)]_{ii}|=D_{ii}(\omega_{c}),
\end{equation}
where $D(\omega_{c})$ is given in Eq. 10 in the main text.
Subsequently, we will apply this method to detect the point gap bound state~\cite{Prb108165}. The non-Hermitian Hamiltonian for this example is given by 
\begin{equation}\begin{aligned}
H_{\rm nH}(k)=&(t_{1}+t_{2}\mathrm{cos}k)\sigma_{x}+t_{2}\mathrm{sin}k\sigma_{y}+(\lambda \mathrm{sin}k+\mu)\sigma_{z}+(t_{3}\mathrm{cos}2k) \sigma_{0}-i(\gamma_{0}\sigma_{0}+\gamma_z\sigma_{z})/2.
\label{M3a}
\end{aligned}\end{equation} 
Impurities are introduced as
\begin{equation}
	V_{I}=V\sum\limits_{i\alpha}\delta_{i1}|i,\alpha\rangle \langle i,\alpha|.
	\label{M3b}
\end{equation}
The OBC spectra and the complex frequency LDOS are depicted in Fig.~\ref{F4} (a) and (b), respectively, vividly demonstrating the existence of a point gap bound state on the complex frequency plane.

\section{Application of CFF in general systems}
The above derivation demonstrates that the CFF is an experimentally feasible framework for open quantum systems. Its core principle, the full matrix element measurement as detailed in the main text, is highly insightful, making the CFF a broader and more general concept. This approach inspires experimental designs across a wide range of physical platforms without imposing restrictive limitations.

The central insight of the CFF can be summarized as follows: Starting with a detectable Green’s function at frequency $\omega$, the CFF approach involves measuring all matrix elements of the Green’s function. For any system, the retarded Green’s function takes the general form:
\begin{equation}
G^{R}_{S}(\omega)=\frac{1}{\omega+i\eta-H_{S}-\Sigma_{S}(\omega+i\eta)},
\end{equation}
where $H_{S}$ is the non-interacting Hamiltonian of a system, and the self-energy $\Sigma_{S}(\omega)$ is arbitrary without limitations, encompassing diverse physical interactions such as many-body effects, disorder, subsystems' coupling, and more. Crucially, the CFF shifts focus from the poles of $G^{R}_{S}(\omega)$ to its full matrix structure. The CFF can be applied to detect the following observable:
\begin{equation}
\begin{split}
G^{\mathrm{CFF}}_{\omega_{0}}(\omega_{c})&=\frac{1}{\omega_{c}-\omega_{0}+[G^{R}_{S}(\omega_{0})]^{-1}}\\
&=\frac{1}{\omega_{c}+i\eta-H_{S}-\Sigma_{S}(\omega_{0}+i\eta)},
\end{split}
\end{equation}
which generally results in a double frequency Green's function with an additional $\omega_{0}$-dependence arising from the self-energy. This formulation coincides with steady-state results in driven-dissipative frameworks but extends beyond them, remaining applicable to arbitrary physical systems supporting the detection of Green's functions.

Furthermore, this extension, specifically the double-frequency Green's function, plays a powerful role in detecting novel responses by characterizing complete quasiparticle information across the entire complex plane. An example of many-body effects has been discussed in the main text, with its detailed calculation procedure provided in the next section.

\section{Numerical calculation of the example model for quasiparticle resolution}

This section provides a detailed analysis of Eq. 15 in the main text. First, we derive the concrete form of the effective frequency-dependent non-Hermitian Hamiltonian. The retarded Green's functions are defined as
\begin{equation}
\begin{split}
G^{R}_{S,\sigma\sigma^{\prime}}(t)&=-i\theta(t)\langle \{\hat{d}_{\sigma}(t),\hat{d}^{\dag}_{\sigma^{\prime}} \}\rangle_{T}\\
G^{R}_{BS,\sigma\sigma^{\prime}}(k,t)&=-i\theta(t)\langle \{\hat{c}_{k\sigma}(t),\hat{d}^{\dag}_{\sigma^{\prime}} \}\rangle_{T}\\
D_{\sigma\sigma^{\prime}}(t)&=-i\theta(t) \langle \{\hat{n}_{\bar{\sigma}}(t)\hat{d}_{\sigma}(t),\hat{d}^{\dag}_{\sigma^{\prime}} \}\rangle_{T}.
\end{split}
\end{equation}
Then, according to the concrete form of $\hat{H}_{sys}=\hat{H}_{S}+\hat{H}_{B}+\hat{H}_{S\text{-}B}$ introduced in the main text, the commutation relations are derived:
\begin{equation}
\begin{split}
[\hat{d}_{\sigma},\hat{H}_{sys}]&=\sum\limits_{\sigma^{\prime}}[H_{S}]_{\sigma\sigma^{\prime}}\hat{d}_{\sigma^{\prime}}+U\hat{n}_{\bar{\sigma}}\hat{d}_{\sigma}+\frac{1}{\sqrt{N}}\sum\limits_{k\sigma^{\prime}}t_{\downarrow}\delta_{\sigma \downarrow}\delta_{\sigma\sigma^{\prime}}\hat{c}_{k\sigma^{\prime}}\\
[\hat{c}_{k\sigma},\hat{H}_{sys}]&=\sum\limits_{\sigma^{\prime}}(2t_{b}\mathrm{cos}k-\mu)\delta_{\sigma\sigma^{\prime}}\hat{c}_{k\sigma^{\prime}}+\frac{1}{\sqrt{N}}\sum\limits_{\sigma^{\prime}}t^{*}_{\downarrow}\delta_{\sigma \downarrow}\delta_{\sigma\sigma^{\prime}}\hat{d}_{\sigma^{\prime}},
\end{split}
\end{equation}
which leads to the equations of motion:
\begin{equation}
\begin{split}
\omega G^{R}_{S}(\omega)&=\boldsymbol{I}+H_{S}G^{R}_{S}(\omega)+UD(\omega)+\frac{1}{\sqrt{N}}\sum\limits_{k}t_{\downarrow} P_{\downarrow} G^{R}_{BS}(k,\omega)\\
\omega G^{R}_{BS}(k,\omega)&=(2t_{b}\mathrm{cos}k-\mu)G^{R}_{BS}(k,\omega)+\frac{1}{\sqrt{N}}t^{*}_{\downarrow}P_{\downarrow} G^{R}_{S}(\omega),
\end{split}
\end{equation}
with $P_{\downarrow}=\frac{\sigma_{0}-\sigma_{z}}{2}$ indicating the projection onto the spin-down space.

Using the mean-field approach: $D_{\sigma\sigma^{\prime}}(t)\approx -i\theta(t)\langle \hat{n}_{\bar{\sigma}}\rangle_{T}\langle\{\hat{d}_{\sigma}(t),\hat{d}^{\dag}_{\sigma^{\prime}} \}\rangle_{T}=\langle \hat{n}_{\bar{\sigma}}\rangle_{T} G^{R}_{S,\sigma\sigma^{\prime}}(t)$, we obtain
\begin{equation}
G^{R}_{S}(\omega)=\frac{1}{\omega+i\eta-H_{S}-U\mathcal{D}-\frac{1}{N}\sum\limits_{k}P_{\downarrow}\frac{|t_{\downarrow}|^{2}}{\omega+i\eta-2t_{b}\mathrm{cos}k-\mu}P_{\downarrow}},
\end{equation}
where $\mathcal{D}=\mathrm{Diag}(\langle \hat{n}_{\downarrow}\rangle_{T},\langle \hat{n}_{\uparrow}\rangle_{T})$. In numerical calculations, the mean-field values are obtained self-consistently through
\begin{equation}
\begin{split}
\langle \hat{n}_{\sigma}\rangle_{T}&=-\frac{1}{\pi}\int^{\infty}_{-\infty}n_{F}(\omega)\mathrm{Im}~G^{R}_{S,\sigma\sigma}(\omega)d\omega\overset{T_{B}\rightarrow 0 \, \mathrm{K}}{=\joinrel=\joinrel=\joinrel=} -\frac{1}{\pi}\int^{0}_{-\infty}n_{F}(\omega)\mathrm{Im}~G^{R}_{S,\sigma\sigma}(\omega)d\omega,
\end{split}
\end{equation}
where $n_{F}(\omega)$ represents the Fermi-Dirac distribution. The spin-down component for the self-energy is given by
\begin{equation}
\Sigma_{\downarrow\downarrow}(\omega+i\eta)=\frac{1}{2\pi}\int^{\pi}_{-\pi}\frac{|t_{\downarrow}|^{2}}{\omega+i\eta-2t_{b}\mathrm{cos}k+\mu}dk=\frac{2\pi\mathrm{Sign}[\omega+\mu]}{\sqrt{(\omega+i\eta+\mu)^{2}-4t^{2}_{b}}},
\end{equation}
where the second equality is derived using the residue theorem by applying the contour integral, with $t_{\downarrow}=\sqrt{2\pi}$, $\mathrm{Sign}[x\geq 0]=1$ and $\mathrm{Sign}[x<0]=0$.

\bibliography{supple-ref}
\bibliographystyle{apsrev4-1}